\documentclass[12pt,preprint]{aastex}

\usepackage{graphicx}
\defcitealias{Cern00}{CGK}

\shorttitle{SMA line survey of IRC +10216}
\shortauthors{Patel N.A., Young K. H., Thaddeus, P. et al.}

\begin{document}

\title{An Interferometric Spectral-line Survey of\\ IRC+10216 in the 345~GHz Band}

\author{Nimesh A. Patel\altaffilmark{1}, Ken H. Young\altaffilmark{1},  Carl~A.~Gottlieb\altaffilmark{1}, Patrick Thaddeus\altaffilmark{1},\\
        Robert W. Wilson\altaffilmark{1},
        Karl~M.~Menten\altaffilmark{2}, Mark~J.~Reid\altaffilmark{1}, 
         Michael~C.~McCarthy\altaffilmark{1}\\ 
        Jos\'{e} Cernicharo\altaffilmark{3}, Jinhua~He\altaffilmark{4}, Sandra Br\"{u}nken\altaffilmark{5},   Dinh-V.~Trung\altaffilmark{6} and Eric~Keto\altaffilmark{1}}

\email{npatel@cfa.harvard.edu}

\altaffiltext{1}{Harvard-Smithsonian Center for Astrophysics, Cambridge, MA 02138, Contact author: N. A. Patel, email: npatel@cfa.harvard.edu}
\altaffiltext{2}{Max-Planck Institut f\"ur Radioastronomie, Auf dem H\"ugel 69,
  53121 Bonn, Germany}
  \altaffiltext{3}{Laboratory of Molecular Astrophysics, Department of Astrophysics, CAB, INTA-CSIC, Ctra de Ajalvir, km 4, 28850 Torrej\'on de Ardoz, Madrid, Spain}
  \altaffiltext{4}{Yunnan Observatory, Chinese Academy of Sciences, Kunming 650011, China}
\altaffiltext{5}{I.\ Physikalisches Institut, Universit\"at zu K\"oln,
  Z\"ulpicher Street 77, 50937 K\"oln, Germany}
\altaffiltext{6}{Academia Sinica, Institute of Astronomy \& Astrophysics, Taipei
106, Taiwan}

\slugcomment{Accepted for publication in ApJ Supplement Series.}
\begin{abstract}
We report  a spectral-line survey of the extreme
carbon star IRC+10216 carried out between 293.9 and 354.8 GHz with the Submillimeter Array.  
A total of 442 lines were detected, more than 200 for the first time; 149 are unassigned.
Maps at an angular resolution of $~\sim3''$ were obtained for each
line.  A substantial new population of narrow lines with
an expansion velocity of $\sim$4 km s$^{-1}$ (i.e. $\approx 30$\% of the terminal velocity)  was detected. Most of
these are attributed to rotational transitions within vibrationally
excited states, emitted from energy levels above the $v=0, J=0$ ground state with excitation energy of 1000--3000~K.
Emission from these lines appears to be centered on the star
with an angular extent of $<1''$.  We use multiple transitions
detected in several molecules to derive physical conditions in this
inner envelope of IRC+10216. 
\end{abstract}


\section{Introduction}

Understanding the formation of complex molecules and dust grains
in space is a major problem in modern astrophysics.
Stars on the asymptotic giant branch (AGB) efficiently produce C, N, O and s-process elements.
Mass loss leads to the formation of expanding circumstellar envelopes (CSEs) whose molecules and dust grains are major sources of replenishment of the interstellar medium \citep{herwig2005,BussoEtAl1999}.  The radius of the C/O core,  after
the exhaustion of core He burning,  is R$_{c} \approx 10^{9}$ cm, and the  
temperature is $\sim10^{8}$ K.  He burning continues
in a shell around the core and the size of the stellar photosphere
expands to R$_{*}\approx 10^{13}$ cm.  During the phase of intermittent
burning of H and He (in the so-called ``thermal pulse''), products of nuclear burning
are dredged up by convection and brought to the stellar surface.
The zone  within a few stellar radii is dynamically important for
mass-loss; in this zone molecules produced in the stellar atmosphere are moved to a region away from the star that is cool enough ($\sim 1000$~K) so that dust grains can condense. Radiation pressure on the
grains drives the CSE's expansion.  The outer circumstellar region is comparatively
less dense by several orders of magnitude and richer in molecular
gas, steadily expanding outwards with velocities $\sim$ 10 km
s$^{-1}$.  The chemistry in the outermost part of the circumstellar
shell is driven by the interstellar UV radiation \citep{glassgold1996}. Due to the clumpy nature of the shell, photochemistry may be important in the inner regions as well \citep{DecinEtAl2010}.

IRC+10216 (CW Leo) is a  well known AGB carbon star
([C]$>$[O] and presence of s-process elements) with a high mass-loss rate (several $\times 10^{-5}$
M$_{\odot}$ yr$^{-1}$) at a distance of
150 pc  \citep[e.g. ][]{YoungEtAl1993, CrosasMenten1997}.  Owing to its closeness
to the Sun, it has been possible to study the physical
and chemical processes in its large circumstellar envelope in great
detail \citep[e.g. ][]{Olofsson1999}. There are nearly 60 molecules observed
in the circumstellar shell of IRC+10216 as a result of previous
single-dish line surveys \citep{KawaguchiEtAl1995,
CernicharoEtAl2000,  AveryEtAl1992,
GroesbeckEtAl1994,HeEtAl2008, ZiurysEtAl2002, CernicharoEtAl2010,TenenbaumEtAl2010}.

Mapping the spatial distribution of molecules in the circumstellar
envelope of IRC+10216 is important for several reasons: (1) molecular
(and isotopic) abundances can be accurately determined, since the
excitation temperature can be inferred from the spatial location
of the molecules in the envelope; (2) such data can be readily and
quantitatively compared with chemical models predicting abundances
as a function of radial distance from the star ; (3) parent molecules
can be distinguished from product molecules given their distribution
in the envelopes; (4) molecules important for the creation of dust
can be identified (e.g., the distribution of SiO, SiS, SiN and
silicon carbides appear to be centrally concentrated near the region
of dust formation and probably are dominant constituents of grains
forming in this region,  their abundance decreases with distance from the star);  (5) multiple transitions of the same molecule allows mapping of physical conditions (e.g., temperature) in the envelope.

Interferometric maps of  NaCN, SiO, SiS, CS, HC$_{5}$N, SiCC, NaCl, MgNC, CN,
HNC, C$_{2}$H, C$_{3}$H, C$_{4}$H have been presented  by  \cite{GuelinEtAl1996} and  \cite{DayalAndBieging1995}. 
Except for the SiO J=5--4 line at 217 GHz  \citep{SchoierEtAl2006}
 and the CS J=14--13 line at 685 GHz \citep{YoungEtAl2004}
mapped with the SMA, all the other maps were obtained using the
IRAM Plateau de Bure Interferometer (PdBI) or the Berkeley-Illinois-Maryland Array (BIMA) at around 100 GHz.

We selected the 345 GHz band for our survey primarily because very
little data exists in this frequency range, and it contains transitions from  many
astrochemically  important molecules,  including various salts,
the cyanopolyyne HC$_{3}$N, and cyclic molecules such as C$_{3}$H$_{2}$.
Two line surveys in the 345 GHz band have been published by
 \cite{AveryEtAl1992} in the frequency range of 339.6--364.6 GHz with a 
sensitivity of 0.3 K rms made with the 15 m diameter James Clerk-Maxwell Telescope (JCMT) and by \cite{GroesbeckEtAl1994},
in the frequency range of 330.2-358.1 GHz, with the Caltech Submillimeter Observatory (CSO) 10.4 m telescope with an rms noise level of 65 mK (4.6 Jy). 
Our frequency range goes well beyond these surveys, including the range of 
300--330 GHz which is almost unexplored.

\section{Observations and data reduction}
\begin{figure*}[tbH]
\centering
\includegraphics[angle=-90,width=6in]{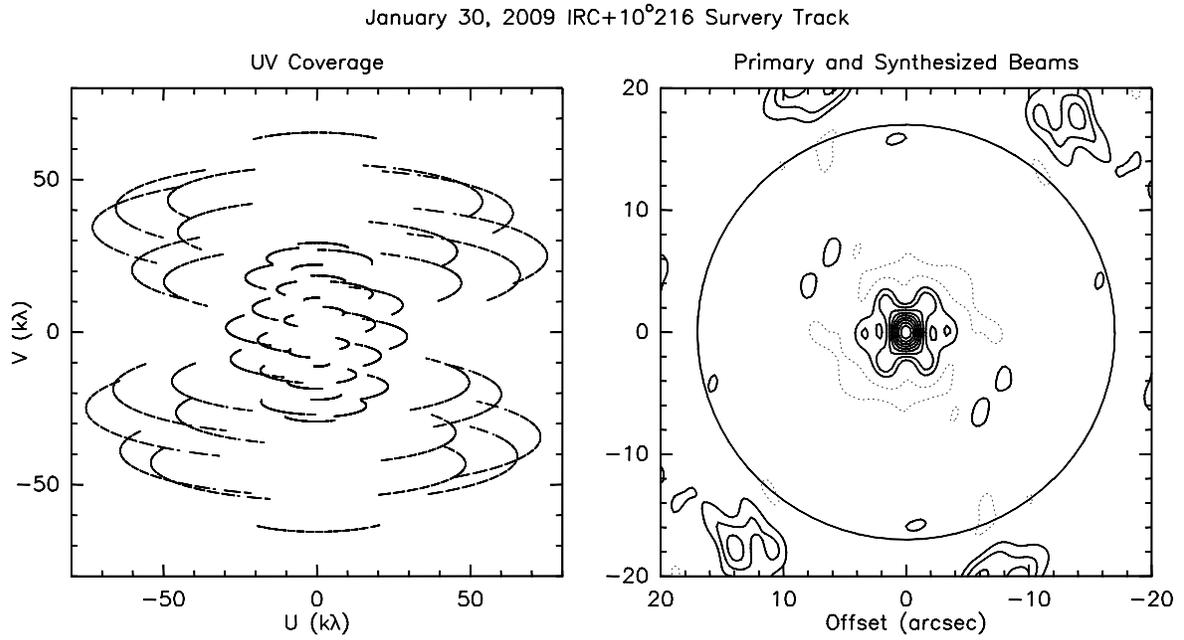}
\caption{UV coverage and beam pattern. Emission extended over angular
scales greater than 15'' is expected to get resolved out significantly.
\label{uvAndBeam}}
\end{figure*}

The SMA observations of IRC$+$10216 were done in two periods,
each about one week long.  The first was  2007 January and February 
and the second  2009 February.  All observations were
made with the array in the subcompact configuration, with baselines
from 9.5 m to 69.1 m.  The typical synthesized beam size was
$3''\times2''$.  Table \ref{table1} summarizes the observational parameters
for all observations.  The duration of each track was from 7
to 9 hours. The phase center was at $\alpha(2000)=09^{h}47^{m}57.38^{s},
\delta(2000)=+13^{\circ}16'43.''70$ for all observations. All
tracks in the first phase of observation (in 2007) were carried
out in mosaiced mode, with 5 pointings with offsets in right ascension and
declination of  $(0'',0'')$ and $(\pm12'',\pm12'')$.  The 2009 epoch
observations were done with a single pointing toward IRC+10216.
The u-v coverage and synthesized beam in one of the single-pointing
tracks are shown in Figure \ref{uvAndBeam}. Titan and the quasars
0851+202 and 1055+018 were observed every 20 minutes for gain
calibration.  The spectral band-pass was calibrated by observations
of Mars and Jupiter. Absolute flux calibration was determined by observations of 
Titan and  Ganymede.

The visibility data were calibrated with the {\it Miriad} package
\citep{SaultEtAl1995}. For the 2007 data, the mosaiced images were
de-convolved using the Miriad task {\it mossdi};  the resulting
synthesized beams are summarized in Table \ref{table1}. 
The single-pointing 2009 data were calibrated with the MIR-IDL package \footnote{See 
\url{http://cfa-www.harvard.edu/$\sim$cqi/mircook.html}.}
and imaged in {\it Miriad} using the standard tasks, {\it invert}, {\it clean} and {\it restor}.
 Maps of continuum emission show the
peak to have a position offset of
($\Delta\alpha,\Delta\delta)\approx(0.''7,0.''2)$ from the phase
center position.  The absolute position measurements
for the continuum emission are estimated to be accurate to $\sim
0.''1$.  Taking into account the proper motion of IRC$+$10216 of
($\dot{\alpha},\dot{\delta})\approx(26,4)$ mas yr$^{-1}$  determined
by Menten et al. (2006), our position is  consistent with theirs.  
The continuum emission was unresolved
at the highest angular resolution of $\sim 0.''8$.  The integrated
continuum flux density was $~\sim$650 mJy at 300 GHz and $\sim$1
Jy at 350 GHz, with an uncertainty of about 15\% in the absolute
flux calibration (see Figure \ref{contvsfreq}).  The frequency resolution
was 0.812 MHz per channel. 

\section{Results}
\label{results}

We detected a total of 442 lines.  Of these, 297 could be assigned to
known molecular transitions. Table \ref{database} summarizes  all
detections, with fitted parameters and molecular assignments.

\begin{figure*}[tbH]
\centering
\includegraphics[angle=-90,width=6in]{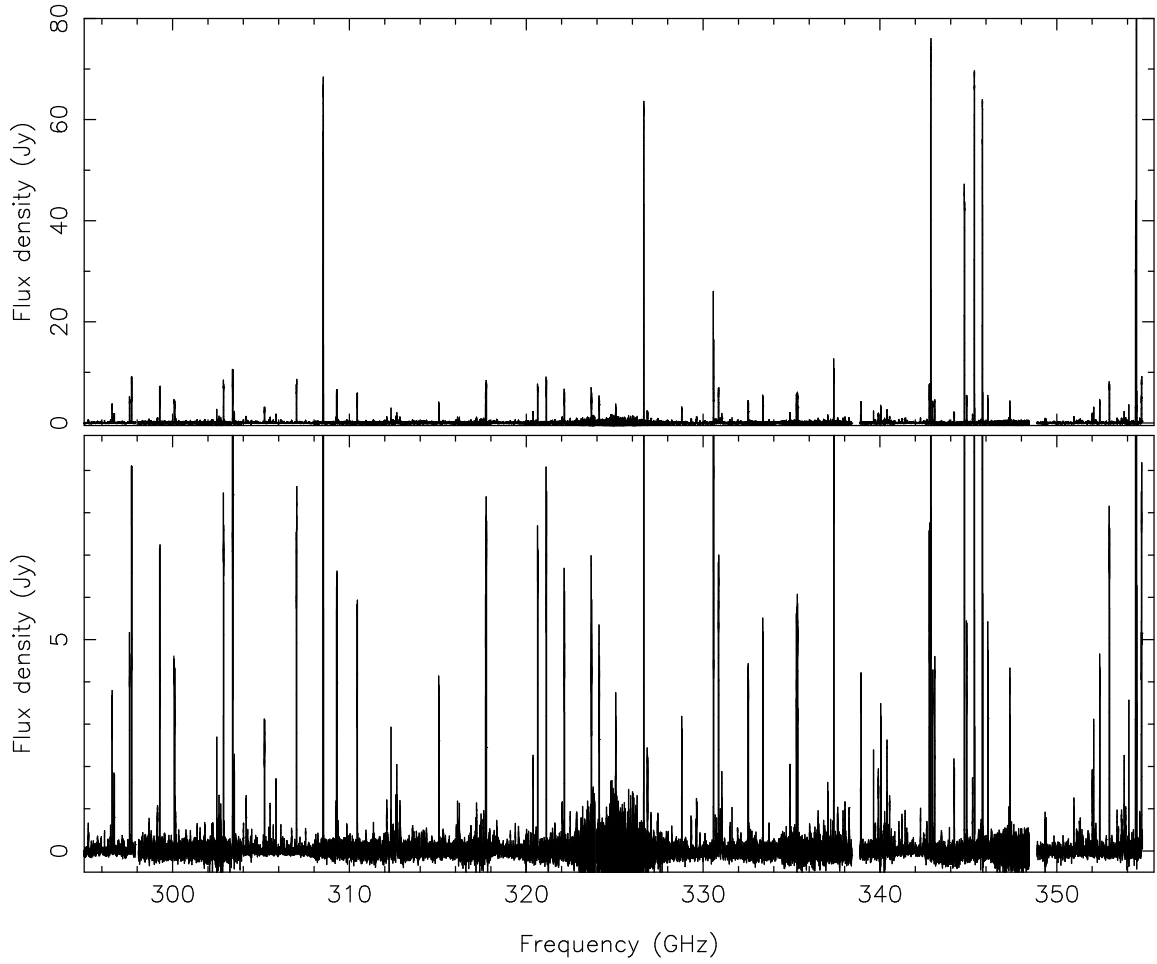}
\caption{Overview of all the detected lines. Top and bottom panels show the
same spectrum with different intensity scale. \label{overviewspectrum}}
\end{figure*}

\begin{figure*}[tbH]
\centering
\includegraphics[angle=-90,width=6in]{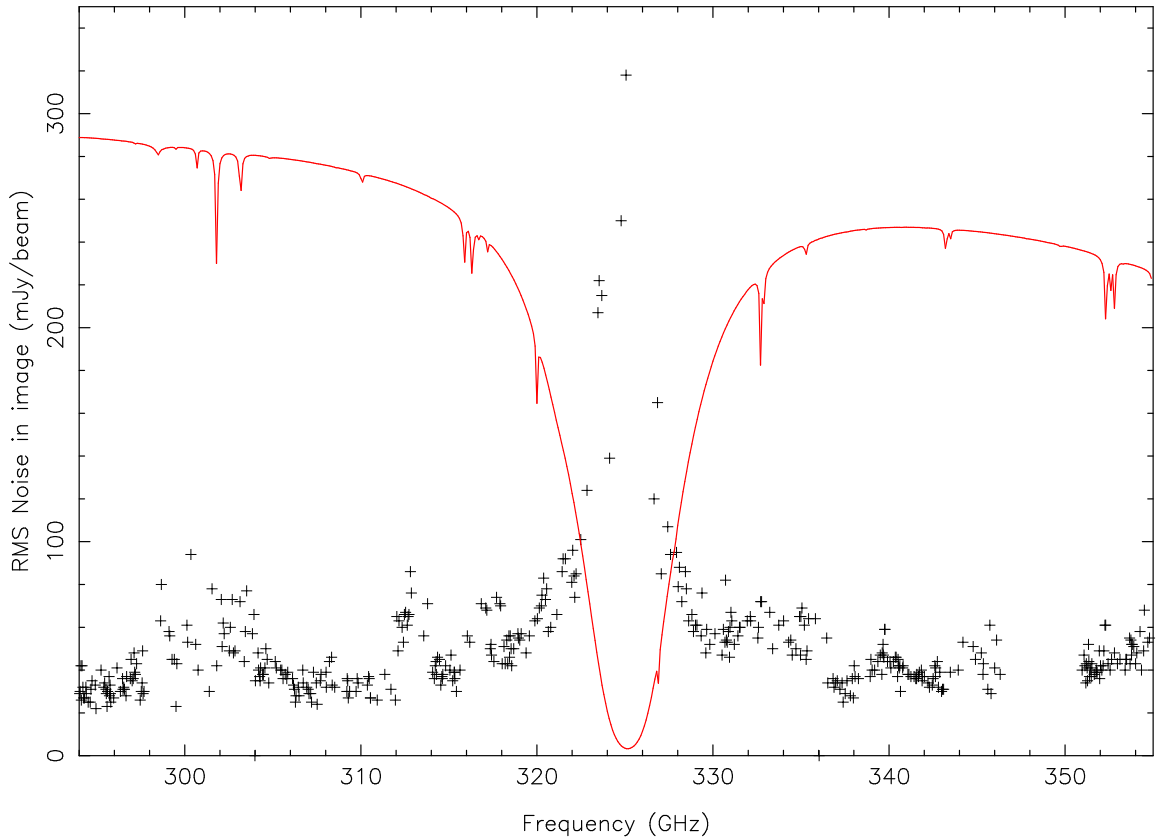}
\caption{RMS noise in image channels near detected lines as a function of
frequency. The red curve is the zenith atmospheric transmission over Mauna Kea (from a model by \cite{PardoEtAl2001}) on a scale of 0 to 100\%, assuming 2 mm of precipitable water vapor.\label{noisevsfreq}(See online for color).}
\end{figure*}

Figure \ref{overviewspectrum} is a summary of the lines
detected in the survey.  The strongest is HCN, J=4--3, at 354503.8
MHz, followed by SiS J=17--16 and J=18--17, and various SiCC lines,
48 of which were detected.  The rms noise level
is shown as a function of frequency in Figure \ref{noisevsfreq}.
The rms noise is calculated from line-free channels  using the Miriad task {\it imstat}.  The
noise peaks at 325 GHz, as expected, owing to the deep
absorption caused by terrestrial pressure broadened H$_{2}$O. The
atmospheric transmission curve is also shown in Figure \ref{noisevsfreq}.

Figure \ref{spectra} presents the spectra with each line numbered
following Table \ref{database}.   All spectra
shown here  were made by integrating the continuum-subtracted
intensity in a $2''\times2''$ rectangle centered on the continuum peak
(using the Miriad task {\it imspec}). These spectra were converted to {\it Gildas CLASS}\footnote{See {\it http://www.iram.fr/iramfr/gildas} for more information about GILDAS software.} format after re-interpolating onto a 1 MHz/channel grid. To help locate the data files of raw or calibrated visibilities, Table \ref{dates} lists the dates of observations for a given range of frequencies.  Table \ref{molecules} summarizes the molecules
and their isotopic species, with the number of transitions detected
in each. 

\subsection{Continuum emission}

Continuum data were obtained from the line-free channels in the lower
and upper sideband spectra from each night of observation.  The
line density toward IRC+10216 in the 345 GHz band is low
enough to allow a good selection of line-free regions.  Images of
the continuum emission are point-like. Results of 2D
Gaussian fits to these images are summarized in Table \ref{continuum}. The continuum
flux densities are plotted as a function of frequency in Figure
\ref{contvsfreq}.   Measurements made during 2007 show a higher continuum flux density by about 15\% compared to that of 2009. The spectral energy distribution is consistent
with a black-body curve in the Rayleigh-Jeans approximation, as observed at cm wavelengths by \cite{MentenEtAl2006}. The continuum emission agrees well with the extrapolated values from cm wavelength measurements \citep{ReidAndMenten1997} and most likely represents photospheric
optically thick blackbody emission ($S\propto \nu^{2}$), with little contribution from circumstellar dust,  as shown as a solid line in Figure \ref{contvsfreq2}. The dashed line is for $S\propto \nu^{3.2}$ following \cite{GroesbeckEtAl1994} with a value of dust emissivity spectral index $\beta=1.2$.  

Figure \ref{weaklinesFlux} shows a distribution of the detected lines with respect to integrated intensities, to estimate the total flux in the lines weaker than our detection limit, following \cite{SuttonEtAl1984, GroesbeckEtAl1994}. The slope of the fitted line shown in Figure \ref{weaklinesFlux} is -0.4. Integrating the emission below the detection limit of 0.5 Jy beam$^{-1}$ km s$^{-1}$, we estimate an integrated flux of 195 Jy beam$^{-1}$ km s$^{-1}$ in un-detected lines in our survey. 

\section{Line identification}
\label{lineiden}

The systemic velocity of IRC+10216 is $-$26.5 km
s$^{-1}$ \citep{HeEtAl2008}. Since most detected lines show the
emission to be spatially centered on the continuum peak, the
fitted centroid frequency of the line is assumed to be at the
source velocity.  Applying this velocity correction to obtain the
spectrum as a function of rest frequency, each detected line was
fitted to the parameterized shell using the {\it CLASS} package.
The fitted observed frequencies for known transitions of SiCC agreed
well with the published values of rest frequencies in the spectral
line catalogs. The mean difference in frequencies was about 0.5 MHz,
(less than the spectral resolution of 0.81 MHz), with a standard
deviation of 0.9 MHz. We referred to the following spectral line
catalogs: 1) Cologne Database for Molecular Spectroscopy\footnote{The
  CDMS catalogue at \url{http://www.ph1.uni-koeln.de/vorhersagen/}} \citep[CDMS,
see][]{Muel01,Muel05},
2) Molecular Spectroscopy database of Jet Propulsion Laboratory
\citep[JPL, see][]{Pick98}\footnote{The JPL catalogue at
  \url{http://spec.jpl.nasa.gov/home.html}} and the online Lovas
  line
list \citep{Lova04}\footnote{The Lovas line list at
  \url{http://physics.nist.gov/PhysRefData/Micro/Html/contents.html}}.
We used the website http://www.splatalogue.net which provides a
convenient interface to these line catalogs (Remijan et al. 2008).
In addition to the published line catalogs, we have also referred to a line catalog developed at the IRAM 30m telescope \citep{CernicharoEtAl2000b}.
We also
used the Cassis software for analysis of spectra \footnote{CASSIS
has been developed by CESR-UPS/CNRS(http://cassis.cesr.fr)}. Using
Cassis, we looked for a series of lines for each of the molecules
known to be present in IRC+10216, based on previous line-surveys.

A close match in frequency between an observed and cataloged line
is a necessary, but not sufficient, condition for
identification. Several such matches were found for molecules which
have a large number of lines (e.g. methanol and acetone), even though
these molecules are dubious in the envelope of IRC+10216.
Several exotic identifications were ruled out when the line's entry was grossly inconsistent with its prediction from the 
molecule's rotational  temperature diagram.  We have
adopted  a conservative standard and labeled such lines unassigned.
Information on spatial distribution from the maps of these lines
may provide additional clues,  to help  the identification.
For example, a majority of the unassigned lines have narrow
line-widths and compact emission, suggesting they arise from a region close to the
star with  high excitation energies. These could be vibrationally
excited lines of known simple molecules, not yet measured in the
laboratory.

It is not surprising that the carriers of many of the narrow lines
arising from the inner envelope remain unassigned, because for many species
laboratory measurements of lines from high lying rotational levels or
vibrationally excited states is  spotty at best.  In general it is
often difficult to assign the rotational spectrum of a polyatomic molecule
with 3 or more atoms to a particular vibrational state.  Usually a unique
assignment requires either independent rotationally resolved IR
measurements, or vibration-rotation coupling constants that may in
principle be obtained from high-level quantum theoretical calculations but
which frequently are unavailable.  Notable examples in which the
laboratory spectroscopy needs to be extended include C$_6$H, SiCC, and
even the well-studied stable molecules HCN and HCCCN --- all well known
constituents of IRC+10216, and all with many accessible transitions in the
radio band.

\section{Discussion}
\subsection{Comparison with previous line surveys}
All previous line surveys of IRC+10216 were done with single-dish
telescopes, although selected lines have been mapped with the IRAM PdBI
and BIMA interferometers. Interferometric mapping of lines in the 345 GHz band presently is
only possible with the SMA. Two of the previous line-surveys \citep{GroesbeckEtAl1994, AveryEtAl1992}
have overlapping frequency ranges with the present survey. Since the
sensitivity in the \cite{AveryEtAl1992} survey is much poorer than that of CSO's, 
we compare our results only with the latter. In
the overlapping frequency ranges, we find that all of the narrow
lines detected in the present SMA  survey are missed
in the CSO line survey by \cite{GroesbeckEtAl1994}. This is mainly owing to our four-fold better
sensitivity, produced by  
the greater collecting area of the 8$\times$6 m SMA antennas, 
but also because of the difference in on-source
integration time at a given frequency. Multiple tunings per night
were used in the observations for the CSO  line survey. In our 2007 observations,
we had four times greater bandwidth compared to the CSO survey, and in 2009, eight times greater bandwidth.

All the lines detected in the CSO line survey are also detected in our SMA line survey in the overlapping frequency range, with the exception of  one of $^{29}$SiS: 338447.3 MHz,  and a U line at 338821 MHz, which fall in one of the gaps in our frequency coverage.  All the SiCC lines in the CSO survey show a flat-topped profile, whereas in the SMA survey they show the double-horn shape indicating that we are spatially resolving the emission in the expanding shell.  A comparison of the peak intensities of the SiCC lines suggests that we are also missing a significant amount of flux due to the interferometer's response to extended emission compared to the $\approx 20''$ diameter of the CSO beam. Similar findings  hold for the line profiles and peak intensities of CO, $^{13}$CO, C$^{34}$S and $^{30}$SiO lines that are common between the two surveys. The CN emission near 340.3 GHz, and C$_{2}$H near 349.4 GHz,  show very different shapes compared to the CSO observations, again primarily owing to missing flux from extended emission. Channel maps of these lines show the emission to be arising from an expanding shell of radius $\sim15''$. As noted above, most of the lines in our survey that show spatially compact emission (angular size$<5''$) and narrow line-widths ($<$10 km s$^{-1}$) are absent in the CSO line survey. Surprisingly, some of the broader lines, with characteristic expansion velocity of $\sim$14 km s$^{-1}$,  are also missed in the CSO survey (e.g., C$^{17}$O  at 337061 MHz). The integrated intensity map of this line shows a compact as well as shell distribution,  but the weaker emission in the shell may fall at the half-power radius of the $20''$ beam of the CSO telescope.

One unassigned line in the CSO survey at 339911 MHz is detected with high S/N in our survey and we assign it to Si$^{33}$S J=19--18 emission. Note that in Figure 1 of Groesbeck et al., the line labeled as ``U'' at around 345.24 GHz is the H$^{13}$CN line at 345238.7 MHz,  as listed in their Table 3.

The Arizona Radio Observatory line surveys of IRC+10216 cover the frequency ranges $130-160$ GHz, $219.5-267.5$ GHz \citep{HeEtAl2008} and $214.5-285.5$ GHz \citep{TenenbaumEtAl2010}.  These surveys are more sensitive compared to the CSO survey and even though the frequency range is different from our survey, it is informative to compare the detection of various molecules.
Table \ref{molecules} summarizes all the detected lines in our survey,  listed by the molecule or isotopologue. The entries in this table are for assignments based on either detection of several lines, or from well-known molecules in IRC+10216 from previous observations.  Some of these assignments are based on a single line, but for many of these, only one transition falls in our frequency range. For example, HCP J=8--7 is possibly a new assignment. Emission from this molecule  in IRC+10216 was reported by \cite{AgundezEtAl2007} and by \cite{TenenbaumEtAl2010}. Several molecules detected in the survey by \cite{TenenbaumEtAl2010} are absent in our line survey.  These molecules are: CP, HNC, l-C$_{3}$H, c-C$_{3}$H, C$_{3}$N, PH$_{3}$, H$_{2}$CO, H$_{2}$CS and HCO$^{+}$.  It is possible that emission from these molecules arises from spatially extended regions, filtered out by the interferometer, or the lines are simply too weak for our sensitivity (or both).

Figure \ref{maps} shows the integrated intensity maps, radial intensity profile (averaged azimuthally), and coarse channel maps for each detected line having spatially resolved emission. Maps of lines which show point-like emission, are omitted. 
Molecules such as C$_{4}$H, C$_{2}$H, CN and SiC show spatially extended emission in  a ring-like distribution, with a radius of
$\sim15''$,  as seen in previous interferometric observations, e.g.,  \citep{GuelinEtAl1996, CernicharoEtAl1989}.  Emission from NaCN, C$^{17}$O and  C$^{18}$O appears in both ring and a compact source at the center. 
Emission from other lines typically  shows extended structure of diameter
$\sim5''-10''$. Some of the unassigned lines also show extended emission, and the size of the emission may provide a clue on physical and chemical conditions to help with the identification of their carriers. A more detailed study of the spatial structure in selected
lines will be presented in a future paper. We note  that imaged
data cubes for all the detected lines, as well as calibrated visibilities
data, are available online on the SMA website.

We note that the spectra shown in Figure \ref{spectra} have weak features at the level of 2--3$\sigma$, which are not tabulated in Table \ref{database}, but the emission in such lines is seen in integrated intensity maps over selected relevant frequency ranges. An example of such a weak line is the $^{13}$CO $v=1$ J=3--2 emission at 327645.5 MHz, shown in Figure \ref{vibex13co3-2}. This line was expected based on the detection of vibrationally excited CO emission \citep{PatelEtAl2009a}.
Unassigned lines are included in the complete table of detected lines (Table \ref{database}). 

\subsection{Contribution of line emission to total submillimeter flux density}

One of the features of our line survey is that we have also obtained line-free continuum emission. Thus, we can attempt to address the question of what fraction of the total flux density as measured by a bolometer is due to line emission. 
To obtain the total integrated line emission over the frequency range of the survey, we need to estimate the total flux in the lines that were below our detection limit. Following \cite{SuttonEtAl1984} and  \cite{GroesbeckEtAl1994}, we plot a distribution of the detected lines with respect to integrated intensities in Figure \ref{weaklinesFlux}.  The slope of the fitted line shown in Figure \ref{weaklinesFlux} is -0.4. Integrating the emission below the detection limit of 0.5 Jy beam$^{-1}$ km s$^{-1}$, we estimate an integrated flux of 49 Jy beam$^{-1}$ km s$^{-1}$ due to lines within the noise floor in our survey. This is a small correction to the total integrated intensity of all the detected lines of 66798 Jy km s$^{-1}$ (column 13 in Table \ref{database}).  Averaging over the band of 61 GHz (dividing by 56394 km s$^{-1}$ at 324.5 GHz), we obtain a total line flux of 1.2 Jy over the range of 294 to 355 GHz. The average value of line-free continuum flux density (Table \ref{continuum}) is 0.9 Jy. Thus the line emission contributes a fraction of 57\% of the total flux. This is comparable to the value of 65\% obtained by \cite{GroesbeckEtAl1994}. We point out three caveats in our comparison of line and continuum emissions.  (i) We are likely to be missing a significant flux due to lines arising from extended regions over 20$''$, (ii)  in our continuum measurement,  we are probably measuring only the compact photospheric emission and missing the flux from extended dust emission and (iii) out continuum flux may still include weak lines, which may be revealed by a more sensitive line survey. Regarding the last point above, it should be noted that a recent very sensitive line survey (1$\sigma=3$mK) has detected more than 700 lines over a frequency range of $\sim70$ GHz \citep{TenenbaumEtAl2010}. The bolometric measurements of line + continuum flux values at submillimeter wavelengths reported in literature are all typically higher than our value of 2.1 Jy (1.9 --- 9 Jy; \cite{GroesbeckEtAl1994,SopkaEtAl985}).

Our measurements of continuum flux from the 2007 data show slightly greater values compared to the 2009 measurements (see Figure \ref{contvsfreq}). According to the light-curve of IRC+10216 at 900$\mu$m measured with the SCUBA bolometer at JCMT \citep{JennessEtAl2002}, our 2007 measurements were at the pulsation phase of 0.32 (with phase 0 at maximum), and the 2009 data were obtained at the phase of 0.47 (very near minimum). However, we see a much smaller than predicted amplitude change. 

\subsection{Isotopic ratios}
Assuming  optically thin lines in both the main and rare
isotopic species, we can estimate the isotopic abundances from the
observed intensities. From the $v=1$, $J=19-18$ lines of SiS, $^{29}$SiS
and Si$^{34}$S, we previously reported near-solar values of isotopic
abundance ratios for [$^{28}$Si/$^{29}$Si]=15.1 and
[$^{32}$S/$^{34}$S]=19.6 (\cite{PatelEtAl2009b}).  From the J=18--17
lines of SiS and $^{30}$SiS, we find [$^{28}$Si/$^{30}$Si]=14.8$\pm$0.2
(19.1---29.0) and [$^{32}$S/$^{33}$S]=53.5$\pm$1.3 (85---103). The
values in parentheses are from Kahane et al. (1988). Multiple lines
from NaCl, Na$^{37}$Cl, AlCl and Al$^{37}$Cl were detected. We find, 
from AlCl and Al$^{37}$Cl J=23--22 lines,
[$^{34}$Cl/$^{37}$Cl]=4.3$\pm$0.6, which is close to the
terrestrial/solar value of 4.13 \citep{CernicharoEtAl2000}.

The isotopic ratios obtained from observed peak intensities are
listed in Table \ref{isotoperatios}. There is a significant
disagreement between the ratios obtained from various species and
transitions. The source of the discrepancy  is most likely the effects
 of optical depth and lines suspected to have large optical depths are noted in the
table. The ratio of N/$^{15}$N  of 1.8$\pm$0.1 is very discrepant with respect to
the solar value of 272 as well as with the known high value of 5200
in IRC+10216, suggesting that the main transition of HCN
used to derive this ratio is very optically thick.

\subsection{Rotational temperatures} The present
survey has resulted in the detection of many rotational
transitions in some molecules, including KCl, NaCl, SiS, AlCl,
HC$_{3}$N and SiCC. Assuming optically thin emission, we can derive
the rotational temperature and column densities of the upper levels
of the transitions, using the standard formula \citep[e.g. ][]{QinEtAl2010, GoldsmithAndLanger1999}:
\begin{equation}\label{rottempeq}
{\rm ln}({N_{u}\over g_{u}})={\rm ln}({N_{T}\over Q_{r}})-{E_{u}\over
T_{r}}={\rm ln}[2.04\times10^{20}\times {F\over
\theta_{a}\theta_{b}\nu^{3}S\mu^{2}}],
\end{equation}
where $N_{u}$ is the column density of the upper energy level with
degeneracy factor $g_{u}$, $N_{T}$ is the total column density,
$Q_{r}$ is the rotational partition function, $E_{u}$ is the upper
level energy in K, $T_{r}$ is the rotational temperature ($\approx$
kinetic temperature, assuming LTE), $F$ is the integrated flux
density in Jy beam$^{-1}$ km s$^{-1}$, $\theta_{a}$ and $\theta_{b}$
are the de-convolved source sizes in arcseconds, $\nu$ is the
frequency of the transition in GHz, $S$ is the line strength and
$\mu$ is the dipole moment of the molecule in Debye.  From a straight
line least-squares fit to the observed values of $ln(N_{u}/g_{u}$) vs $E_{u}$,
the y-intercept gives the column density and the negative reciprocal
of the slope gives the rotational temperature.

Figure \ref{rottemp} shows the rotational temperature diagrams for
selected molecules. The poor fits are due to both 1) optically thick
emission likely in several transitions and 2) missing flux in
transitions with spatially extended emission. For SiCC, lines with
$E_{u}<150$ K were excluded from the fit since the emission is
expected to be resolved out for the low-excitation lines. The lines
of compact emission such as NaCl and KCl appear to have similar
strengths, suggesting a very high temperature of 1000 K or more,
and similar level-populations in the high excitation states. SiCC
lines do not show any clear pattern of cross versus intra K-ladder
transitions, as noted before from observations of the much lower J
transitions \citep{ThaddeusEtAl1984, AveryEtAl1992}. We will report the full results of rotational temperature analysis in a future paper. Here, we find preliminary results for SiCC, AlCl, and HC$_{3}$N,  which are summarized in Table \ref{rottempTable}.

\subsection{Probing the inner circumstellar envelope} %

Figure \ref{histogramvexp} shows a distribution of $V_{exp}$, the
expansion velocities from all the detected lines in the survey,
compared with that from the previous single-dish surveys. The
important conclusion is that there is a new population of narrow
lines, peaking at $\sim$4 km s$^{-1}$, with a continuous variation
of expansion velocity reaching the terminal velocity of 14 km s$^{-1}$
(shown by the maximum number of lines). Previous line surveys have
missed the narrow lines either entirely \citep{GroesbeckEtAl1994,
 AveryEtAl1992}, or have detected very few of them \citep{HeEtAl2008,
 CernicharoEtAl2000}.  More recently, the sensitive line survey by \cite{TenenbaumEtAl2010} has revealed  34 narrow lines.  Our greater angular resolution here, 
by a factor of 4 to 10, and sensitivity to higher excitation
lines have yielded the detections of a large number of these interesting narrow lines. 
Most appear to be spatially unresolved in our $3''$ beam.  Examination of the
channel maps of these lines show that the emission is indeed confined
very close to the central position at the continuum peak ---  
an indication that we are probably probing the inner envelope within a
radius of 50--100 AU from the star \citep{PatelEtAl2009a, PatelEtAl2009b}.

 Many of these narrow lines are identified as vibrationally excited
lines of simple diatomic molecules known to exist in IRC+10216
\citep{PatelEtAl2009b}. The upper energy levels of these lines have
excitation energies of 1000---3000 K, or higher, consistent with
the observation that the emission is spatially compact and close to the
star.  The unassigned lines typically appear to be narrow (Figure \ref{histogramvexpUlines}), suggesting that they are produced by vibrationally excited transitions of polyatomic molecules  whose rest frequencies have not yet been measured in the laboratory. We note that the line widths of the narrow lines are still larger than expected widths due to thermal broadening. For comparison, a CO line would have a FWHM width of 0.9 km/s for a kinetic temperature of 500 K, and 1.8  km/s for 2000 K. 

 It is possible that some of the new lines here may be radiatively
excited (\cite{PatelEtAl2009b}).  Because the two epochs of our observations
 are separated by nearby two years, a time interval
similar to the pulsation period of $\sim$700 days of
IRC+10216, we are unable to determine whether the flux density in
some of the narrow vibrationally excited lines follows the IR or
radio continuum.

Although the narrow line emission is spatially unresolved in the present 
line survey, we can obtain an estimate of the size of the emitting
region (in most cases as upper limits to the actual size),  from the
2D Gaussian fitted de-convolved source sizes. Figure \ref{vexpvsr}
shows the radial velocity profile of the inner envelope up to an
angular radius of $3''$. For many lines, the S/N is insufficient
and the uncertainty in either expansion velocity or the de-convolved
source size is too large. The points plotted in Figure \ref{vexpvsr}
were selected only for stronger lines with integrated intensity $>
0.5$ Jy km s$^{-1}$, with error in $V_{exp}<1$ km s$^{-1}$, and error
in size $< 0.''3$.

Also shown in Figure \ref{vexpvsr} are theoretical velocity profiles.
The black curve is the assumed velocity profile used as a model for
the analysis of observed infrared molecular lines by \cite{KeadyAndRidgway1993}
(see their figure 3a). The red curve shows the
velocity profile given by
\begin{equation}\label{vvsr}
v(r)=v_{\infty}(1-\theta_{0}/ \theta)^{1/2}, 
\end{equation}
where $v_{\infty}$ is the terminal velocity of 14 km s$^{-1}$ and
$\theta_{0}$ is the dust formation radius of $0.''13$ as assumed by
Kwan \& Linke (1982). Figure \ref{vexpvsr} suggests a larger value
of $\theta_{0}\approx 0.''5$.

\section{Conclusions}
The SMA line survey of IRC+10216 has yielded 442 lines, 293 of which
have been assigned to known transitions. Most are from
molecules known to exist in the circumstellar envelope of IRC+10216,
including, SiCC, SiS, SiO, CS, C$_{4}$H, CH$_{3}$CN, HCN, HC$_{3}$N,
and their isotopic species. Also detected are several lines from
salts and metals, including, NaCl, KCl, AlCl, and AlF.  
More than 100 lines remain unidentified. Maps
of these U-lines typically show very compact emission, suggesting
vibrationally excited lines of known simple molecules,  produced very close to the star or even with its 
photosphere.
Assignment of the substantial number of unidentified lines observed in
the inner envelope awaits laboratory measurements of rotational
transitions from high lying levels in the ground and vibrationally
excited states of polyatomic molecules.

\acknowledgments
We are grateful to Raymond Blundell, SMA director,  for his encouragement and support throughout the course of this project.


\begin{deluxetable}{lccclcc}
\rotate
\tablecolumns{7}
\tablewidth{0pt}
\tablecaption{Summary of observations\label{table1}}
\tablehead{
Date&LSB&USB&Synthesized&$\tau_{225 GHz}$\tablenotemark{1}&T$_{sys}$\tablenotemark{2}\\
      &Coverage&Coverage&beam &  &(SSB, K)}
\startdata
2007 January  25 & 338.9--340.9 & 348.9--350.9 & $3.''0\times2.''0$,
P.A.=-4$^{\circ}$  & 0.08 & 180--270 \\
2007 February  5 & 338.4--340.1 & 348.7--350.0 & $3.''0\times2.''0$,
P.A.=-4$^{\circ}$  & 0.08 & 180--270 \\
2007 February  7 & 298.1--300.1 & 308.1--310.1 & $3.''2\times2.''4$,
P.A.=-4$^{\circ}$  & 0.08 & 180--270 \\
2007 February  8 & 300.1--302.1 & 310.1--312.1 & $3.''3\times2.''4$,
P.A.=-6$^{\circ}$  & 0.09 & 160--230 \\
2007 February  9 & 336.4--338.4 & 346.5--348.4 & $3.''0\times2.''4$,
P.A.=-8$^{\circ}$  & 0.08 & 180--350 \\
2007 February 10 & 334.5--336.4 & 344.5--346.4 & $3.''0\times2.''4$,
P.A.=-8$^{\circ}$  & 0.08 & 180--350 \\
2007 February 12 & 332.5--334.5 & 342.5--344.5 & $3.''0\times2.''4$,
P.A.=-8$^{\circ}$  & 0.08 & 180--350 \\
2009 January  22 & 302.0--306.0 & 313.9--317.9 & $3.''3\times2.''3$,
P.A.=-24$^{\circ}$ & 0.05 & 100--200 \\
2009 January  23 & 312.0--316.0 & 323.9--327.9 & $2.''7\times2.''1$,
P.A.=-13$^{\circ}$ & 0.05 & 200--500 \\
2009 January  26 & 315.9--319.9 & 327.9--331.9 & $2.''7\times2.''1$,
P.A.=-13$^{\circ}$ & 0.08 & 200--400 \\
2009 January  30 & 338.9--342.9 & 350.8--354.8 & $3.''0\times2.''2$,
P.A.=-2$^{\circ}$  & 0.06 & 150--300 \\
2009 January  31 & 319.9--323.9 & 331.9--335.8 & $3.''0\times2.''4$,
P.A.=-6$^{\circ}$  & 0.08 & 280--700 \\
2009 February  2 & 293.9--297.9 & 305.9--309.9 & $5.''9\times3.''0$,
P.A.=38$^{\circ}$  & 0.06 & 100--240 \\
\enddata
\tablenotetext{1}{Zenith optical depth measured at 225 GHz with the CSO's radiometer.}
\tablenotetext{2}{T$_{sys}$ range is over all antennas and over the range of elevations covered during the observations of IRC+10216.}
\end{deluxetable}
\normalsize

\clearpage

\clearpage
\begin{deluxetable}{rrrrrrrrrrrrrrrr}
\tabletypesize{\scriptsize}
\tabletypesize{\tiny}
\rotate
\tablecaption{\small Parameters of the lines detected in IRC+10216 in the SMA line survey. \label{database}}
\tablewidth{0pt}
\tablehead{
\colhead{No.} & 
\colhead{Species} & 
\colhead{Vib. state} & 
\colhead{Transition} &
\colhead{$\nu_{\rm obs}$} & 
\colhead{$\sigma$} & 
\colhead{$\nu_{\rm cat}$} & 
\colhead{$\Delta\nu$} & 
\colhead{$V_{\rm exp}$} &
\colhead{$\sigma$} & 
\colhead{Peak} & 
\colhead{$\sigma$} & 
\colhead{$I_{\rm int}$} & 
\colhead{$\theta_{x}$} & 
\colhead{$\theta_{y}$} & 
\colhead{PA} 
\\
 & & & & (MHz) & (MHz) & (MHz) & (MHz) & (km\,s$^{-1}$) & (km\,s$^{-1}$) & (Jy) & (Jy) & (Jy km\,s$^{-1}$) & $('')$ & $(")$ & ($^{\circ}$)
 }
\startdata
1 &  U  &    &    & 293972.6 & 0.8 & & &  5.3  &  0.8  &  3.1  &  0.6  &  3.4  &  0.0  &  0.0  &  0.0 \\
2 &  CH$_{3}$CN  &    &  $16(7)-15(7)$  & 294024.8 & 3.0 & 294024.8 & -0.0 &  14.2  &  4.8  &  4.0  &  0.2  &  4.7  &  2.4  &  0.7  &  -2.2 \\
3 &  U  &    &    & 294056.3 & 0.8 & & &  2.1  &  0.8  &  3.1  &  0.1  &  3.1  &  0.0  &  0.0  &  0.0 \\
4 &  CH$_{3}$CN  &    &  $16(6)-15(6)$  & 294099.7 & 2.5 & 294098.9 & 0.8 &  14.7  &  4.1  &  8.2  &  0.3  &  12.7  &  3.2  &  2.5  &  -16.6 \\
5 &  HCN  &  $v_{2}+v_{3}$   &  $36-36$  & 294121.3 & 5.8 & 294121.3 & -0.0 &  5.4  &  3.9  &  2.1  &  0.3  &  1.7  &  0.0  &  0.0  &  0.0 \\
6 &  CH$_{3}$CN  &    &  $16(5)-15(5)$  & 294162.1 & 2.7 & 294162.1 & -0.0 &  13.5  &  6.0  &  6.6  &  0.3  &  11.0  &  3.7  &  2.6  &  -14.5 \\
7 &  CH$_{3}$CN  &    &  $16(4)-15(4)$  & 294213.0 & 1.6 & 294211.9 & 1.1 &  13.5  &  2.3  &  9.1  &  0.2  &  14.8  &  3.2  &  3.1  &  -2.3 \\
8 &  CH$_{3}$CN  &    &  $16(3)-15(3)$  & 294252.8 & 0.5 & 294252.9 & -0.1 &  14.3  &  0.9  &  30.3  &  0.7  &  61.0  &  4.3  &  3.9  &  11.5 \\
\enddata

\noindent {\bf Notes.}  The designation of the quantum numbers in columns
3 and 4 follow standard spectroscopic convention.  Listed in column~4 are
the rotational quantum numbers, with the upper level denoted first
(i.e., $J_{\rm{upper}} - J_{\rm{lower}}$).  The rotational
levels of symmetric tops (e.g., CH$_3$CN) are designated by $J(K)$,
asymmetric tops (e.g., SiCC) by $J(K_a,K_c)$, and those with electron spin
(e.g., SiN) by $N, J$.  
In column~2, the vibrational states of HCN are designated
by the number of quanta in the $(v_1, v_2, {\rm{and}}~v_3)$ modes,
and the vibrational angular momentum $l$ in the $v_2$ bending
vibration.  The symmetry of the $l$-type doublet is denoted by
$e$ and $f$.  When $e$ and $f$ are not specified, both states are
considered.  Similarly, the vibrational states of HC$_{3}$N are labeled
by $v_4,v_5,v_6, {\rm{and}}~v_7$, and the vibrational angular momentum
as $l = l_5,l_6,l_7~X$, where $X = e~{\rm{or}}~f$ as in HCN.

\end{deluxetable}

\clearpage
\begin{deluxetable}{lllll}
\tabletypesize{\scriptsize}
\tablecaption{\small  Dates of observations\label{dates}}
\tablewidth{0pt}
\tablehead{
\colhead{No.} & 
\colhead{Frequency} & 
\colhead{Date} & 
\colhead{U/L 2 GHz} &
\colhead{Sideband}}
\startdata
1--29 & 293972.6--295792.0 & 2009 Feb 2 & Upper & LSB\\
30--52 & 295960.9--297687.6 & 2009 Feb 2 & Lower & LSB\\
53--60 & 298630.4--299546.6 & 2007 Feb 7 &  & LSB\\
61--68 & 300120.4--301816.7 & 2007 Feb 8 &  & LSB\\
69--83 & 302069.7--303927.7 & 2009 Jan 22 & Upper & LSB\\
84--104 & 303980.9--305849.8 & 2009 Jan 22 & Lower  & LSB\\
105--123 & 306120.0--307735.5 & 2009 Feb 2 & Lower & USB\\
124--137 & 308024.5--309849.2 & 2009 Feb 2 & Upper & USB\\
138--146 & 310368.3--311960.8 & 2009 Feb 8 &  & USB\\
147--160 & 312042.3--313788.1 & 2009 Jan 23 & Upper & LSB\\
161--180 & 314075.6--315635.1 & 2009 Jan 22 & Lower  & USB\\
181--192 & 316050.8--317935.8 & 2009 Jan 22 & Upper & USB\\
193--210 & 318027.0--319882.6 & 2009 Jan 26 & Lower & LSB\\
211--223 & 320028.3--321619.2 & 2009 Jan 31 & Upper & LSB\\
224--233 & 321976.7--323684.5 & 2009 Jan 31 & Lower & LSB\\
234--236 & 324125.1--325060.7 & 2009 Jan 23 & Lower & USB\\
237--242 & 326649.7--327926.4 & 2009 Jan 23 & Upper & USB\\
243--257 & 328027.1--329815.9 & 2009 Jan 26 & Lower  & USB\\
258--273 & 330111.1--331536.2 & 2009 Jan 26 & Upper & USB\\
274--283 & 331948.4--333733.5 & 2009 Jan 31 & Lower & USB\\
284--297 & 334008.3--335818.5 & 2009 Jan 31 & Upper & USB\\
298--315 & 336026.3--338259.7 & 2007 Feb 9 &  & LSB\\
316--340 & 338929.9--340645.7 & 2009 Jan 30 & Upper & LSB\\
341--364 & 340771.9--342805.7 & 2009 Jan 30 & Lower & LSB\\
365--371 & 342881.7--344200.3 & 2007 Feb 12 &  & USB\\
372--382 & 344778.6--346313.1 & 2007 Feb 10 &  & USB\\
383--412 & 350968.2--352802.1 & 2009 Jan 30 & Lower & USB\\
413--432 & 352974.5--354795.9 & 2009 Jan 30 & Upper & USB\\

\enddata
\end{deluxetable}

\begin{deluxetable}{ll}
\tablecaption{Summary of molecules detected in IRC+10216 in the SMA line survey\label{molecules}}
\tablewidth{0pt}
\tablehead{
Molecule/ & 
Row number 
\\
Isotopologue & 
in Table \ref{database}\\

}
\startdata

 AlCl (4) & 105 216 290 387\\
 Al$^{37}$Cl (5)  & 54 159 240 346\\
 AlF (2) & 38 257\\
\\
 CN  & 333\\
 C$^{15}$N & 258\\
\\
 CO (2) & 322 363 380\\
\enddata
\vspace{0.5cm}

\noindent{\bf Notes:} The numbers in parenthesis following the name of the species are the total number of transitions, including vibrational states, detected in that species. Assignments of KCN and some of NaCN lines are tentative as denoted by ``t''.
\end{deluxetable}
\clearpage
\begin{deluxetable}{lccrccc}
\tabletypesize{\scriptsize}
\tabletypesize{\tiny}
\tablecaption{\small Continuum emission\label{continuum}}
\tablewidth{0pt}
\tablehead{
\colhead{Frequency} & 
\colhead{Peak} & 
\colhead{$\sigma$} &
\colhead{Integrated} & 
\colhead{$\theta_{x}$\tablenotemark{1}} & 
\colhead{$\theta_{y}$} & 
\colhead{$PA$}
\\(GHz) & (mJy/beam) & (mJy/beam) & (mJy) & ($''$) & ($''$) & ($^{\circ}$)}
\startdata
294.95 & 566.8 & 10.5& 633.1 & 1.48 & 1.06 & -23.7\\
296.90 & 581.2 & 11.9& 643.6 & 1.43 & 1.02 & -25.9\\
299.10\tablenotemark{2} & 802.3 & 25.1 & 897.2 & 1.01 & 0.89 & -21.2 \\
300.90\tablenotemark{2} & 775.4 & 17.6 & 883.4 & 1.21 & 0.91 & -24.6 \\
305.00 & 636.6 & 17.6& 736.7 & 1.25 & 1.02 & -41.7\\
306.90 & 626.9 & 11.8& 696.0 & 1.39 & 1.04 & -18.1\\
308.90 & 637.2 & 11.8& 705.6 & 1.39 & 0.96 & -15.6\\
311.16\tablenotemark{2} & 826.8 & 16.7 & 906.8 & 0.96 & 0.75 & 4.8 \\
\enddata
\tablenotetext{1}{De-convolved source size from 2D Gaussian fit.}
\tablenotetext{2}{From February 2007 observations. (All other measurements
are from 2009 observations).}
\end{deluxetable}

\clearpage
\begin{deluxetable}{llll}
\tabletypesize{\scriptsize}
\tabletypesize{\tiny}
\tablecaption{\small  Isotopic ratios\label{isotoperatios}}
\tablewidth{0pt}
\tablehead{
\colhead{Element/isotope} & 
\colhead{Source} & 
\colhead{Transition} &
\colhead{Ratio}}
\startdata
S/$^{34}$S	&	$^{29}$Si$^{34}$S	/	$^{29}$SiS	&		J=17--16	&	37.8$\pm$	1.2	\\
	&	Si$^{34}$S	/	SiS	&	v=2	J=17--16	&	10.8$\pm$	0.6\tablenotemark{a}	\\
	&	Si$^{34}$S	/	SiS	&		J=18--17	&	8.5	$\pm$0.3\tablenotemark{a,b}	\\
	&	Si$^{34}$S	/	SiS	&	v=1	J=18--17	&	4.5$\pm$	0.6\tablenotemark{a}	\\
	&	Si$^{34}$S	/	SiS	&	v=1	J=19--18	&	10.9	$\pm$1.0\tablenotemark{a}	\\
	&	Si$^{34}$S	/	SiS	&		J=19--18	&	5.1$\pm$	0.1\tablenotemark{a,b}	\\
	&	C$^{34}$S	/	CS	&		J=7--6	&	6.7$\pm$	0.1\tablenotemark{a,b}	\\
	&	$^{13}$C$^{34}$S	/	$^{13}$CS	&		J=7--6	&	22.1$\pm$	0.9	\\
\enddata
\tablenotetext{a}{At least one of the transitions is likely to be optically thick.}
\tablenotetext{b}{The transition in the main species may have extended emission over scales several times larger than the angular resolution of 3$''$. These transitions may have missing flux leading to an overestimate of the isotopic ratio.}
\end{deluxetable}

\clearpage

\clearpage
\begin{deluxetable}{lll}
\tablecaption{\small  Rotation temperature and column density\label{rottempTable}}
\tablewidth{0pt}
\tablehead{
\colhead{Molecule} & 
\colhead{$T_{rot}$ (K)} & 
\colhead{$N_{T}$ (cm$^{-2})$}}
\startdata
SiCC & 270 & $1.3\times10^{14}$\\
AlCl & 527 & $3.2\times10^{14}$\\
HC$_{3}$N & 206 & $5.0\times10^{11}$\\
\enddata
\end{deluxetable}

\clearpage

\clearpage

\clearpage

\clearpage

\begin{figure*}[tbH]
\centering
\includegraphics[width=6in]{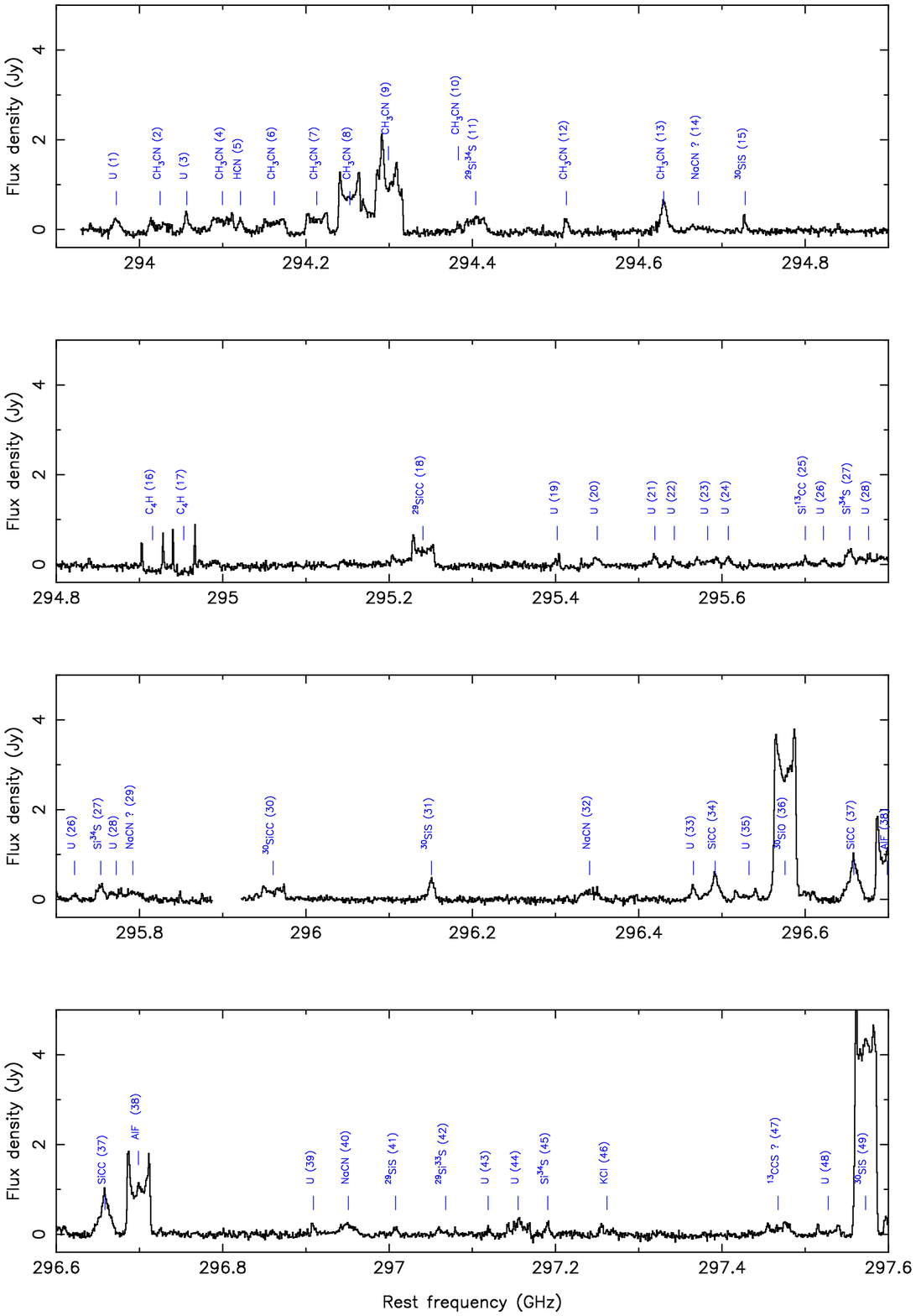}
\caption{ Caption is at the end of this figure. \label{spectra}}
\end{figure*}

\clearpage
\addtocounter{figure}{-1}
\begin{figure*}[tbH]
\centering
\includegraphics[width=6in]{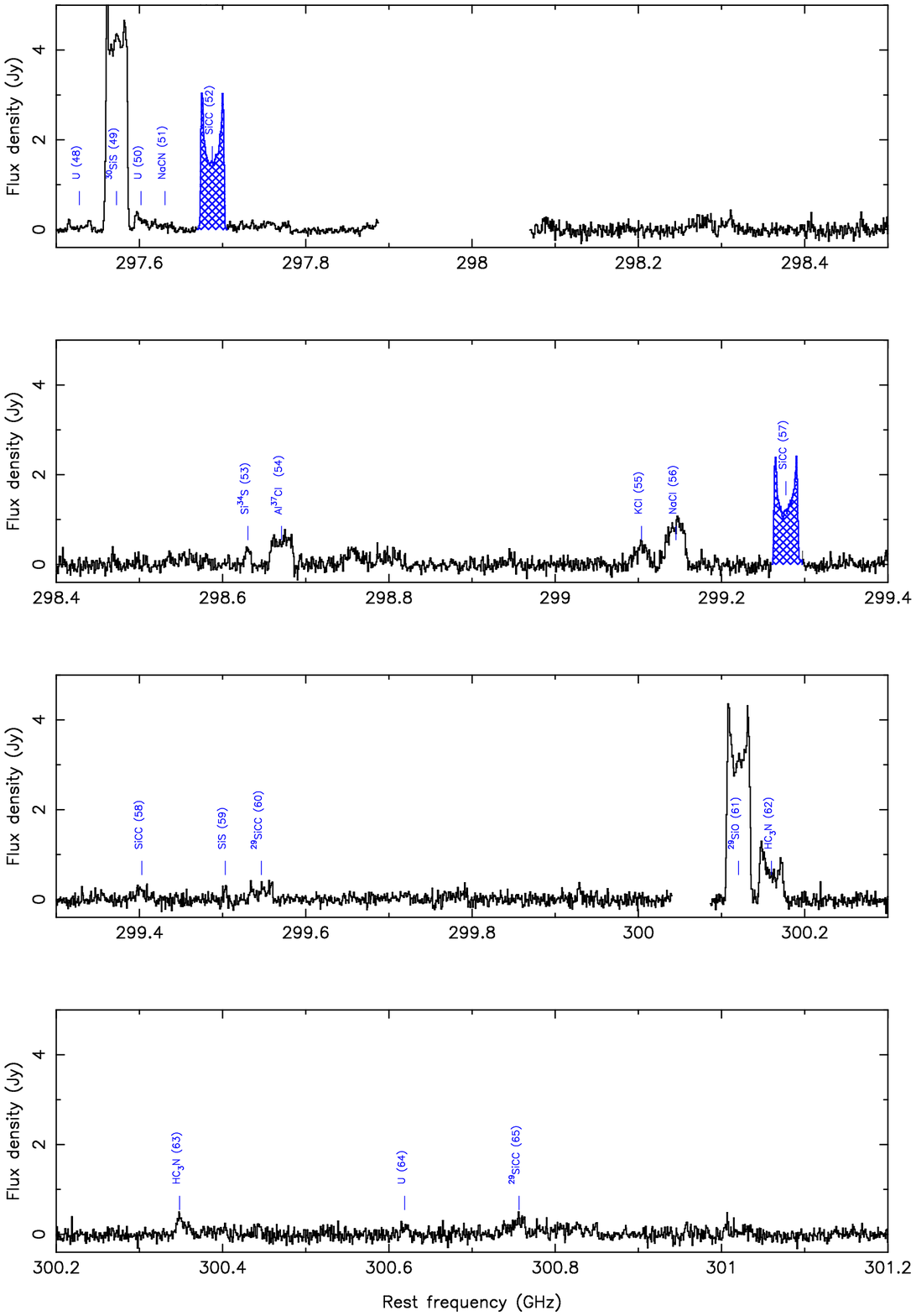}
\caption{continued.  \label{}}
\end{figure*}

\addtocounter{figure}{-1}

\clearpage
\begin{figure*}[tbH]
\centering
\includegraphics[width=6in]{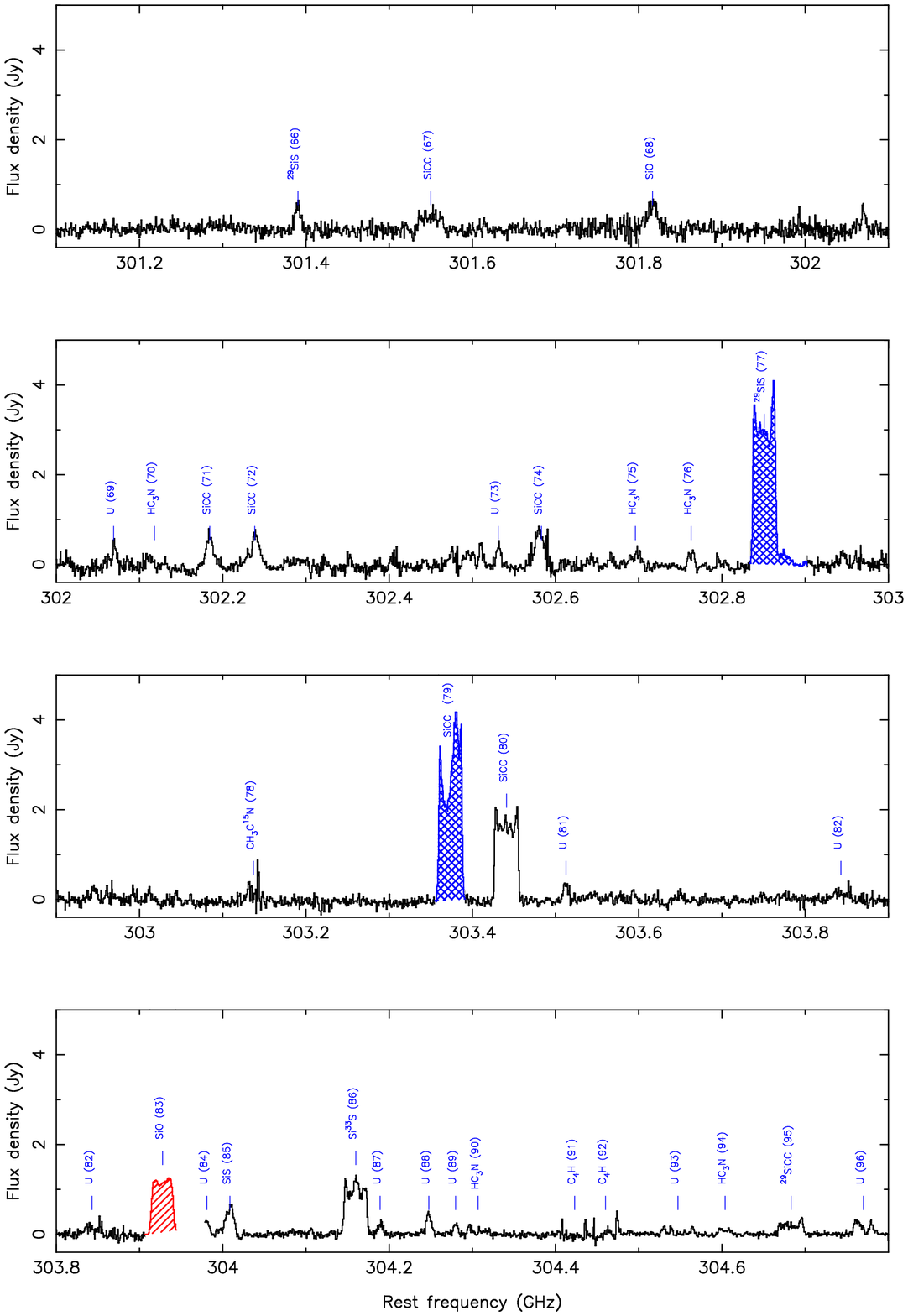}
\caption{continued.  \label{}}
\end{figure*}
\clearpage
\addtocounter{figure}{-1}

\begin{figure*}[tbH]
\centering
\includegraphics[width=6in]{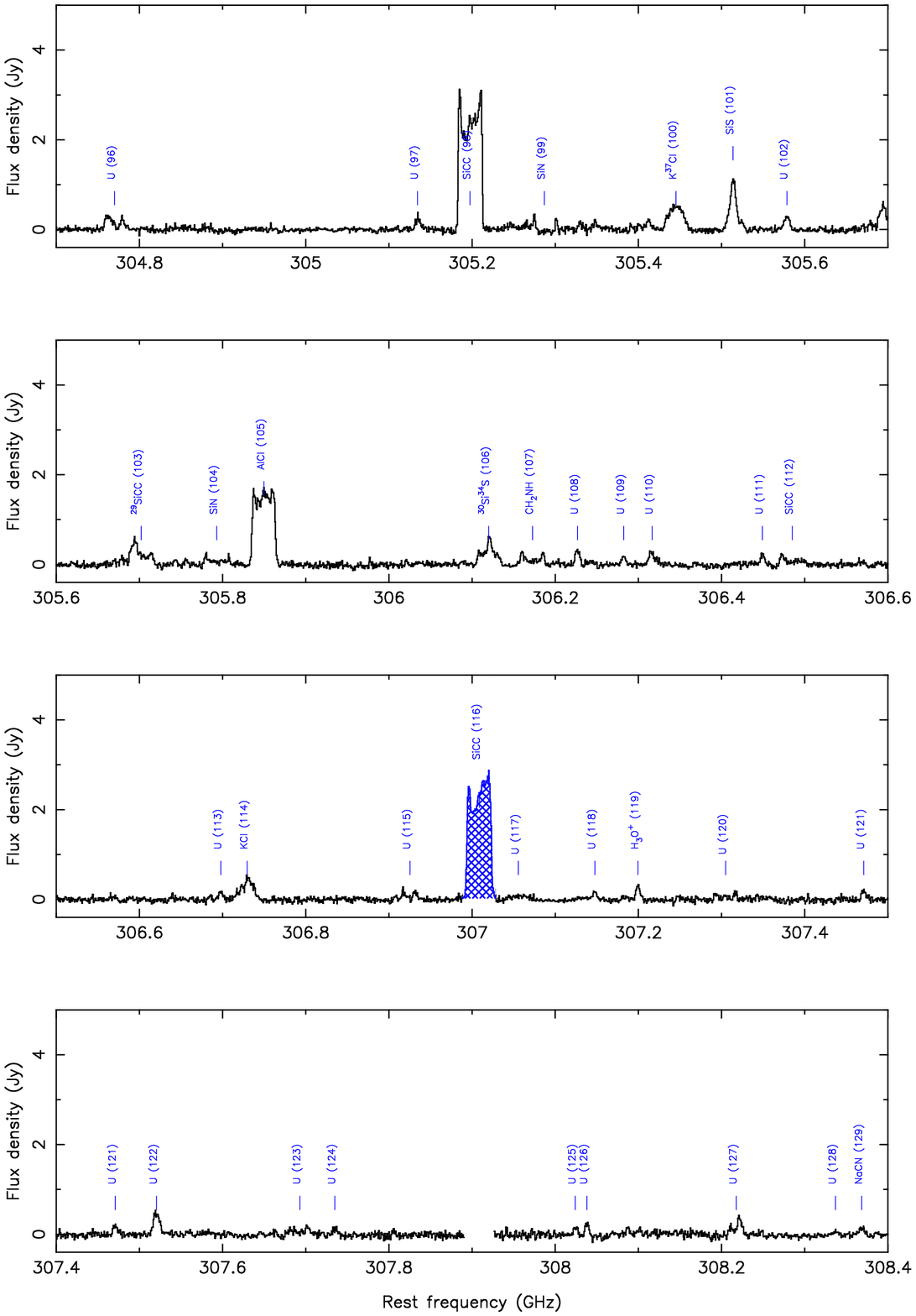}
\caption{continued.  \label{}}
\end{figure*}
\clearpage
\addtocounter{figure}{-1}

\begin{figure*}[tbH]
\centering
\includegraphics[width=6in]{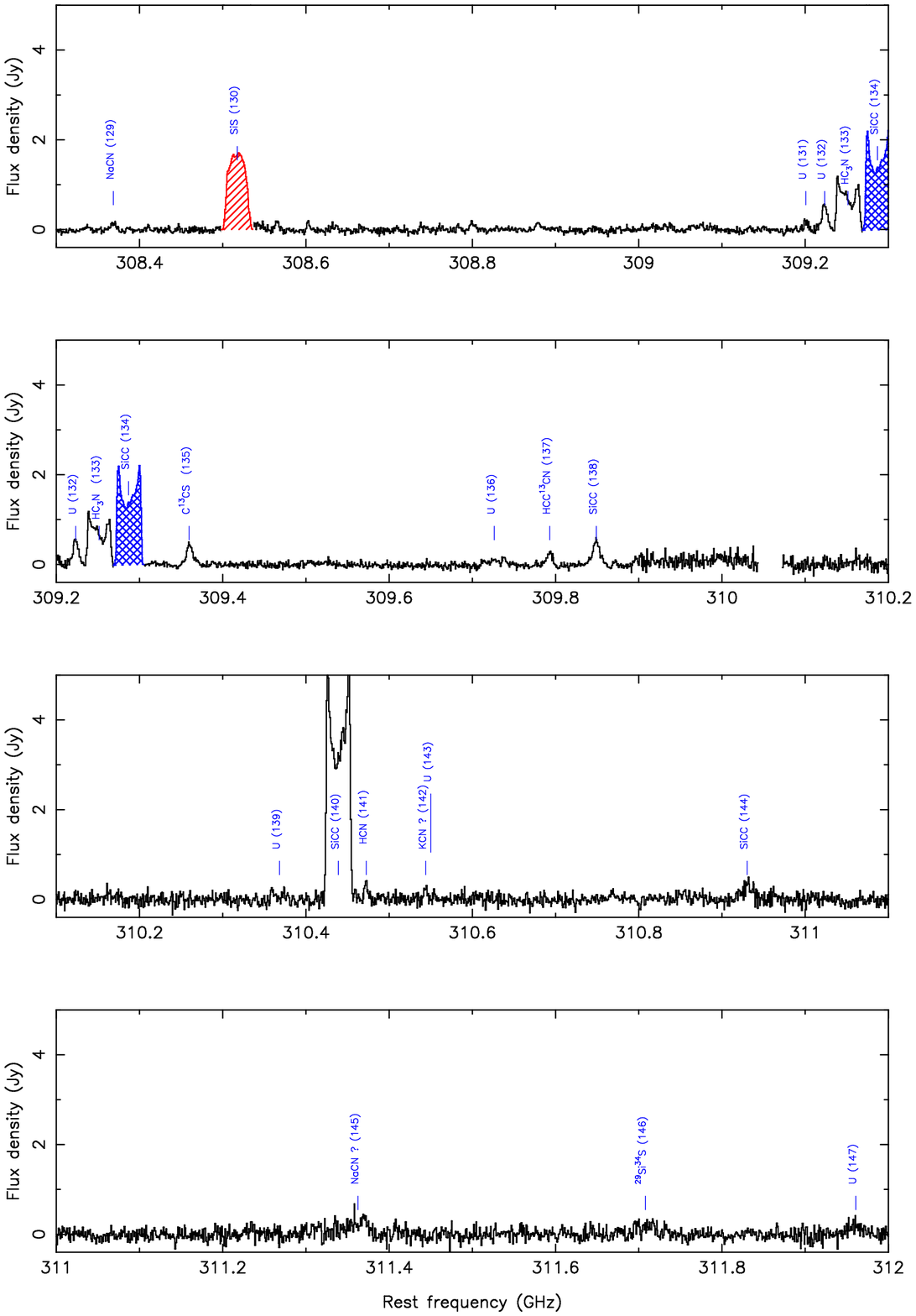}
\caption{continued.  \label{}}
\end{figure*}
\clearpage
\addtocounter{figure}{-1}

\begin{figure*}[tbH]
\centering
\includegraphics[width=6in]{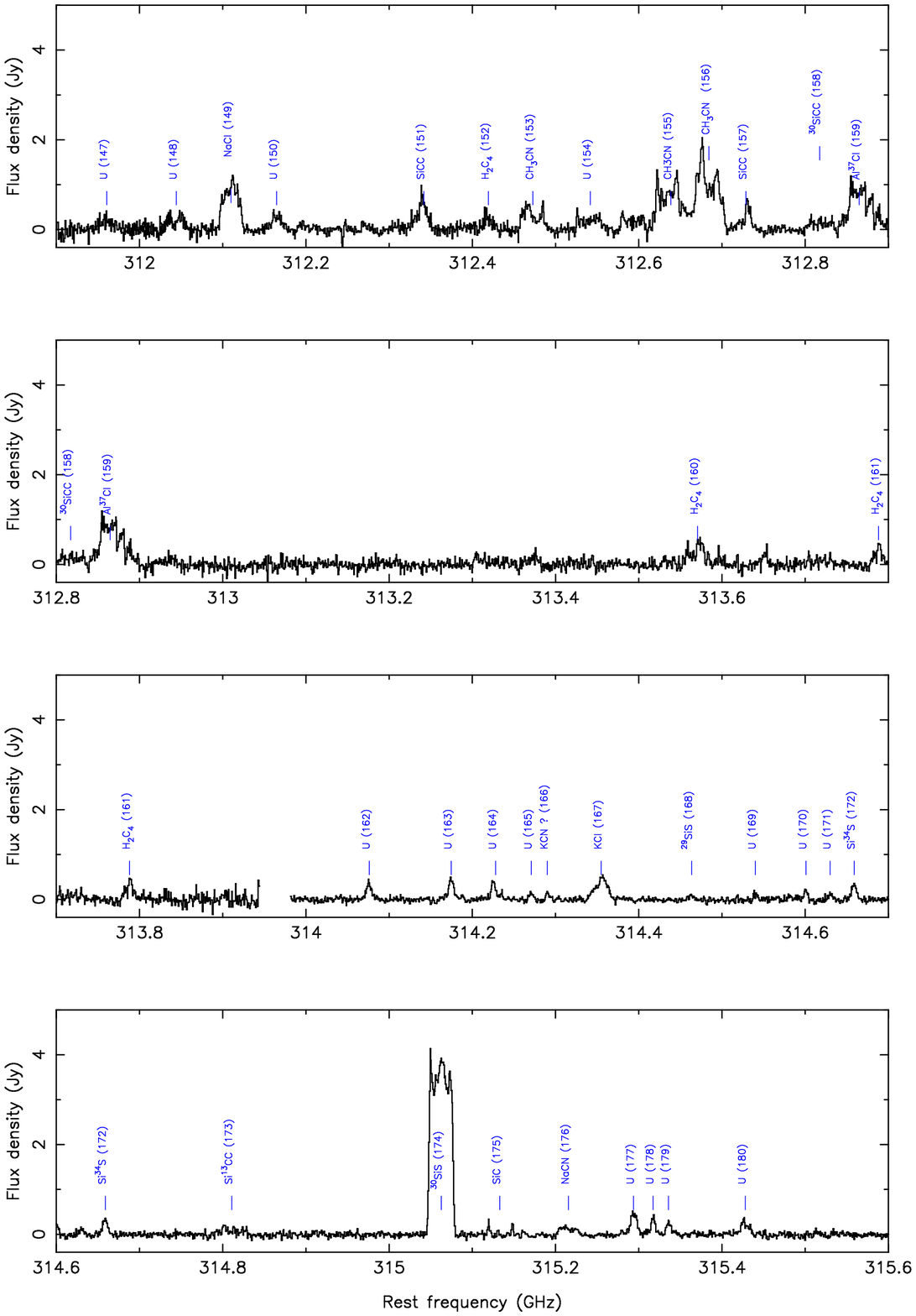}
\caption{continued.  \label{}}
\end{figure*}
\clearpage
\addtocounter{figure}{-1}

\begin{figure*}[tbH]
\centering
\includegraphics[width=6in]{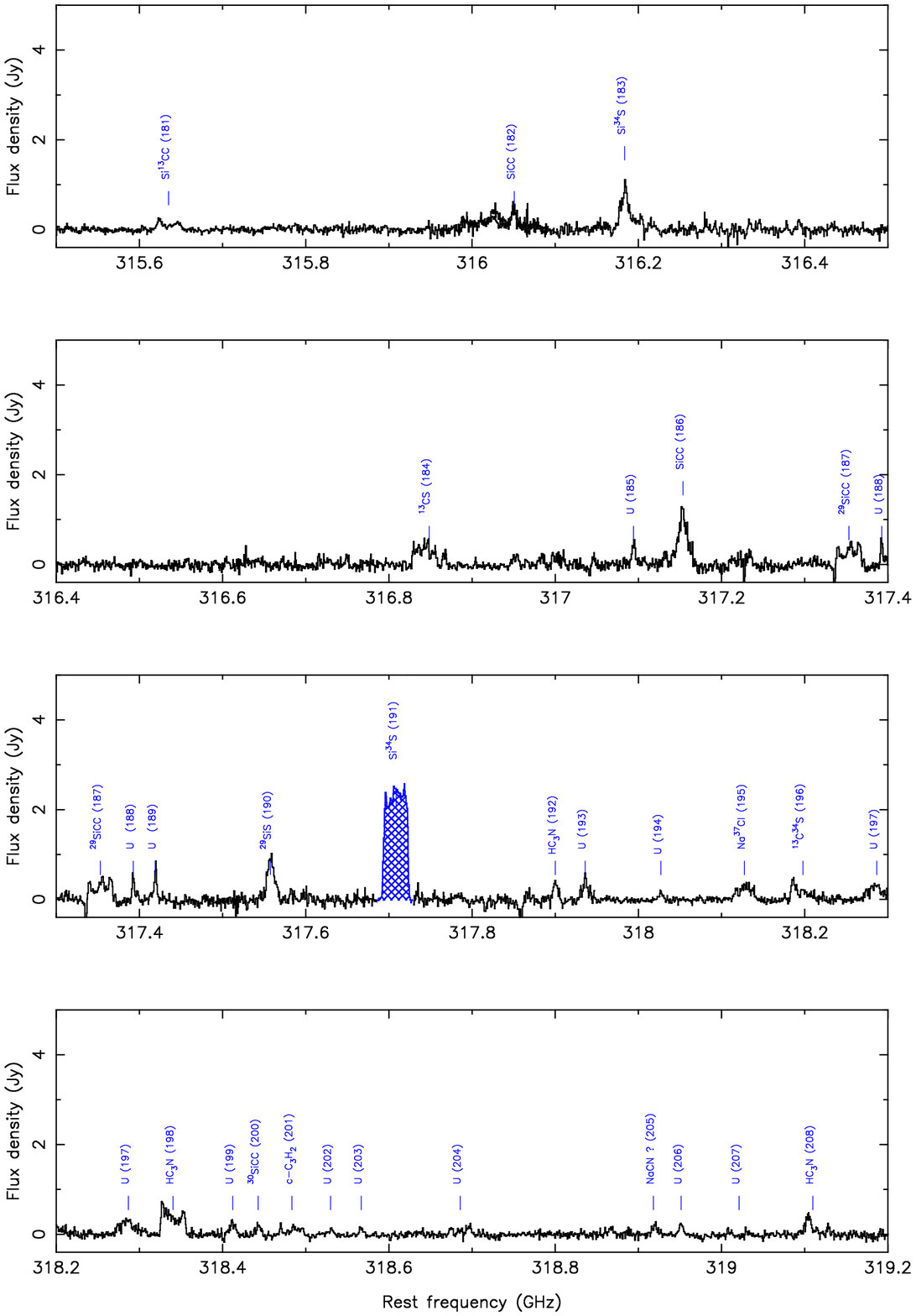}
\caption{continued.  \label{}}
\end{figure*}
\clearpage
\addtocounter{figure}{-1}

\begin{figure*}[tbH]
\centering
\includegraphics[width=6in]{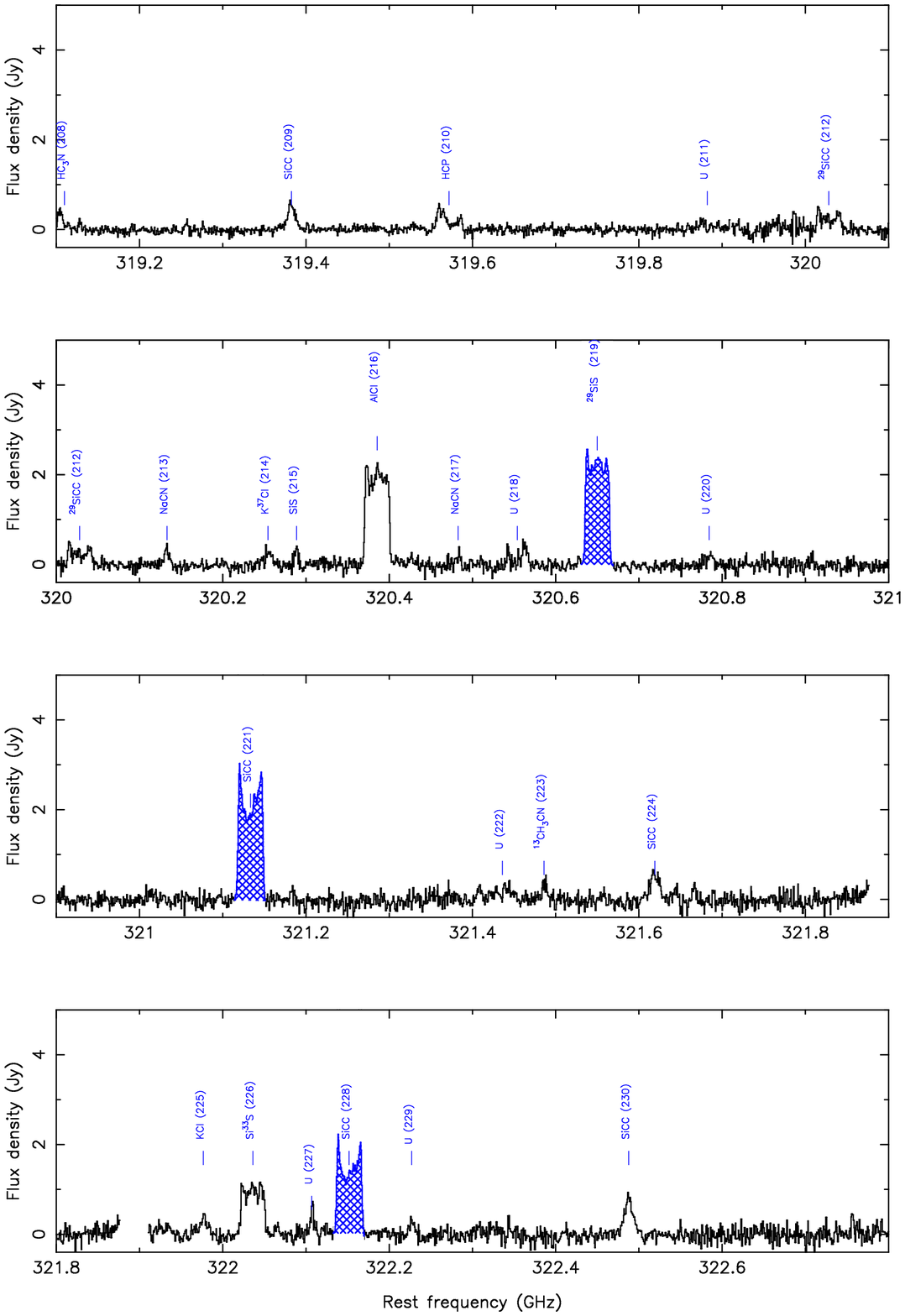}
\caption{continued.  \label{}}
\end{figure*}
\clearpage
\addtocounter{figure}{-1}

\begin{figure*}[tbH]
\centering
\includegraphics[width=6in]{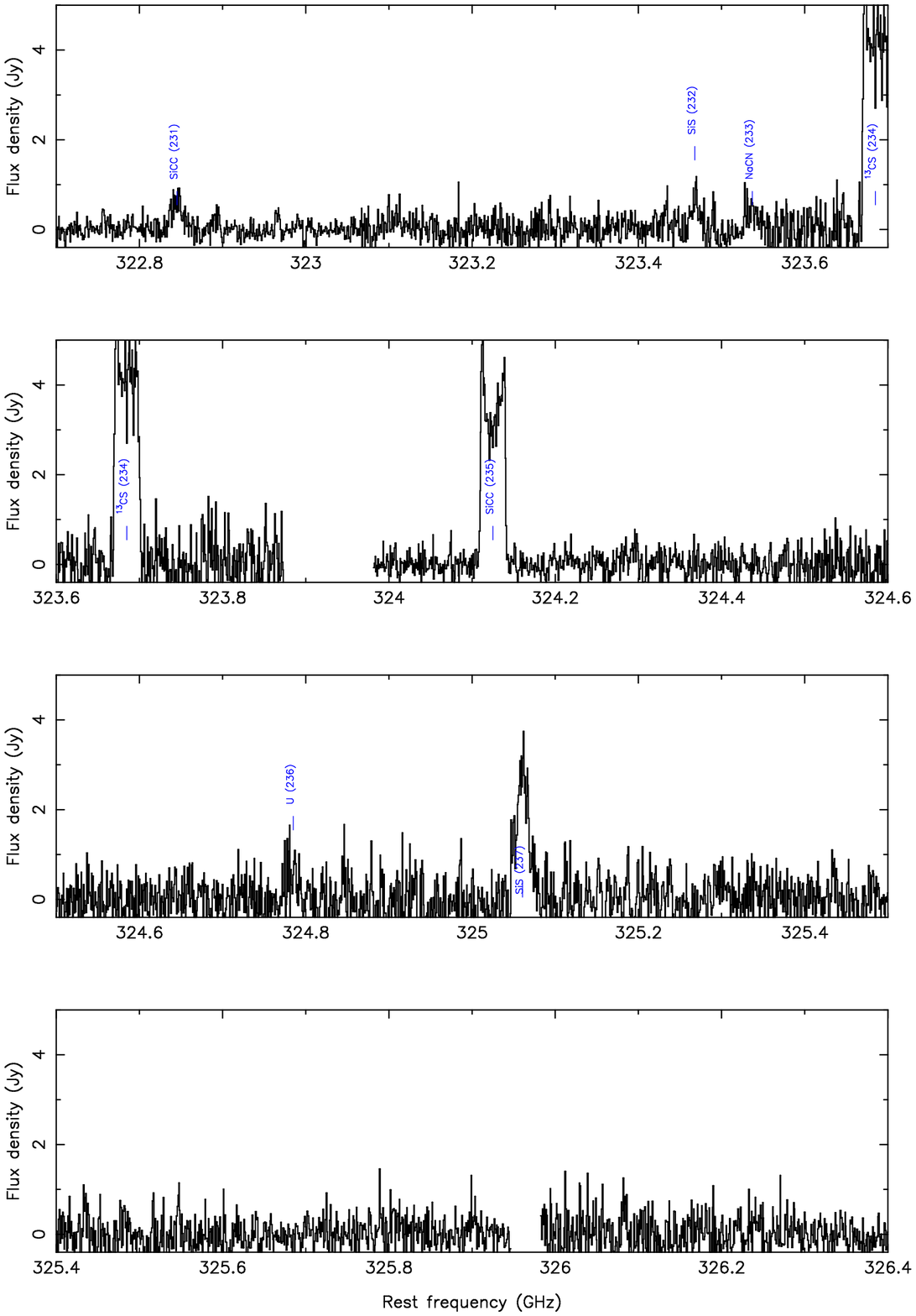}
\caption{continued.  \label{}}
\end{figure*}
\clearpage
\addtocounter{figure}{-1}

\begin{figure*}[tbH]
\centering
\includegraphics[width=6in]{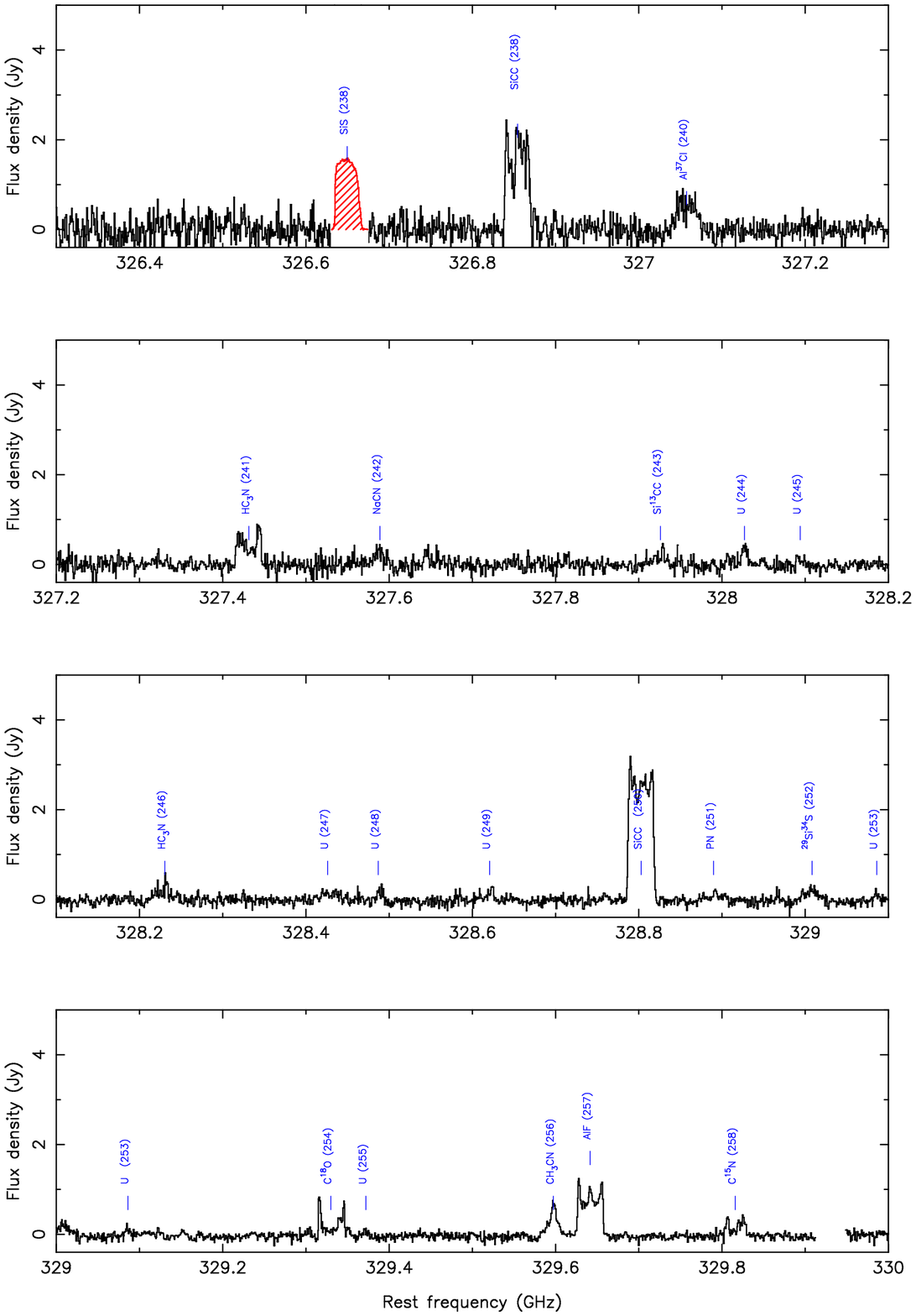}
\caption{continued.  \label{}}
\end{figure*}
\clearpage
\addtocounter{figure}{-1}

\begin{figure*}[tbH]
\centering
\includegraphics[width=6in]{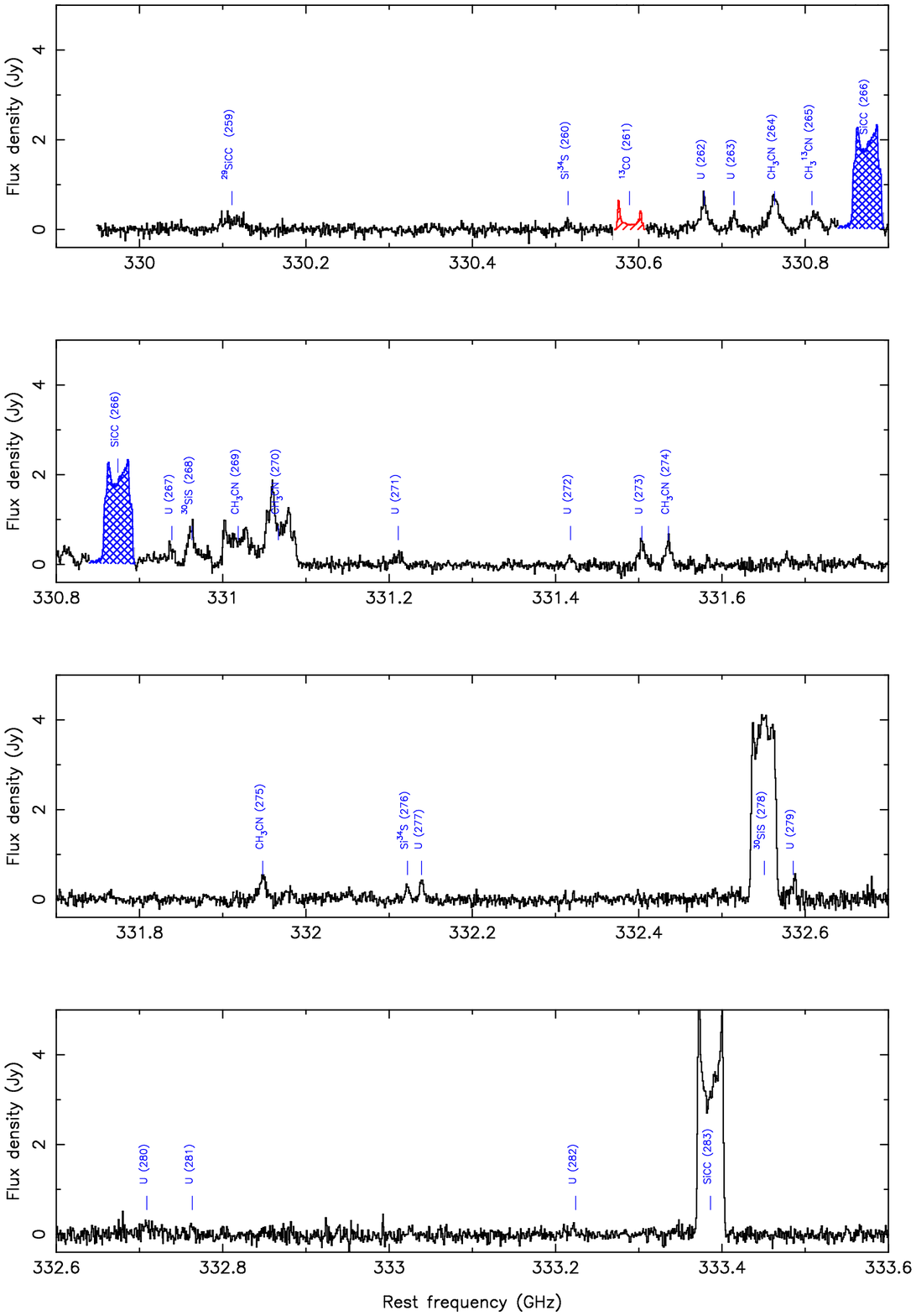}
\caption{continued.  \label{}}
\end{figure*}
\clearpage
\addtocounter{figure}{-1}

\begin{figure*}[tbH]
\centering
\includegraphics[width=6in]{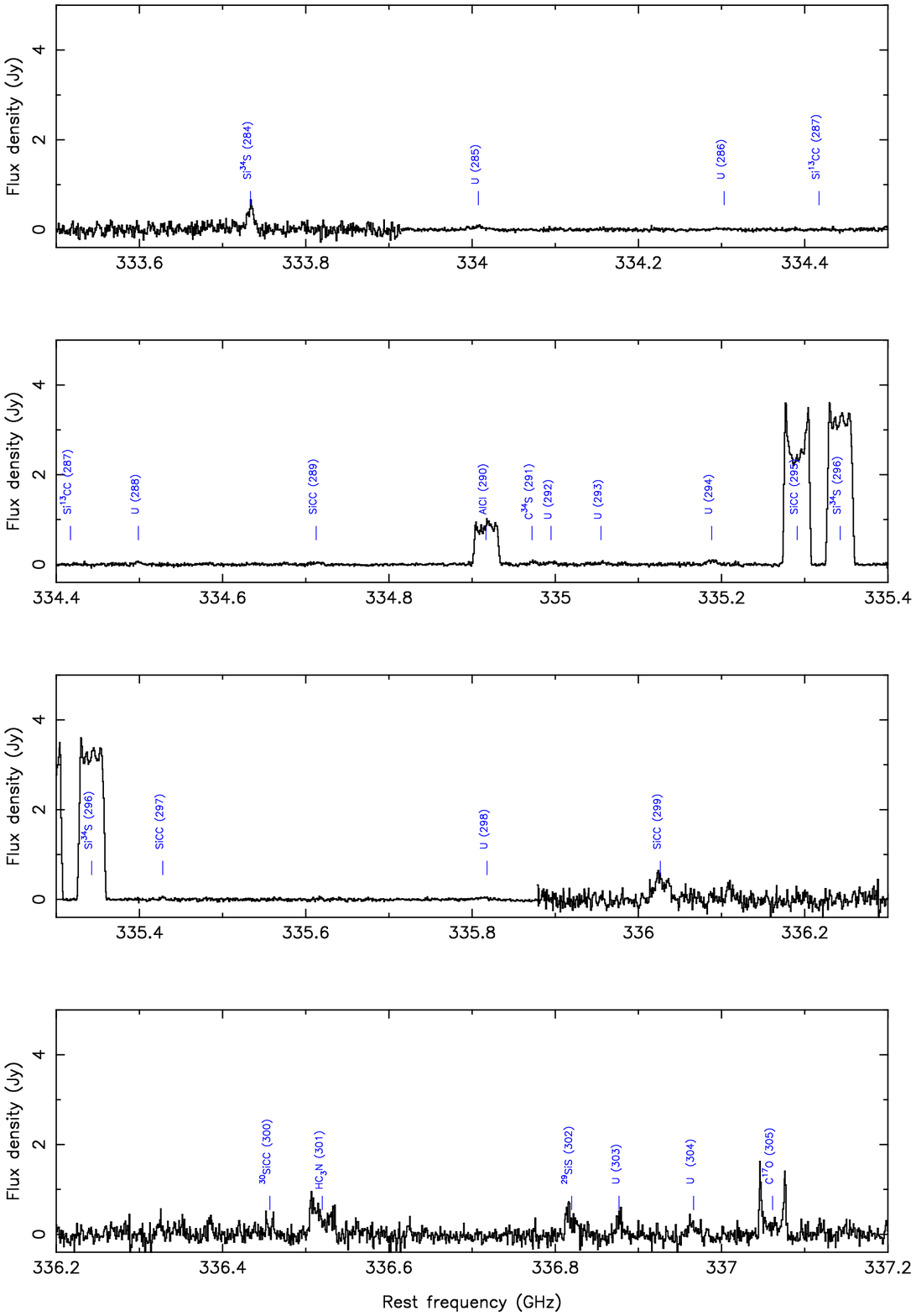}
\caption{continued.  \label{}}
\end{figure*}
\clearpage
\addtocounter{figure}{-1}

\begin{figure*}[tbH]
\centering
\includegraphics[width=6in]{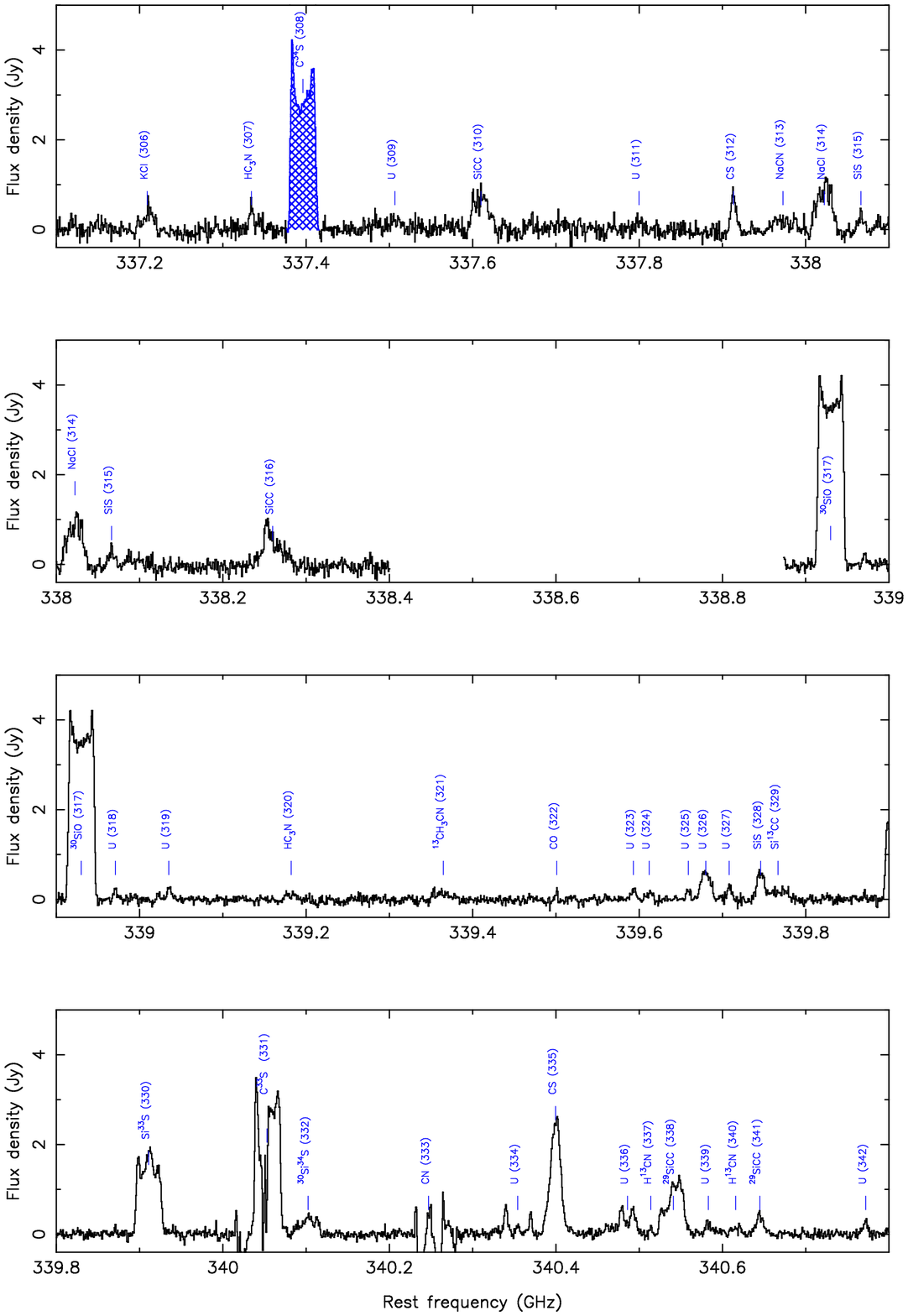}
\caption{continued.  \label{}}
\end{figure*}
\clearpage
\addtocounter{figure}{-1}

\begin{figure*}[tbH]
\centering
\includegraphics[width=6in]{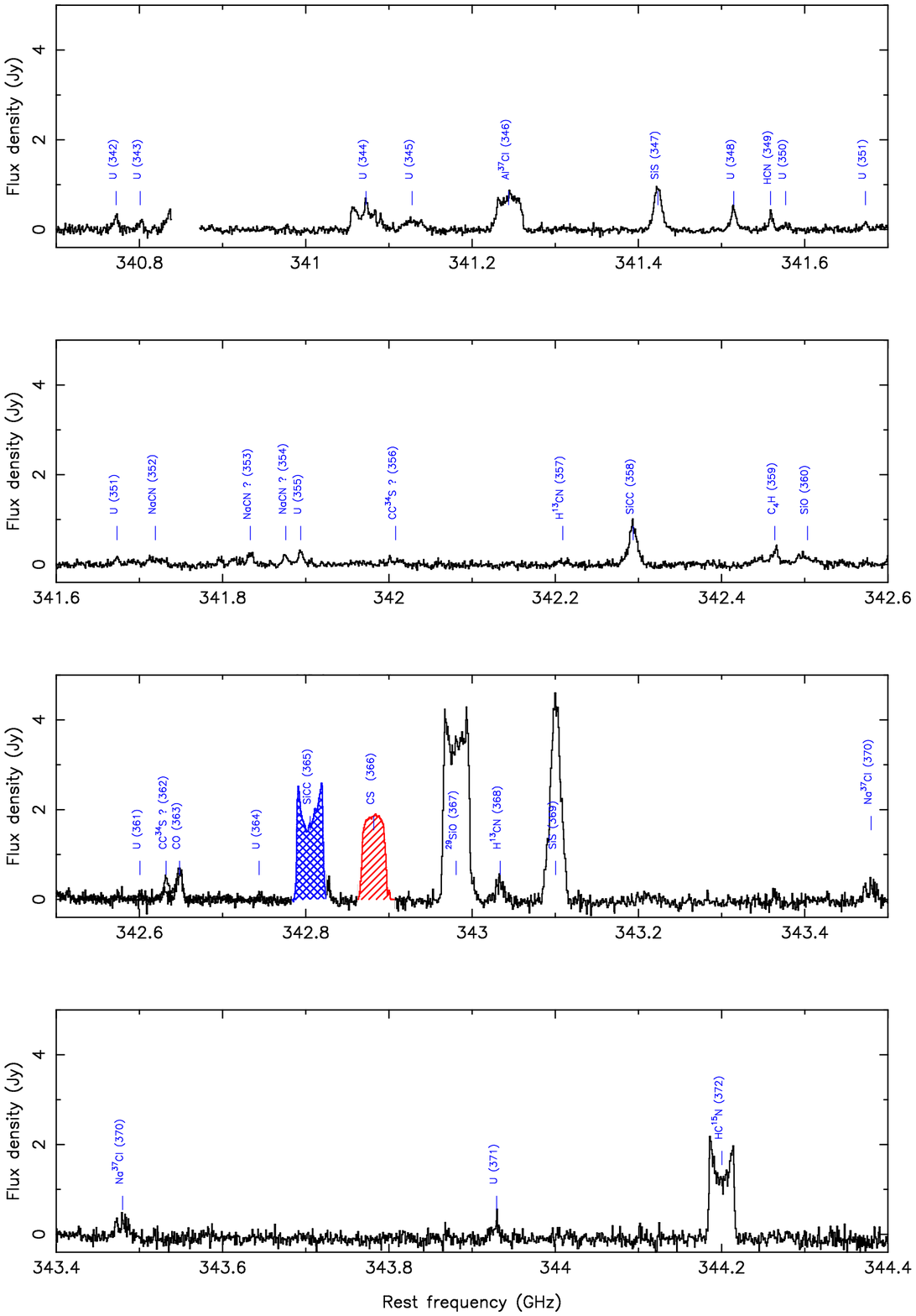}
\caption{continued.  \label{}}
\end{figure*}
\clearpage
\addtocounter{figure}{-1}

\begin{figure*}[tbH]
\centering
\includegraphics[width=6in]{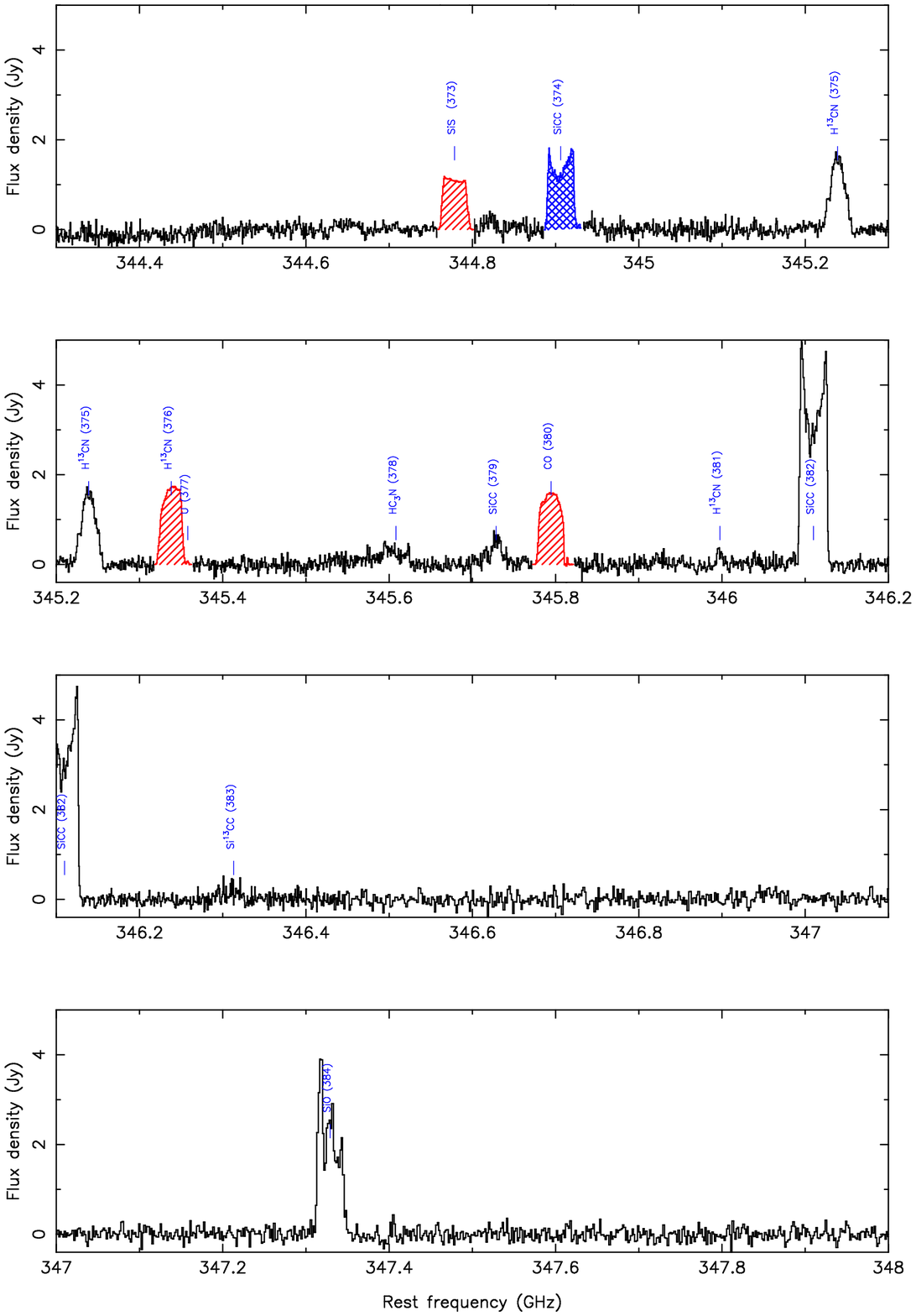}
\caption{continued.  \label{}}
\end{figure*}
\clearpage
\addtocounter{figure}{-1}

\begin{figure*}[tbH]
\centering
\includegraphics[width=6in]{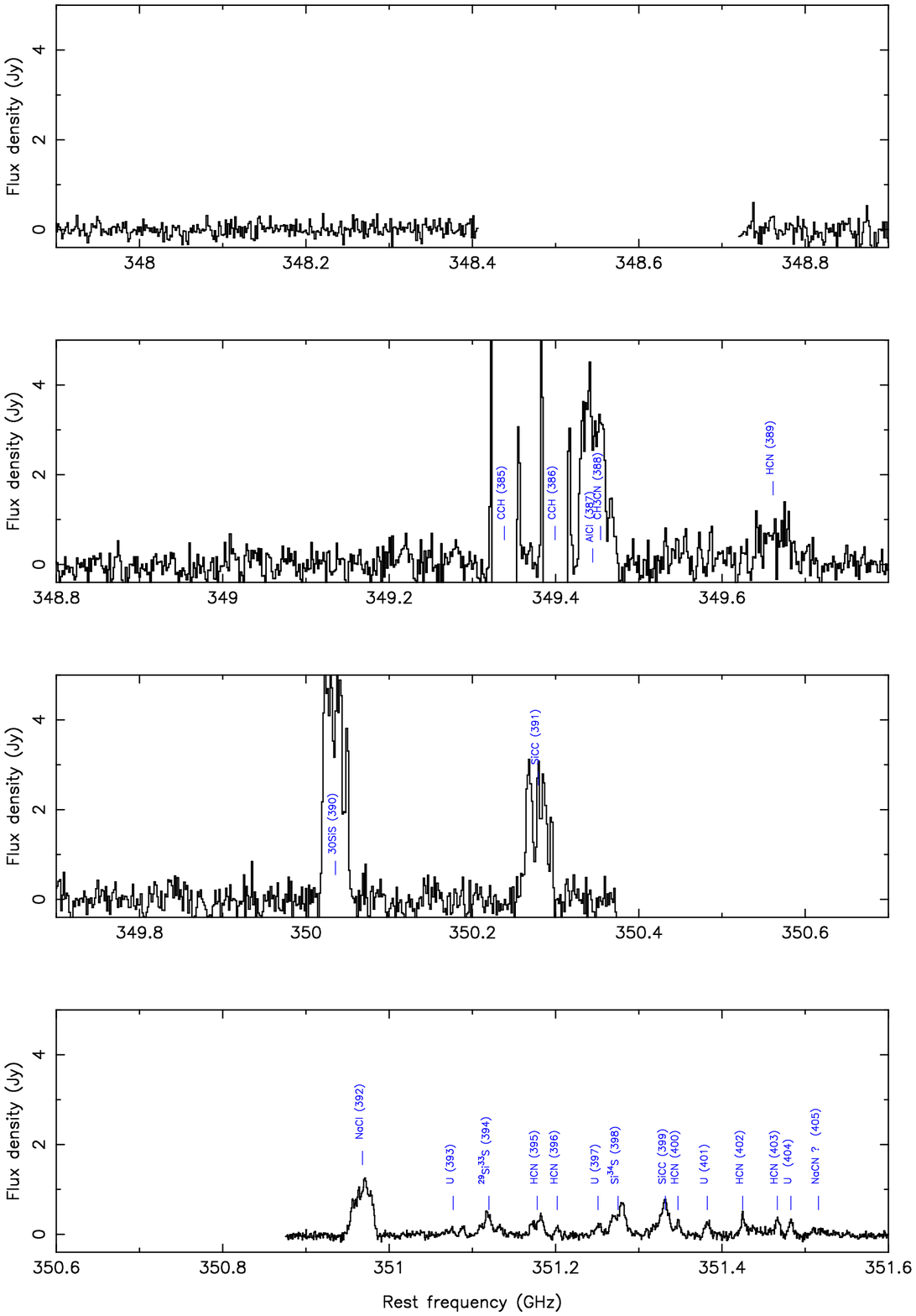}
\caption{ continued.  \label{}}
\end{figure*}
\clearpage
\addtocounter{figure}{-1}

\begin{figure*}[tbH]
\centering
\includegraphics[width=6in]{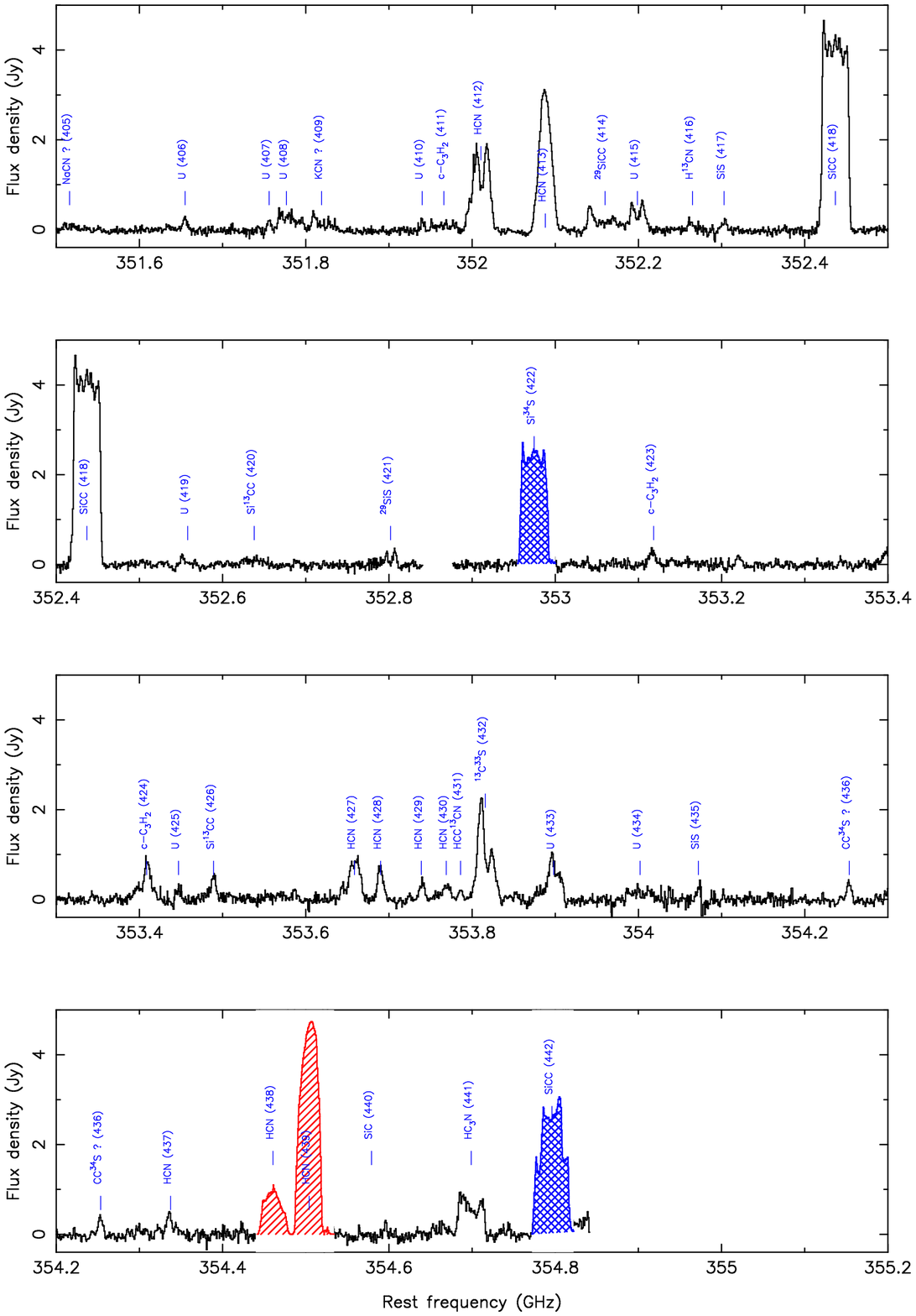}
\caption{ continued. \label{}}
\end{figure*}
\clearpage

{\bf Figure \ref{spectra}:} Spectra obtained from the integrated flux over a $2''\times2''$ region centered on the position of continuum peak (star). Flux densities for the lines shown in blue and cross-hatched filling are 3 times the y-axis scale. Red spectra with hatched filling are 40 times stronger than the value shown on the ordinate. Each line is labeled with the molecular/isotopologue species and the row-number in Table \ref{database}. Features appearing as absorption spikes are artifacts caused by imaging extended emission with limited short u-v short spacings. Maps of selected lines that show spatially resolved emission are shown in Figure \ref{maps}. (See online for color).

\clearpage

\begin{figure*}[tbH]
\centering
\includegraphics[width=6in]{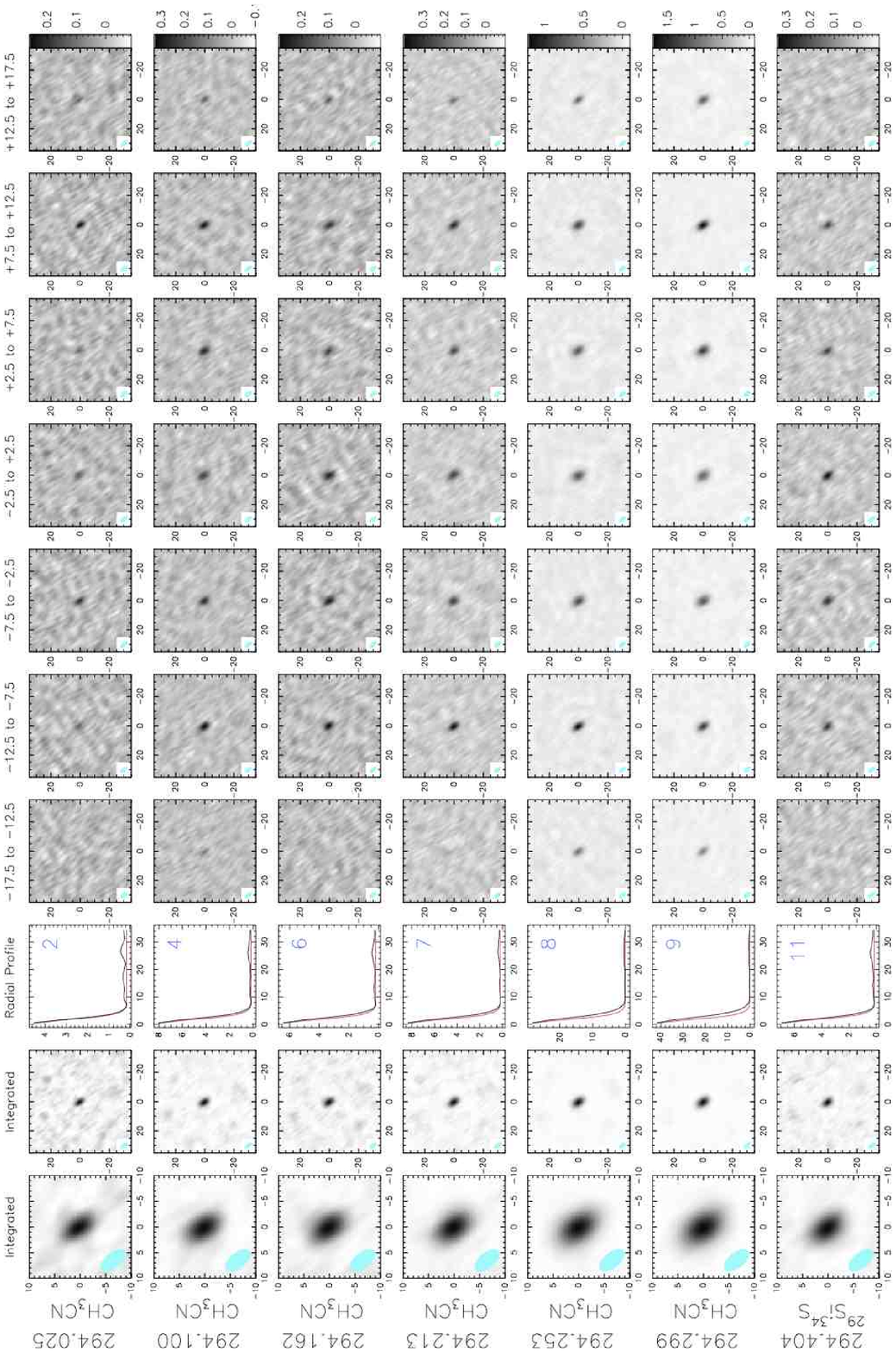}
\caption{Caption is at the end of this figure. \label{maps}}
\end{figure*}
\clearpage
\addtocounter{figure}{-1}

\begin{figure*}[tbH]
\centering
\includegraphics[width=6in]{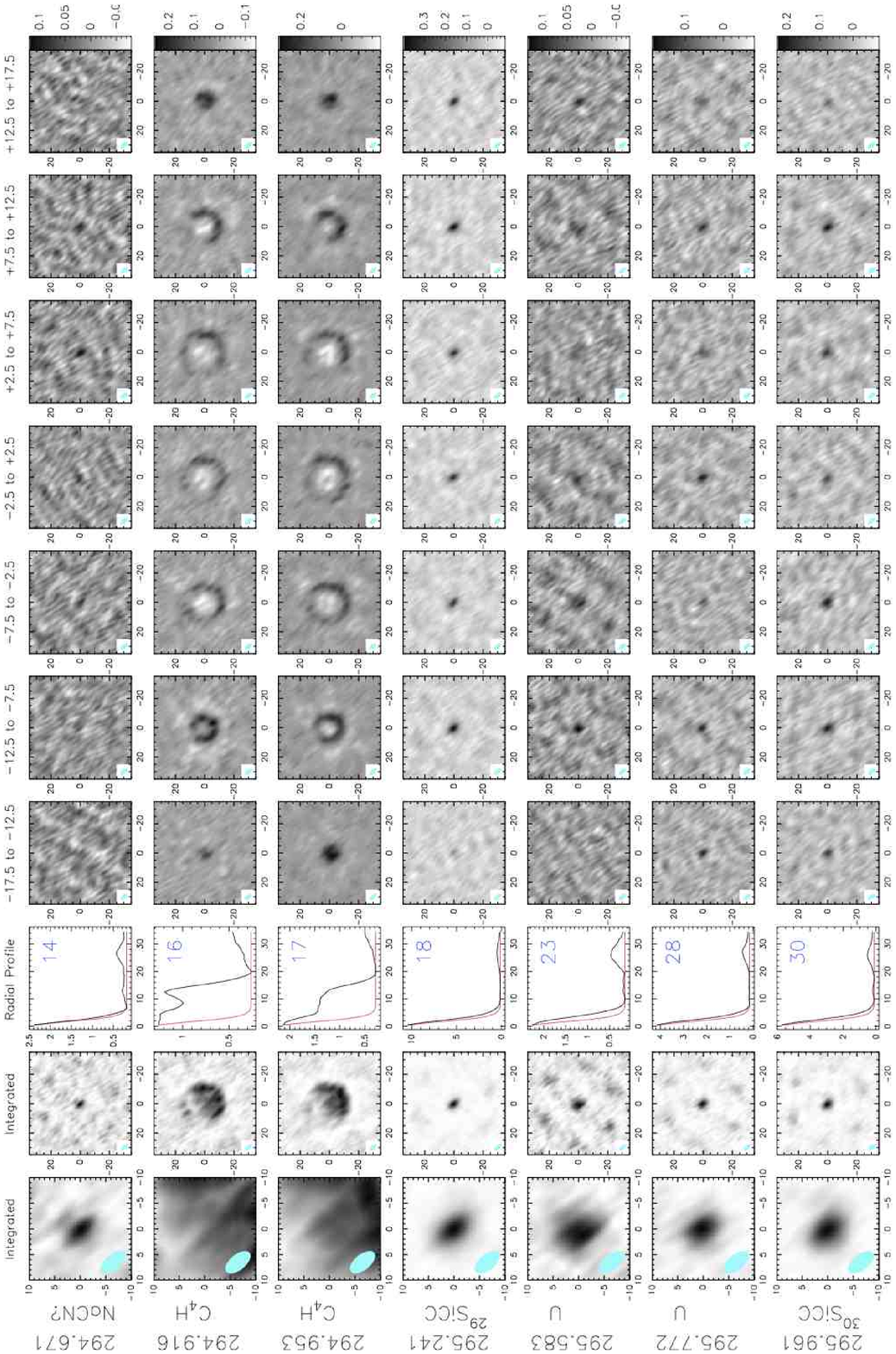}
\caption{continued. \label{}}
\end{figure*}
\clearpage
\addtocounter{figure}{-1}

\begin{figure*}[tbH]
\centering
\includegraphics[width=6in]{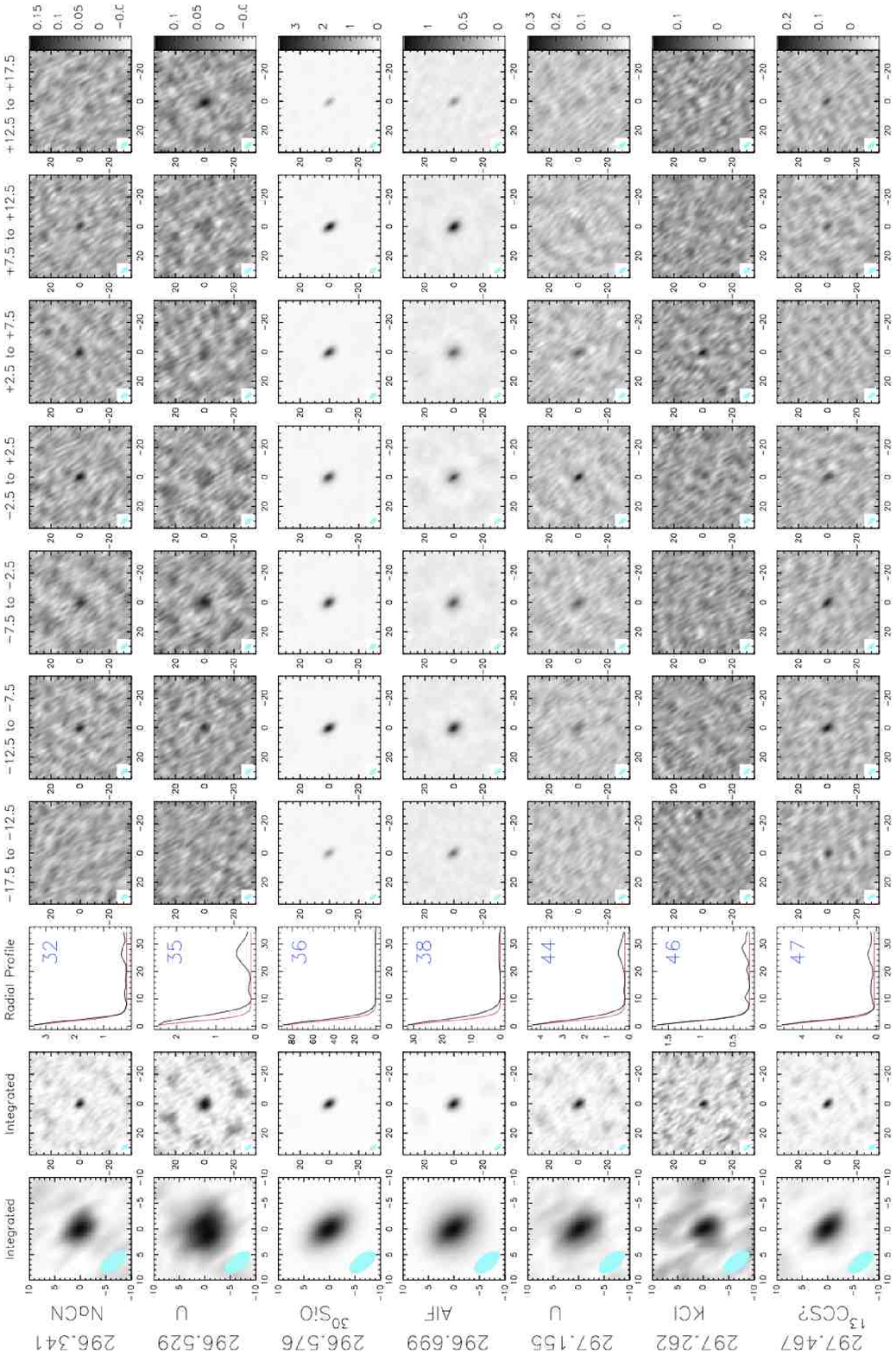}
\caption{continued.  \label{}}
\end{figure*}
\clearpage
\addtocounter{figure}{-1}

\begin{figure*}[tbH]
\centering
\includegraphics[width=6in]{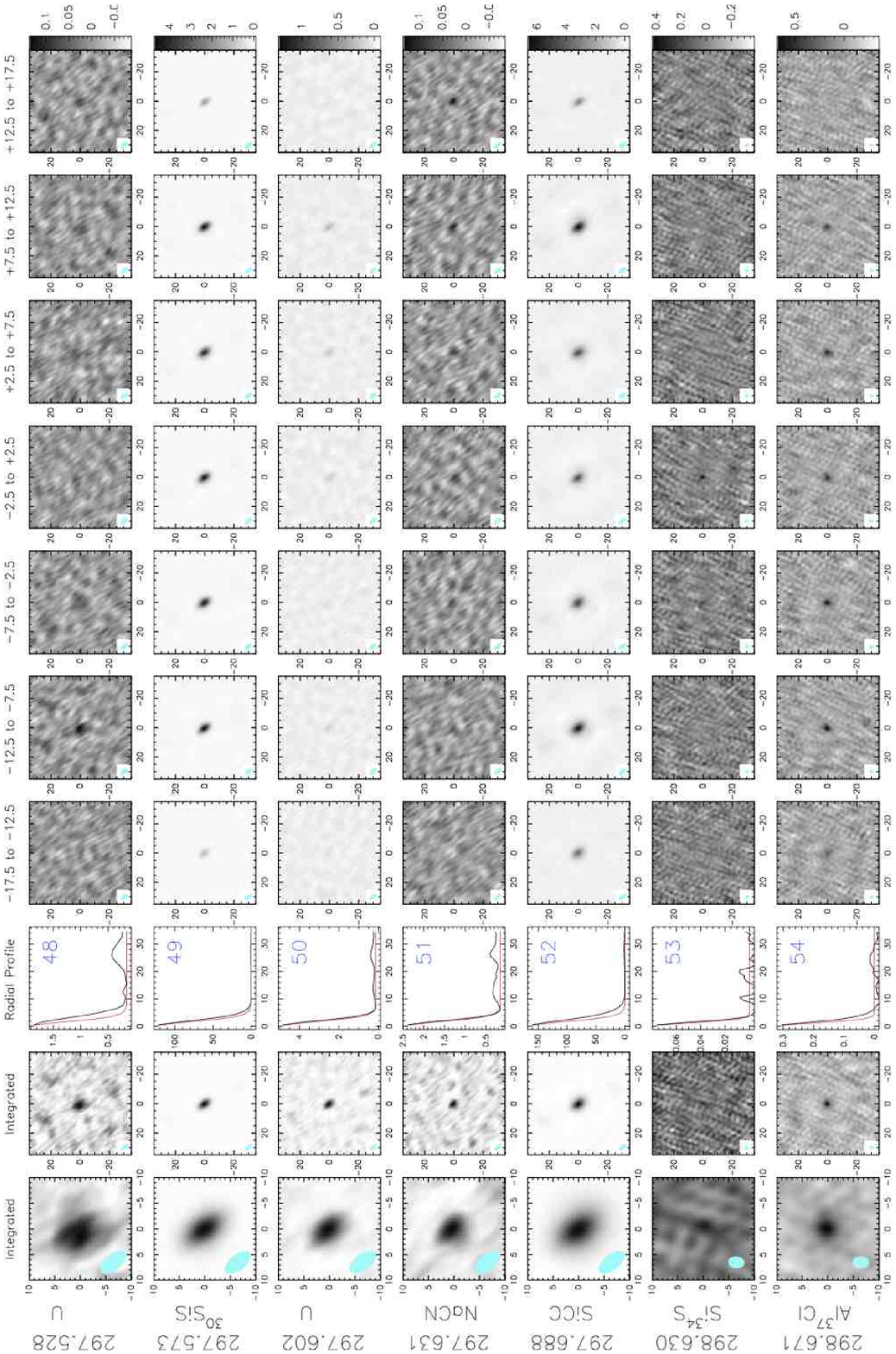}
\caption{continued.  \label{}}
\end{figure*}
\clearpage
\addtocounter{figure}{-1}

\begin{figure*}[tbH]
\centering
\includegraphics[width=6in]{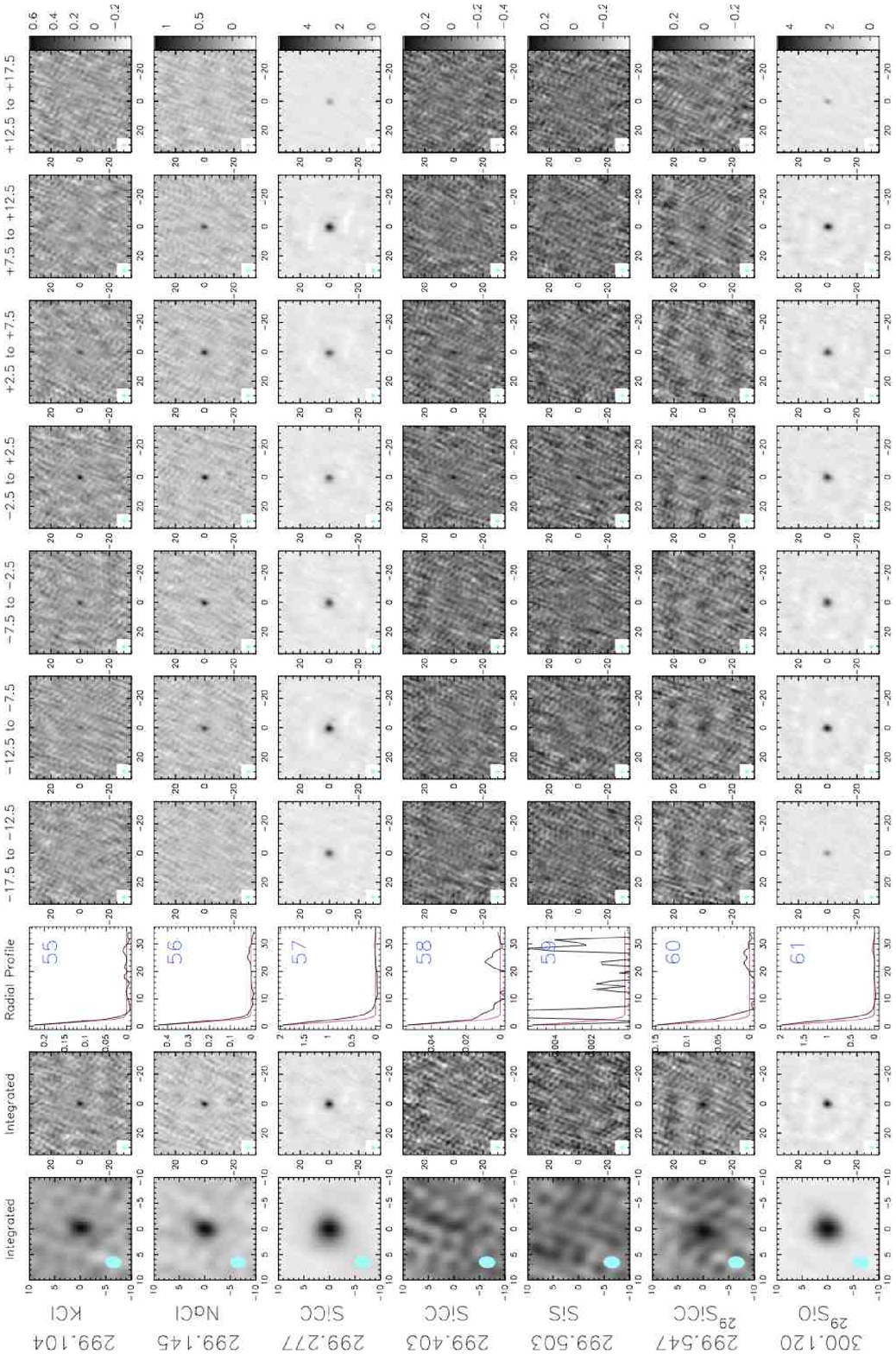}
\caption{continued. \label{}}
\end{figure*}
\clearpage
\addtocounter{figure}{-1}

\begin{figure*}[tbH]
\centering
\includegraphics[width=6in]{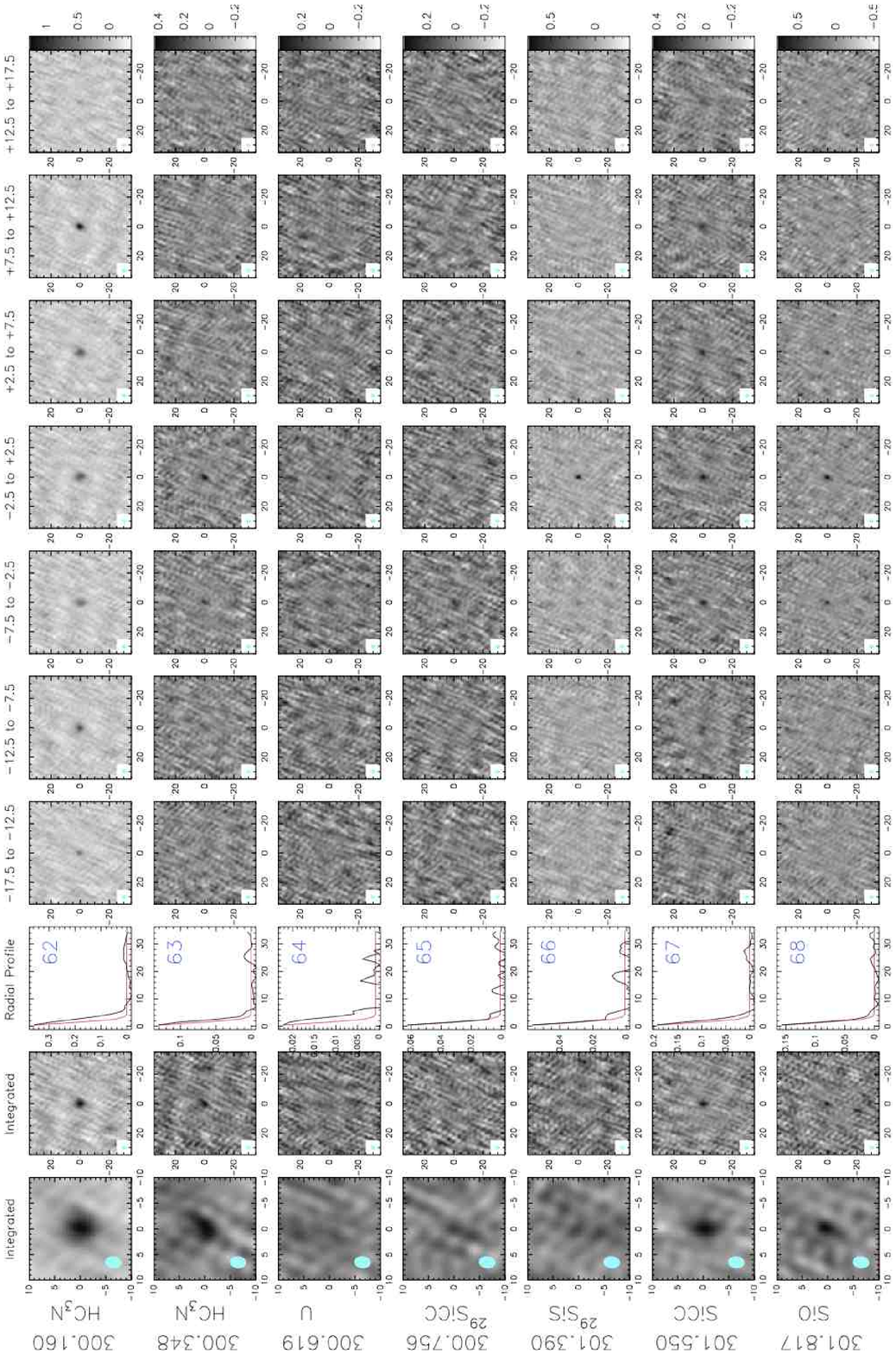}
\caption{continued.  \label{}}
\end{figure*}
\clearpage
\addtocounter{figure}{-1}

\begin{figure*}[tbH]
\centering
\includegraphics[width=6in]{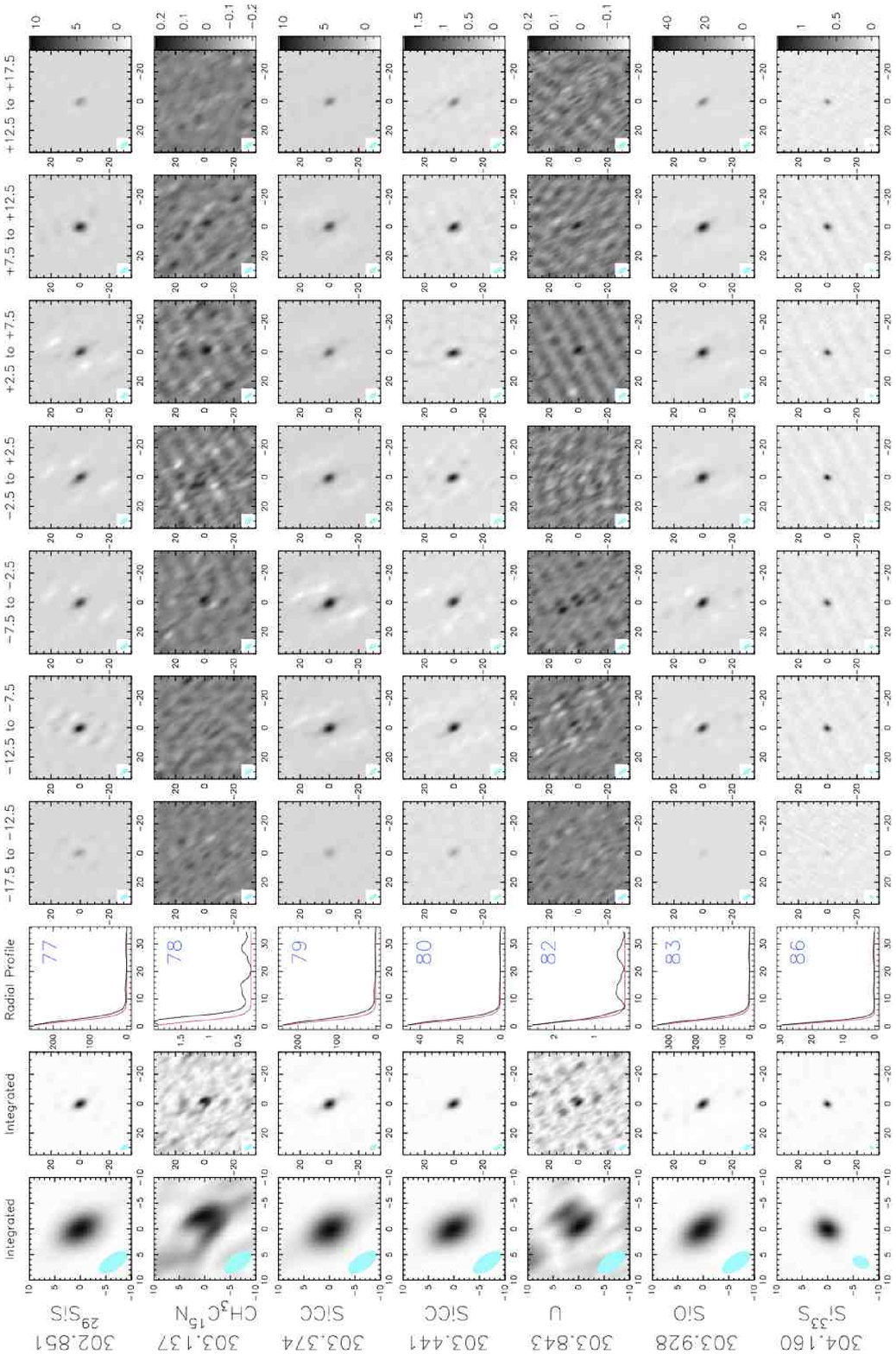}
\caption{continued.  \label{}}
\end{figure*}
\clearpage
\addtocounter{figure}{-1}

\begin{figure*}[tbH]
\centering
\includegraphics[width=6in]{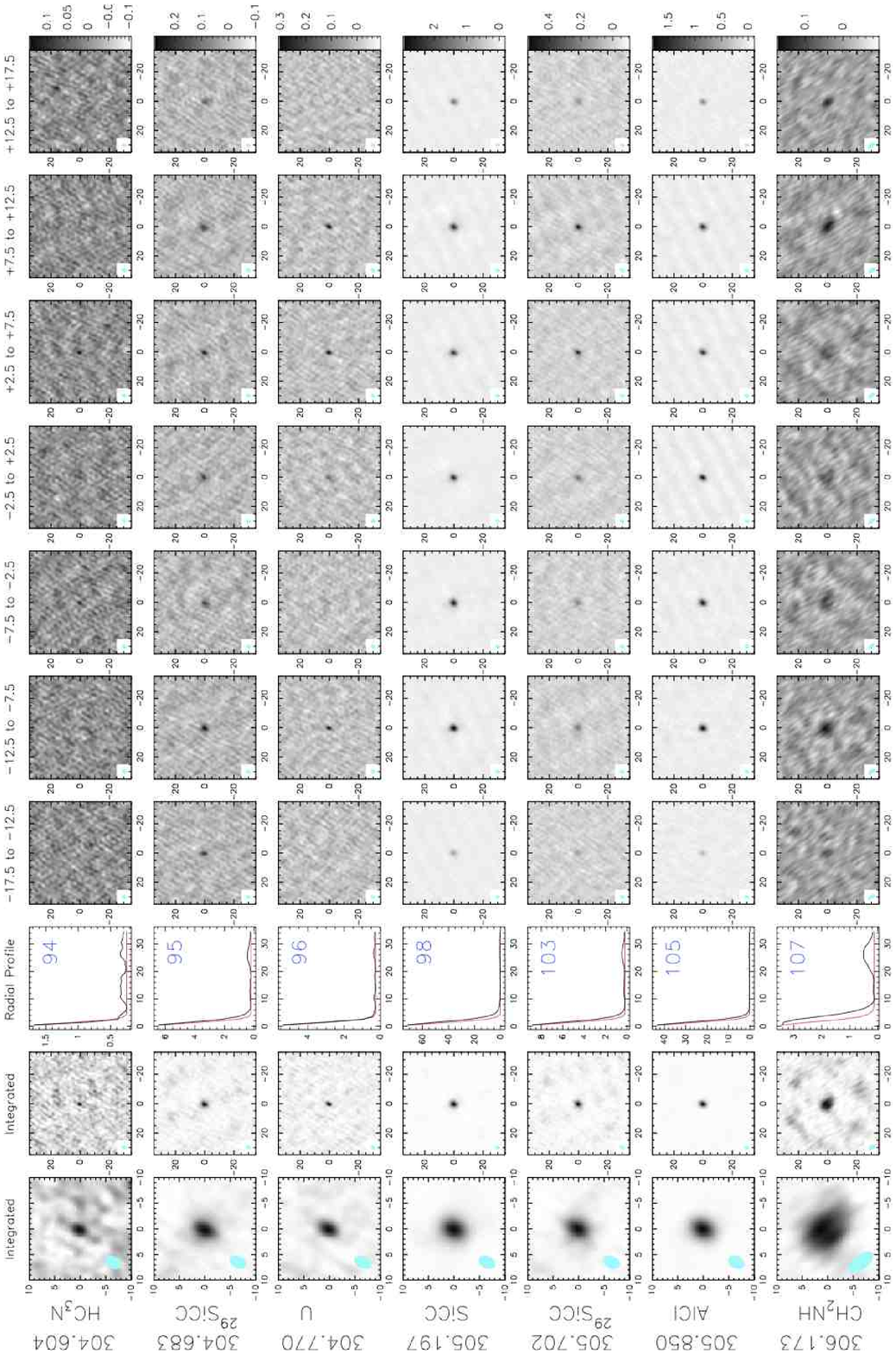}
\caption{continued.  \label{}}
\end{figure*}
\clearpage
\addtocounter{figure}{-1}

\begin{figure*}[tbH]
\centering
\includegraphics[width=6in]{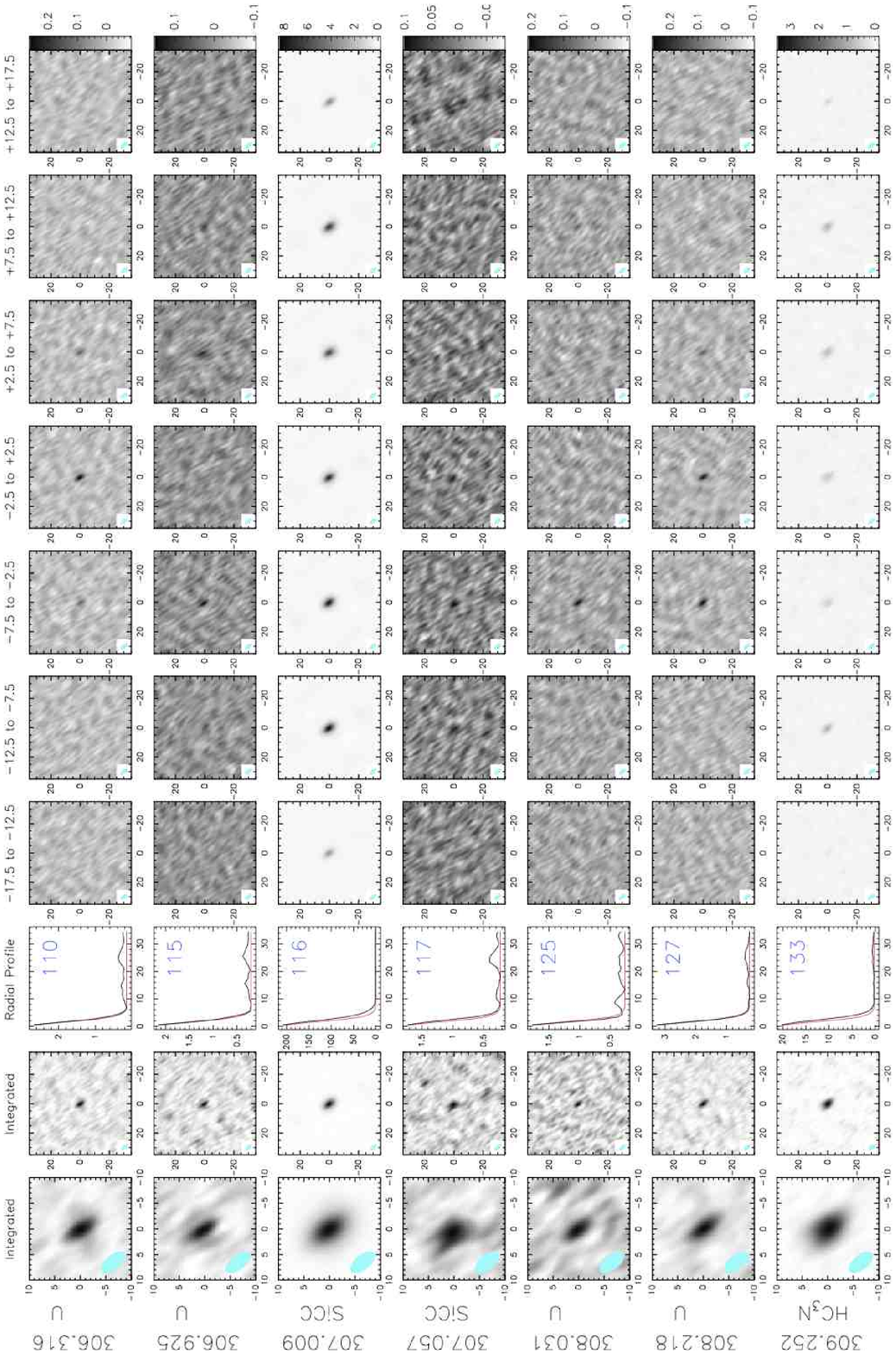}
\caption{continued.  \label{}}
\end{figure*}
\clearpage
\addtocounter{figure}{-1}

\begin{figure*}[tbH]
\centering
\includegraphics[width=6in]{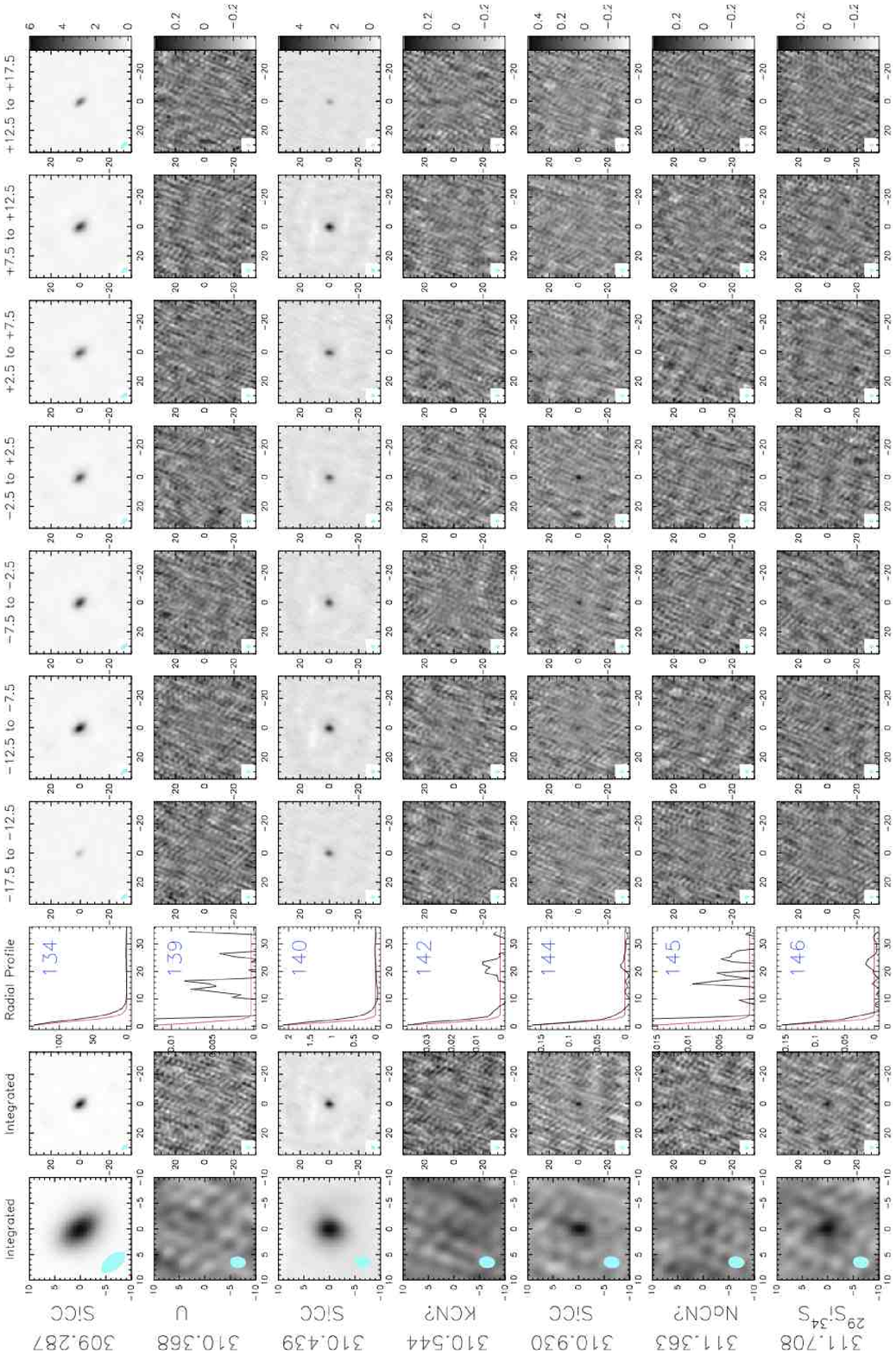}
\caption{continued.  \label{}}
\end{figure*}
\clearpage
\addtocounter{figure}{-1}

\begin{figure*}[tbH]
\centering
\includegraphics[width=6in]{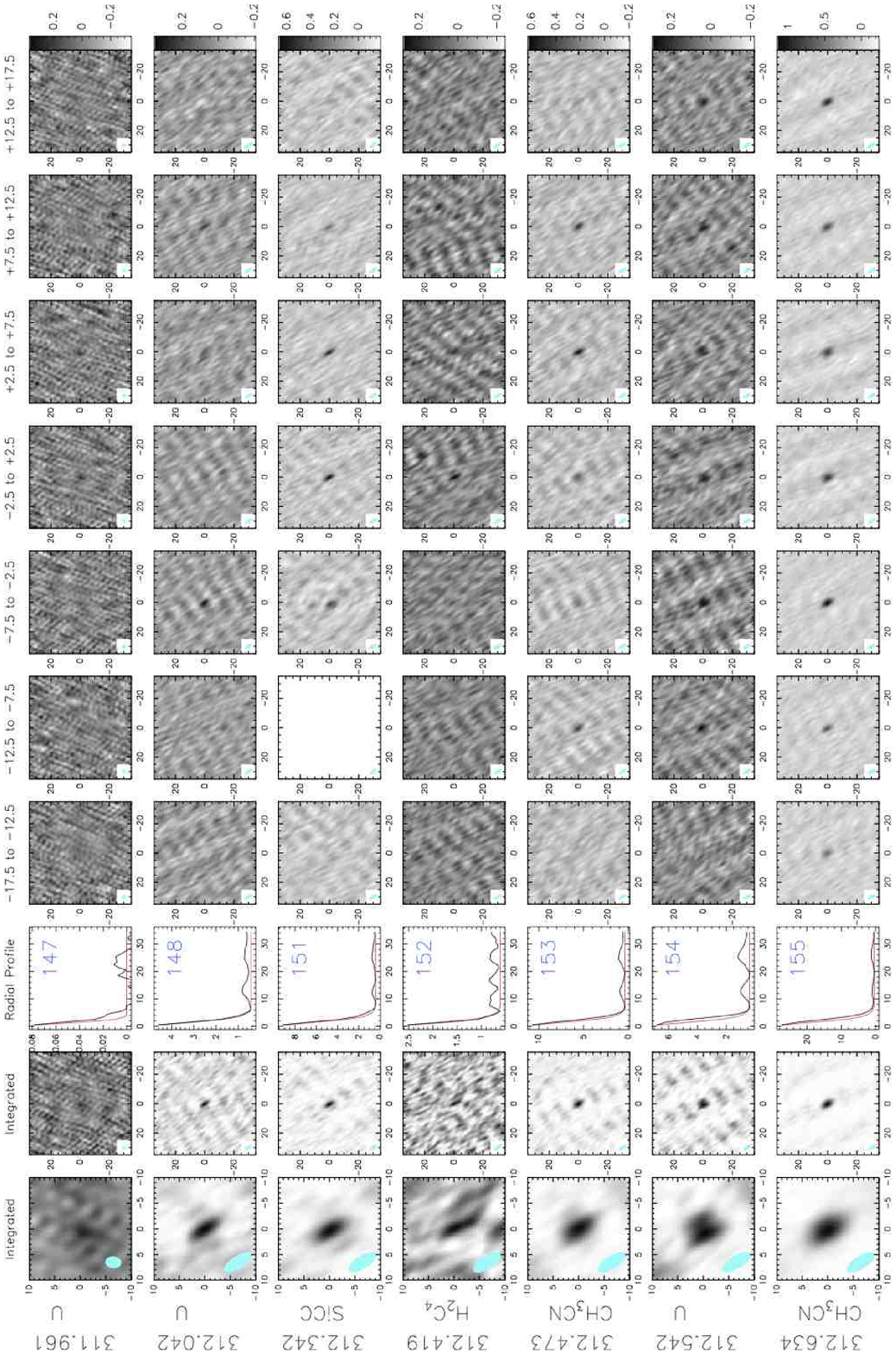}
\caption{continued.  \label{}}
\end{figure*}
\clearpage
\addtocounter{figure}{-1}

\begin{figure*}[tbH]
\centering
\includegraphics[width=6in]{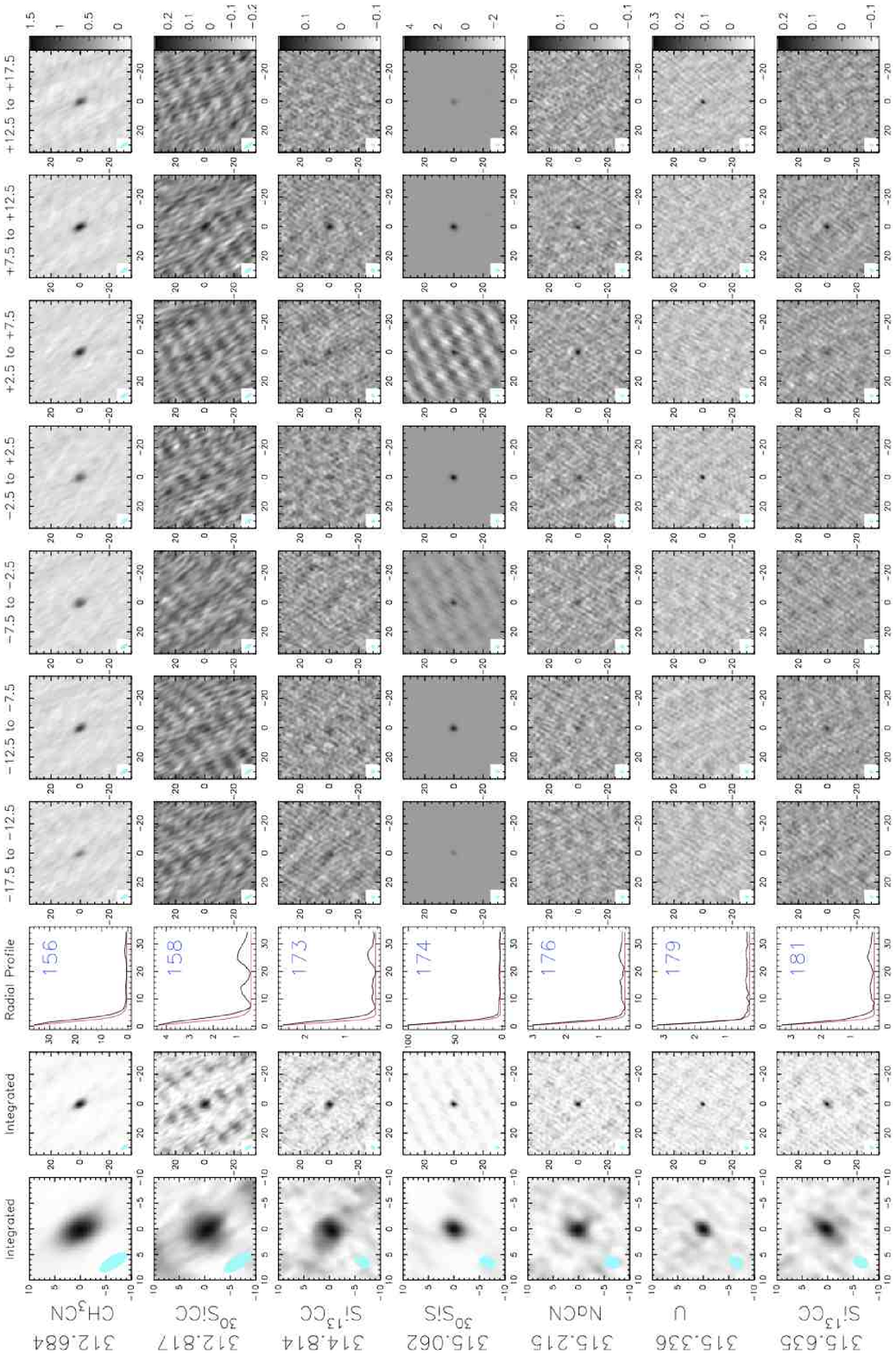}
\caption{continued.  \label{}}
\end{figure*}
\clearpage
\addtocounter{figure}{-1}

\begin{figure*}[tbH]
\centering
\includegraphics[width=6in]{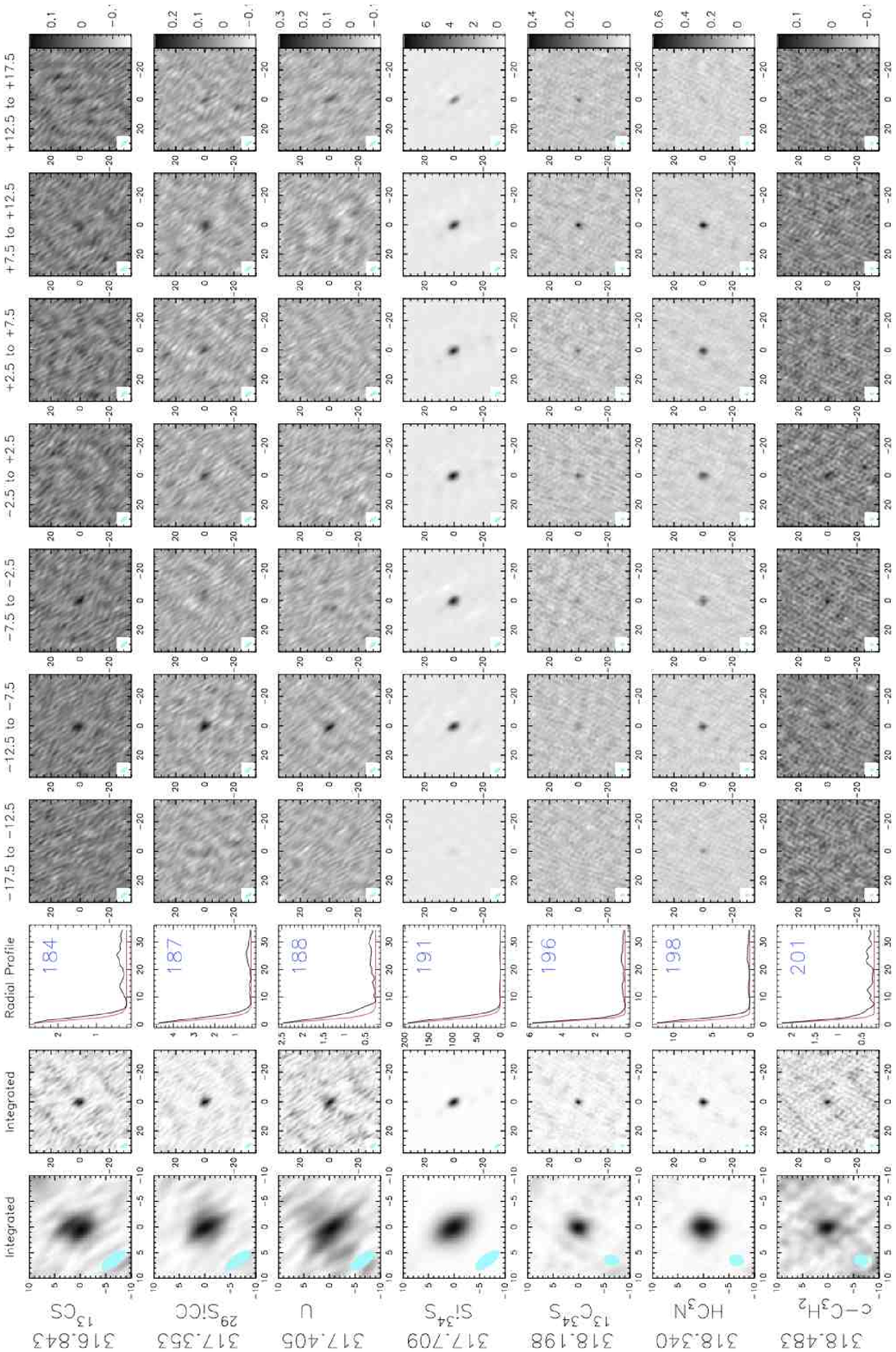}
\caption{continued.  \label{}}
\end{figure*}
\clearpage
\addtocounter{figure}{-1}

\begin{figure*}[tbH]
\centering
\includegraphics[width=6in]{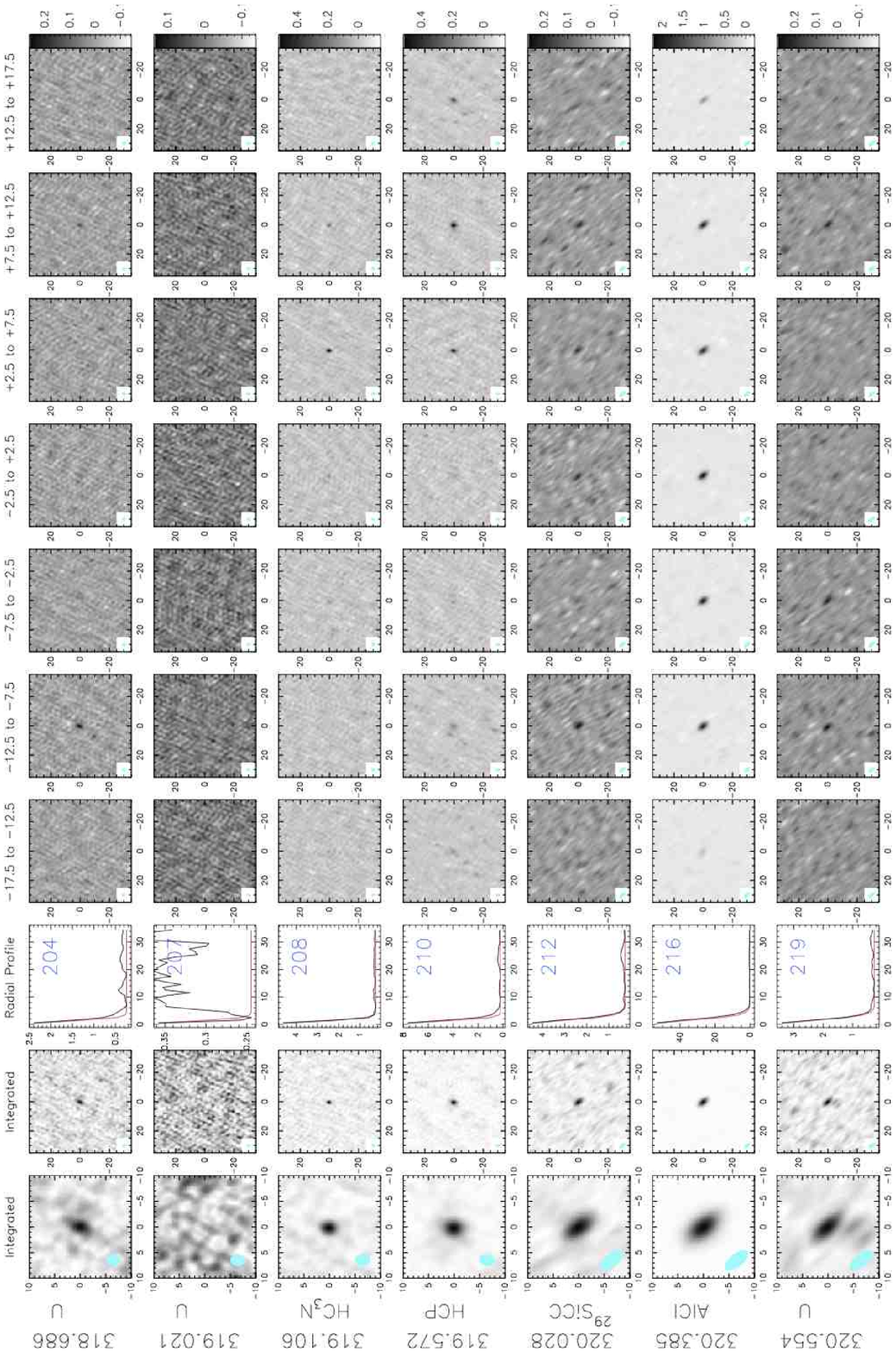}
\caption{continued.  \label{}}
\end{figure*}
\clearpage
\addtocounter{figure}{-1}

\begin{figure*}[tbH]
\centering
\includegraphics[width=6in]{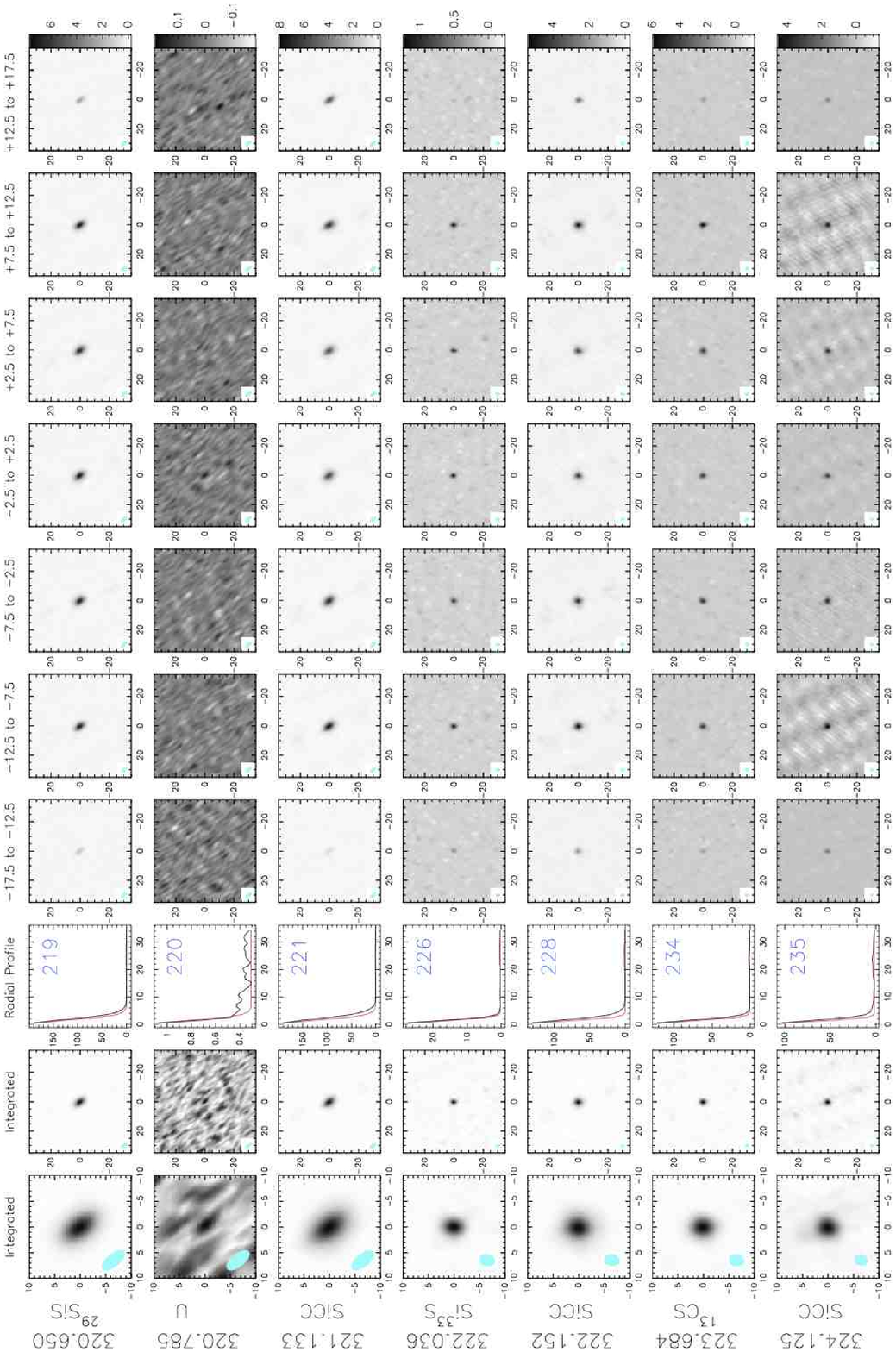}
\caption{continued.  \label{}}
\end{figure*}
\clearpage
\addtocounter{figure}{-1}

\begin{figure*}[tbH]
\centering
\includegraphics[width=6in]{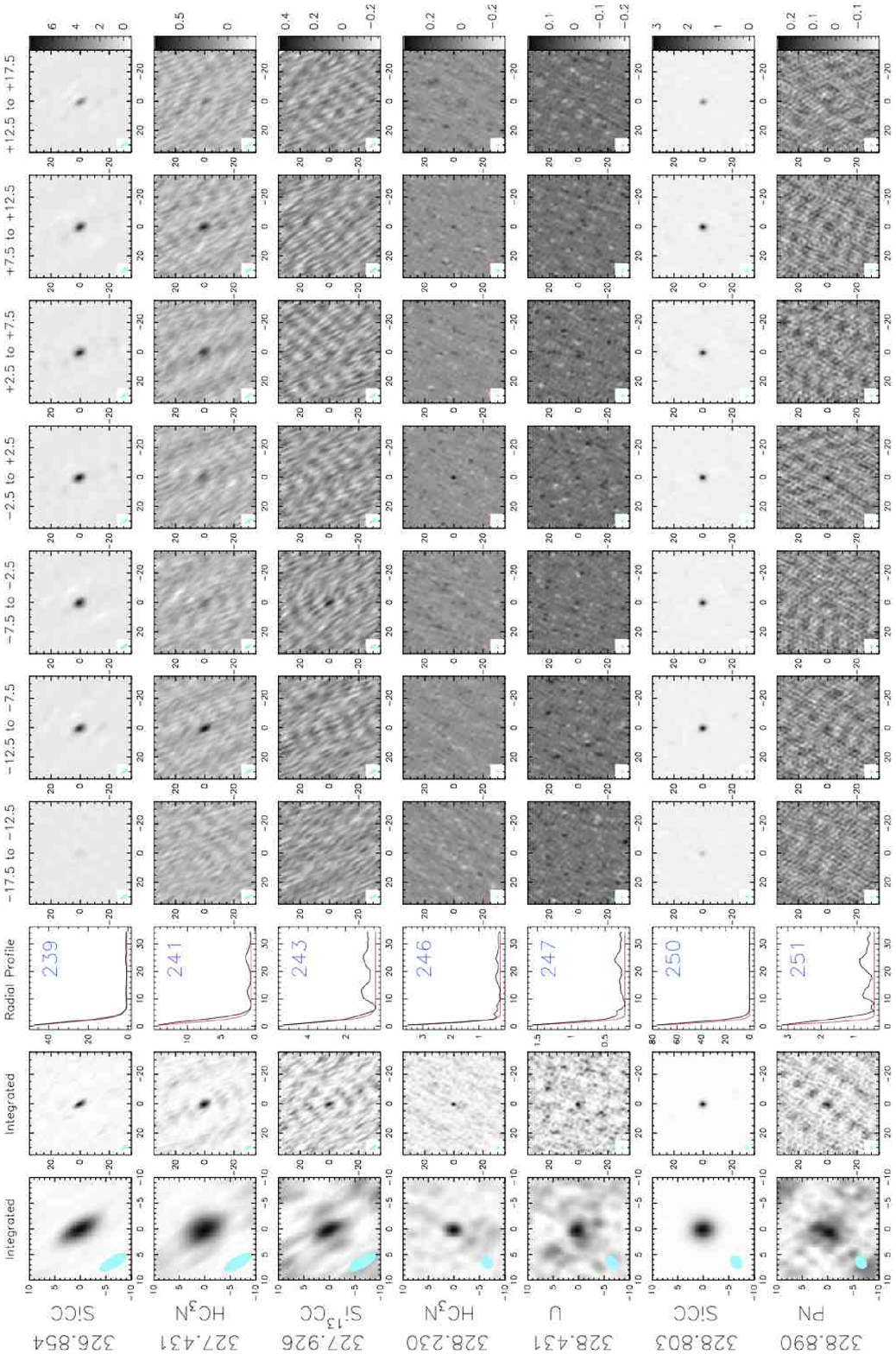}
\caption{continued.  \label{}}
\end{figure*}
\clearpage
\addtocounter{figure}{-1}

\begin{figure*}[tbH]
\centering
\includegraphics[width=6in]{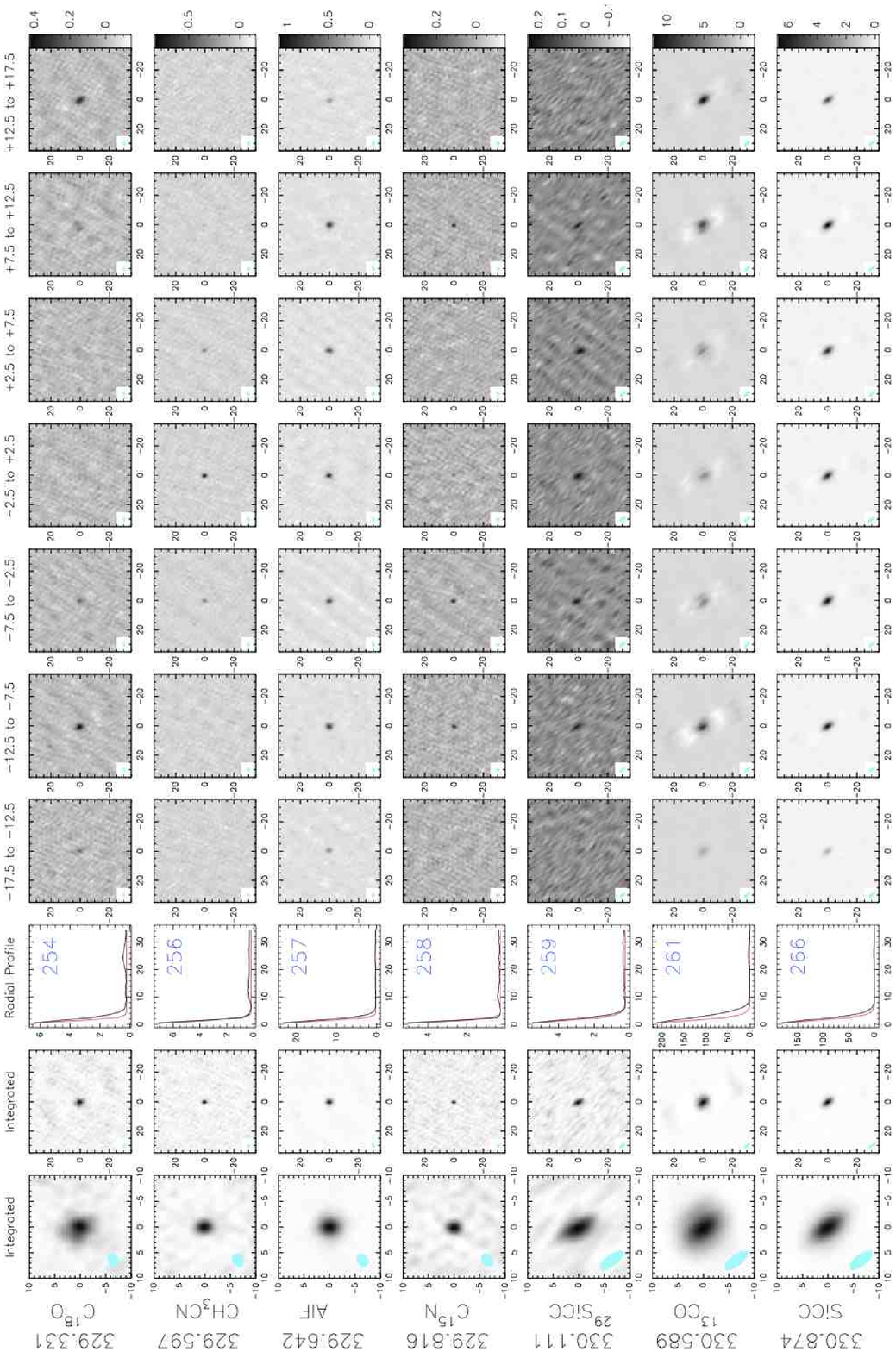}
\caption{continued.  \label{}}
\end{figure*}
\clearpage
\addtocounter{figure}{-1}

\begin{figure*}[tbH]
\centering
\includegraphics[width=6in]{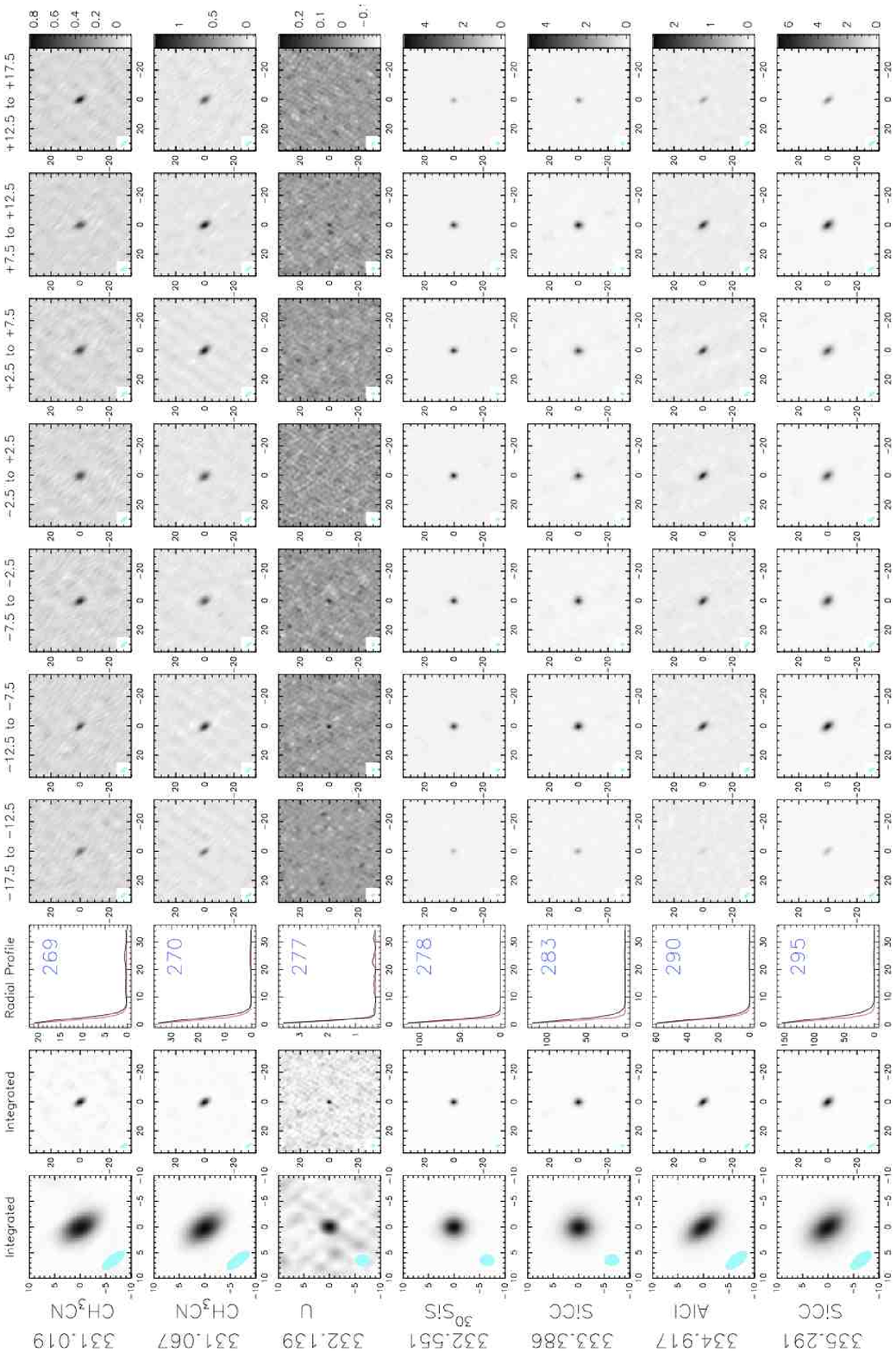}
\caption{continued.  \label{}}
\end{figure*}
\clearpage
\addtocounter{figure}{-1}

\begin{figure*}[tbH]
\centering
\includegraphics[width=6in]{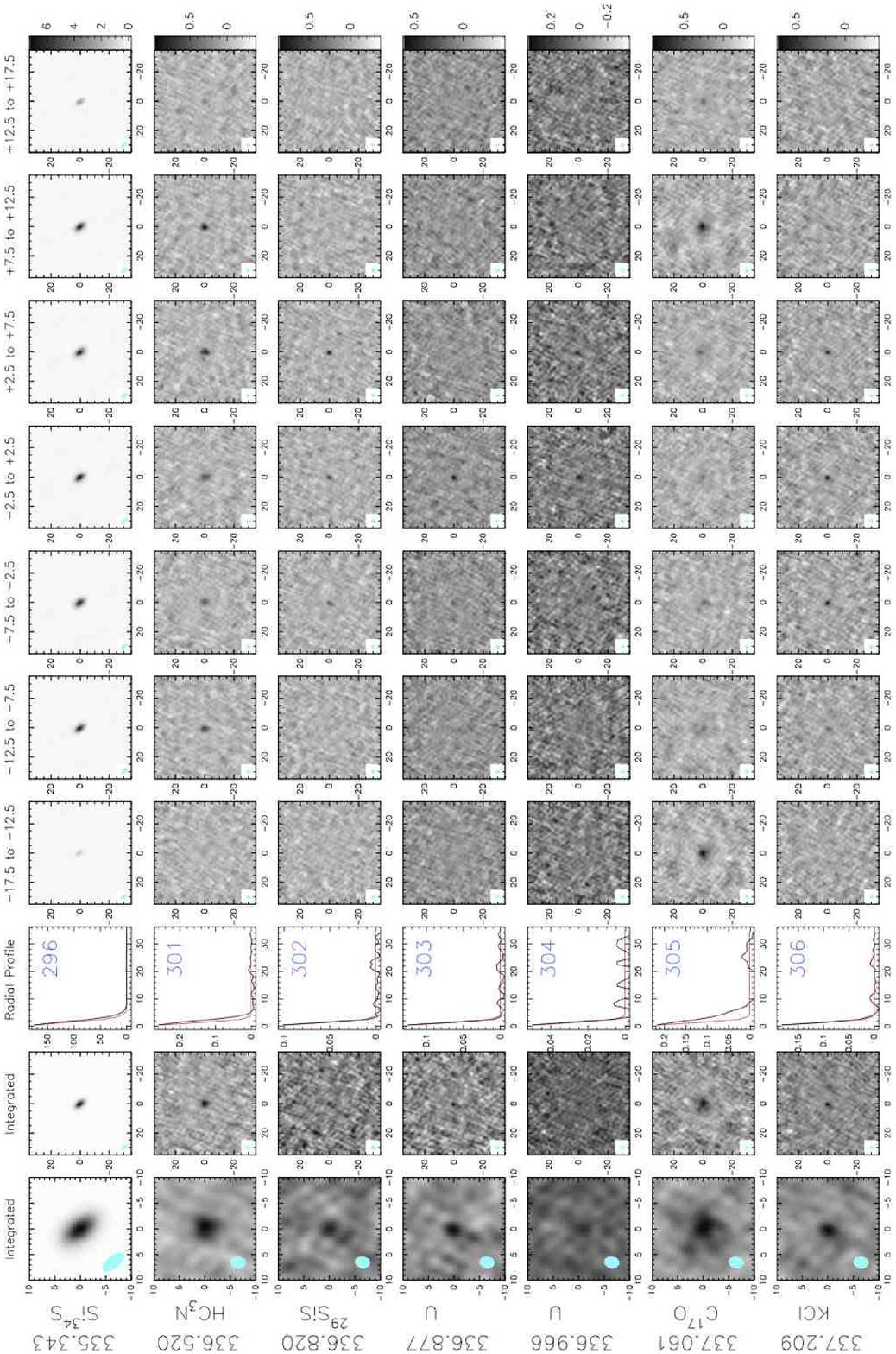}
\caption{continued.  \label{}}
\end{figure*}
\clearpage
\addtocounter{figure}{-1}

\begin{figure*}[tbH]
\centering
\includegraphics[width=6in]{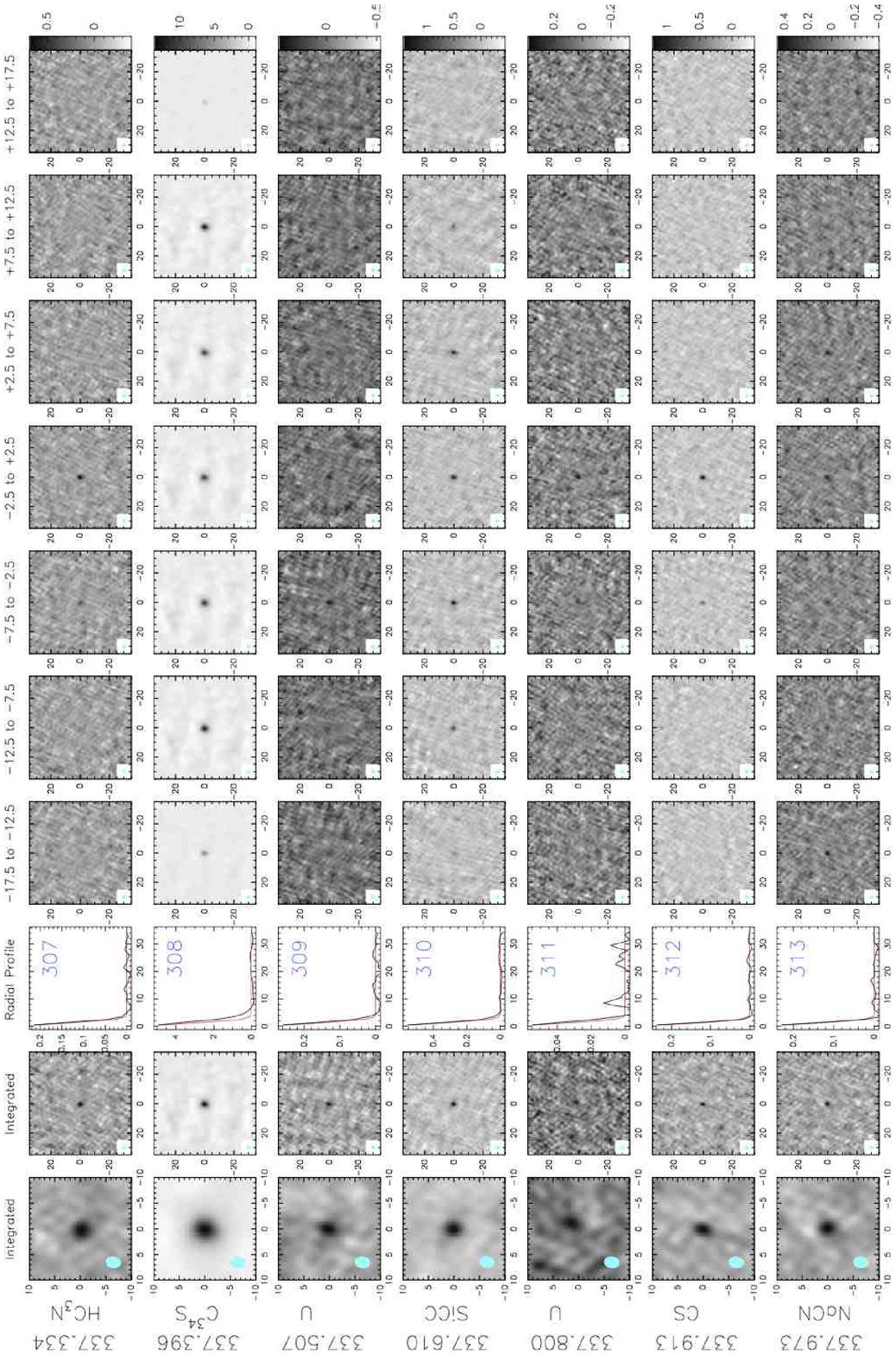}
\caption{continued.  \label{}}
\end{figure*}
\clearpage
\addtocounter{figure}{-1}

\begin{figure*}[tbH]
\centering
\includegraphics[width=6in]{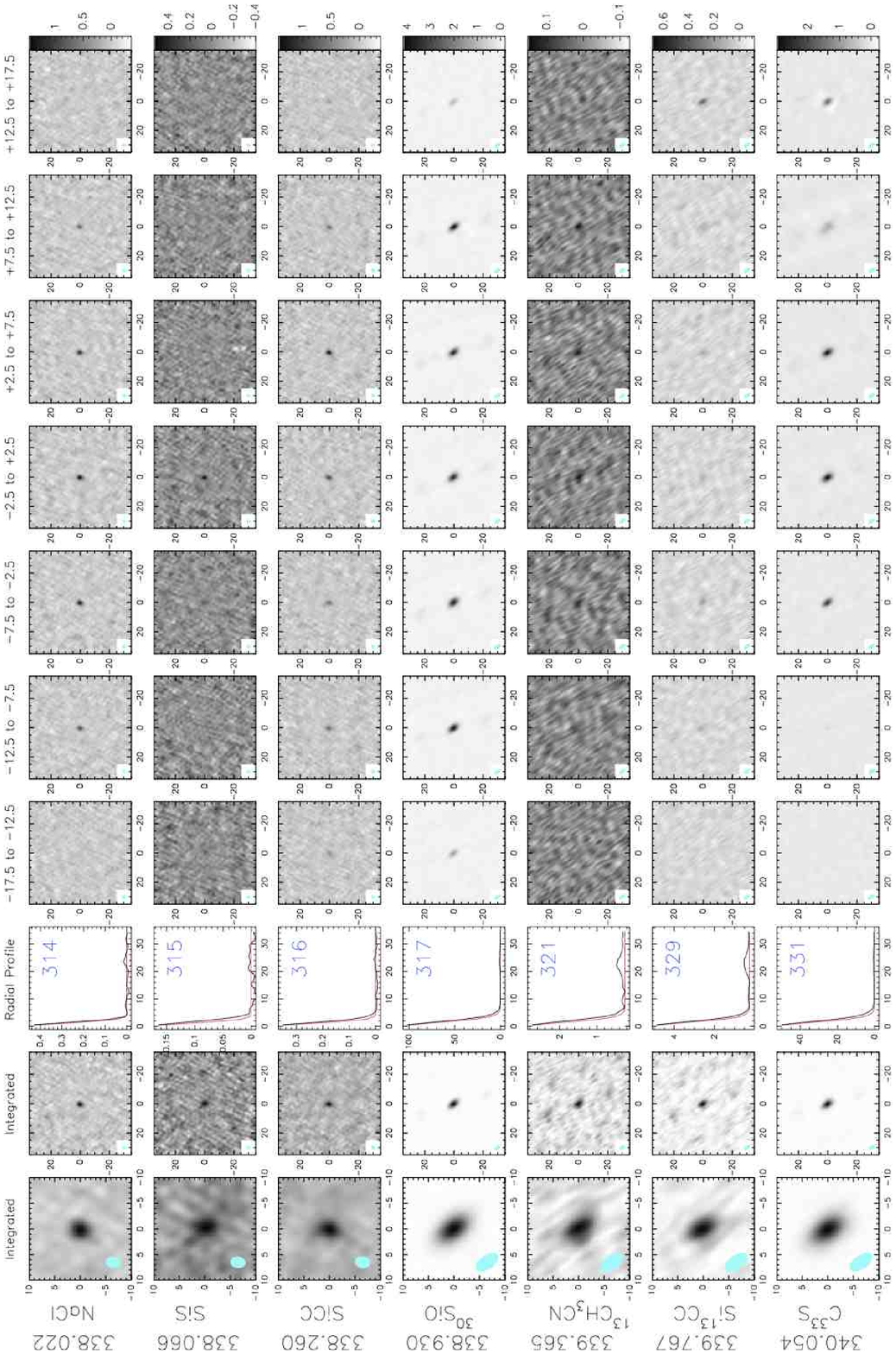}
\caption{continued.  \label{}}
\end{figure*}
\clearpage
\addtocounter{figure}{-1}

\begin{figure*}[tbH]
\centering
\includegraphics[width=6in]{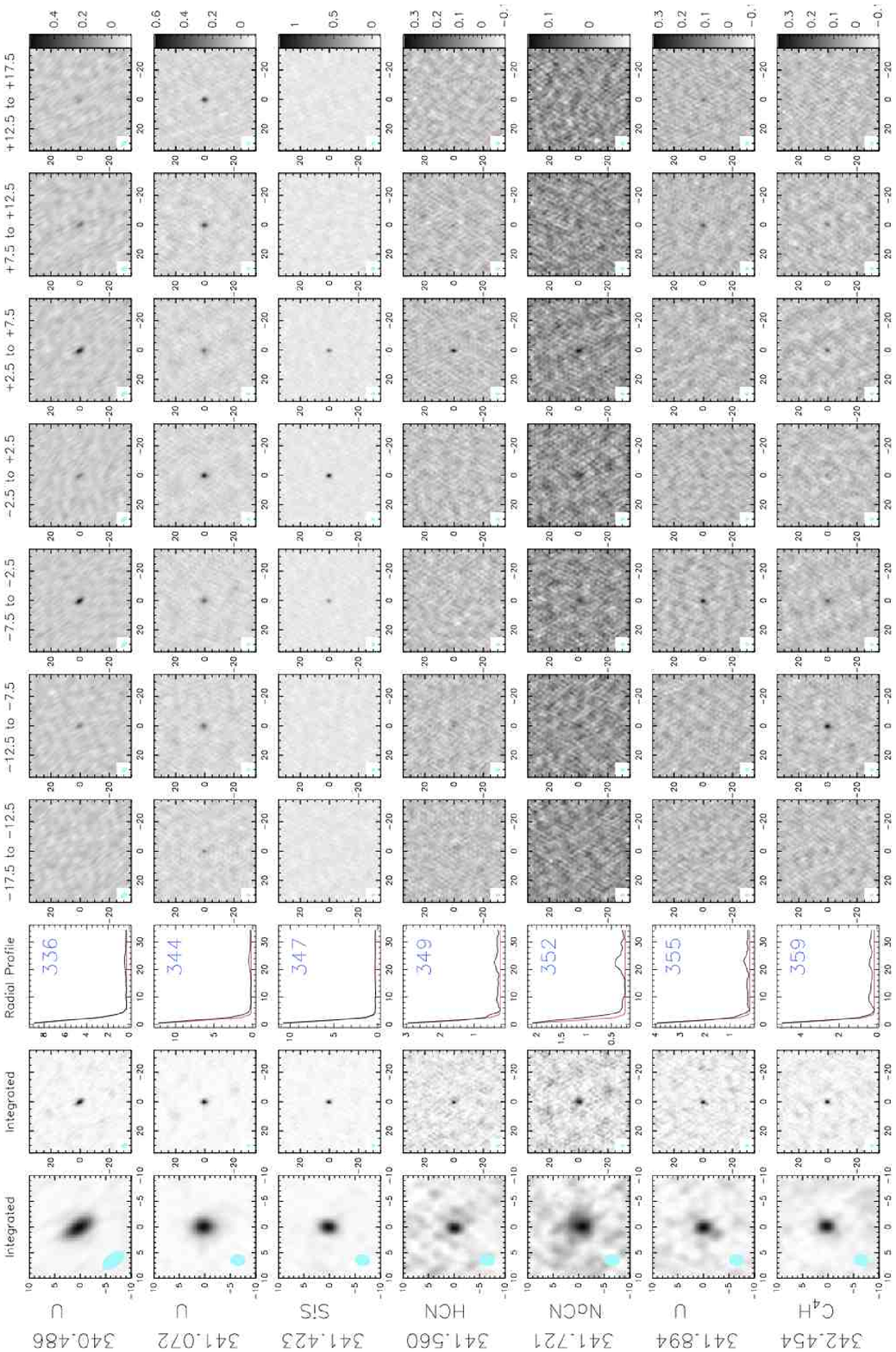}
\caption{continued.  \label{}}
\end{figure*}
\clearpage
\addtocounter{figure}{-1}

\begin{figure*}[tbH]
\centering
\includegraphics[width=6in]{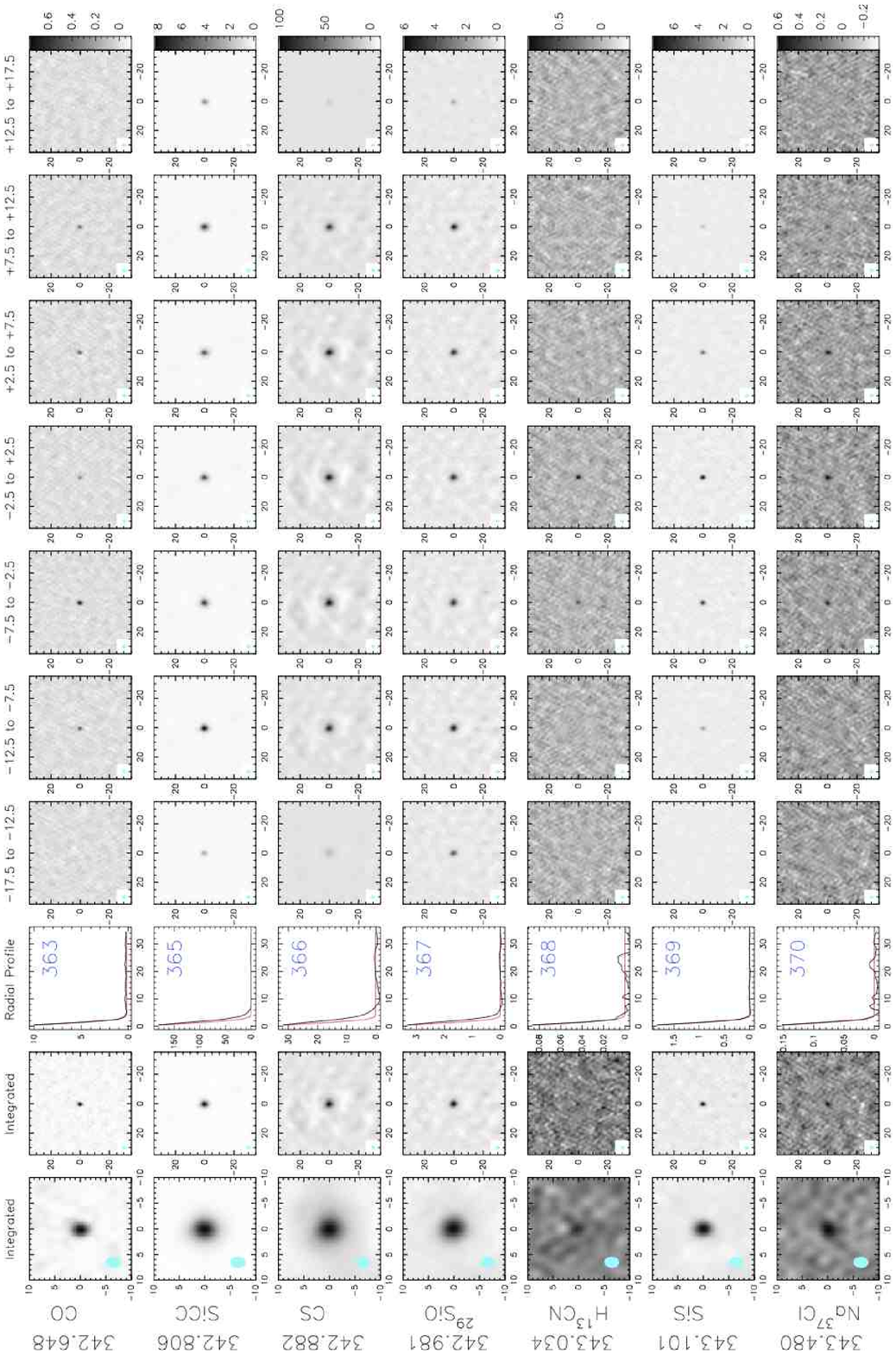}
\caption{continued.  \label{}}
\end{figure*}
\clearpage
\addtocounter{figure}{-1}

\begin{figure*}[tbH]
\centering
\includegraphics[width=6in]{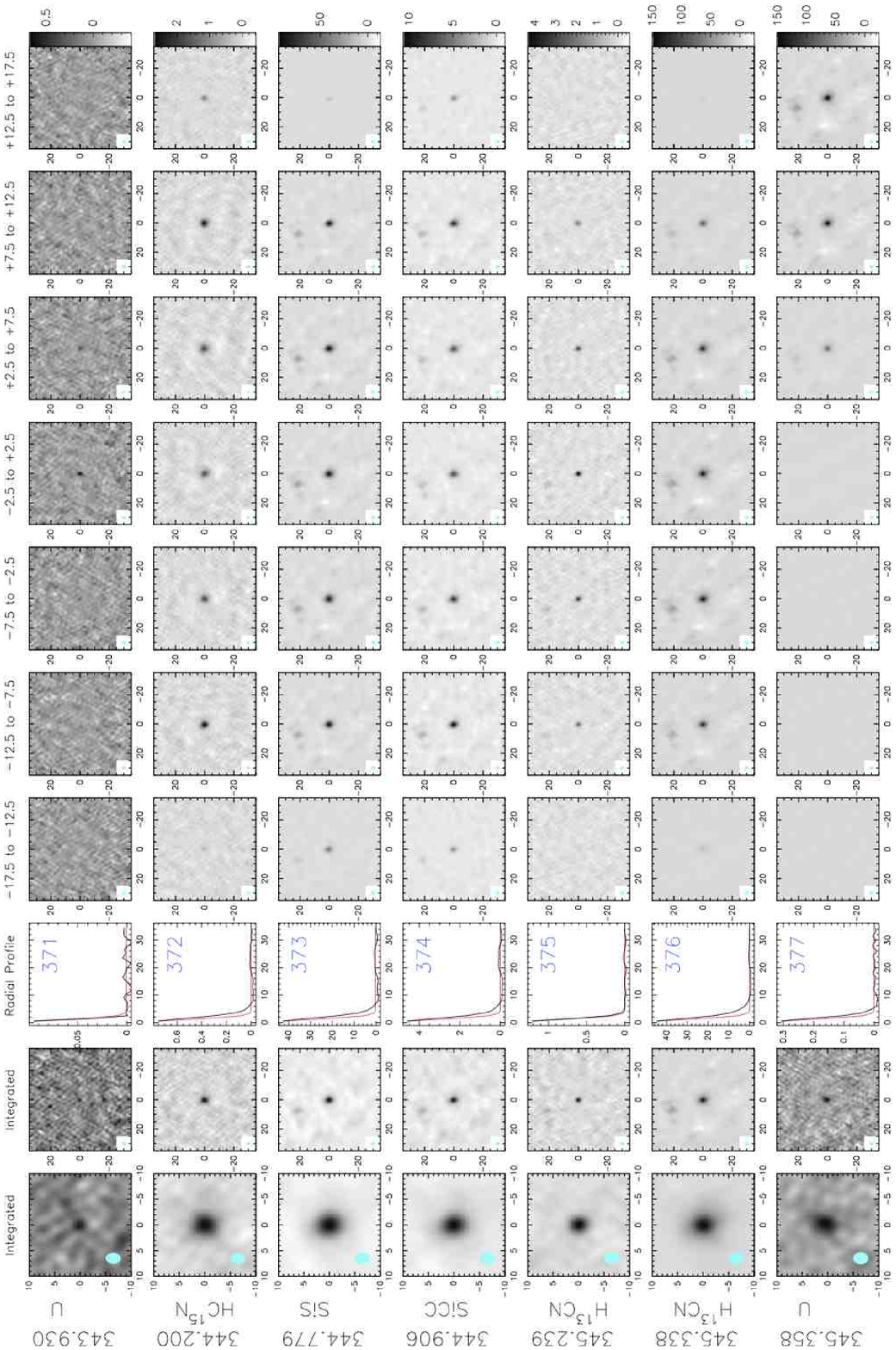}
\caption{continued.  \label{}}
\end{figure*}
\clearpage
\addtocounter{figure}{-1}

\begin{figure*}[tbH]
\centering
\includegraphics[width=6in]{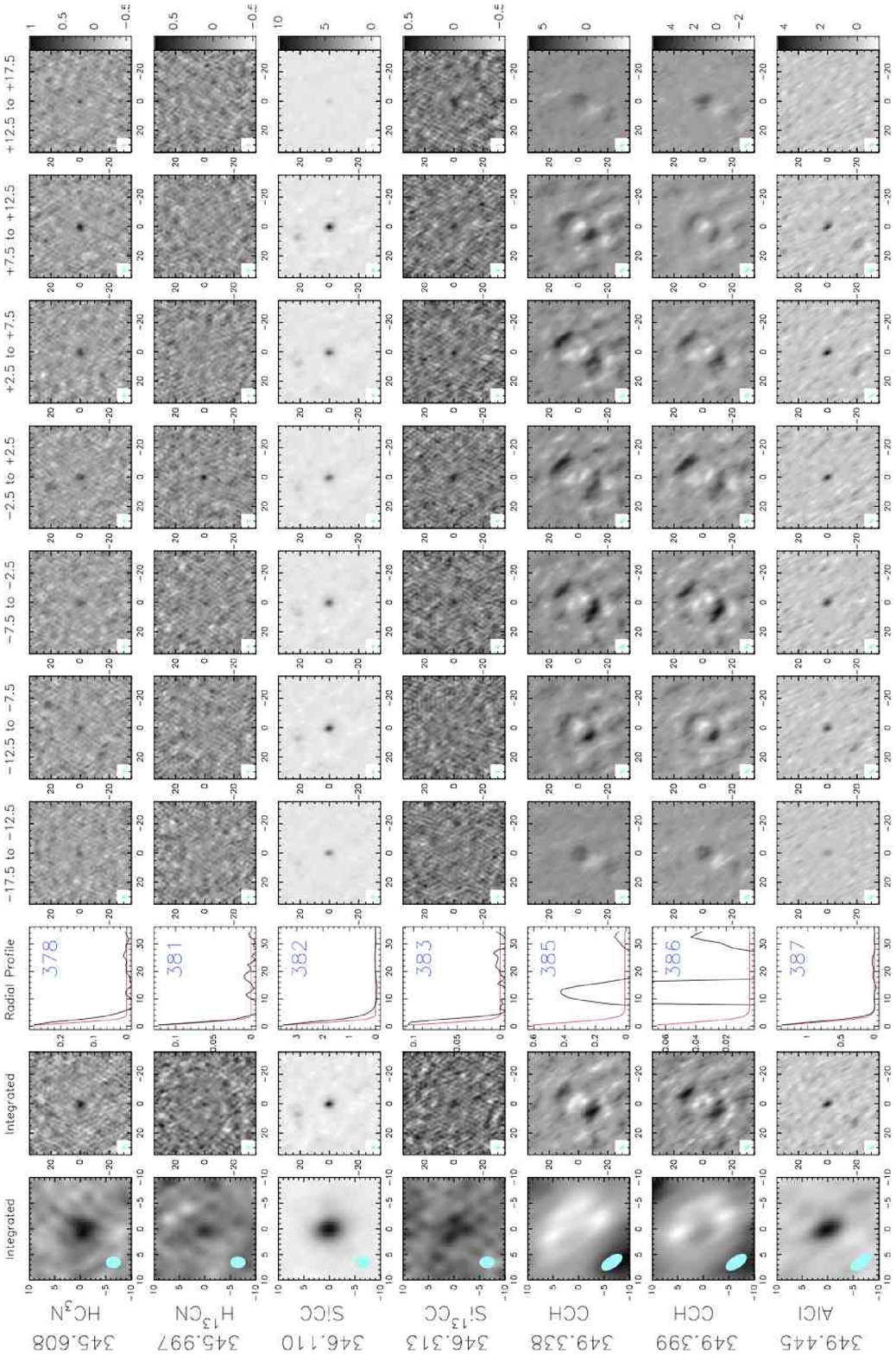}
\caption{continued.  \label{}}
\end{figure*}
\clearpage
\addtocounter{figure}{-1}

\begin{figure*}[tbH]
\centering
\includegraphics[width=6in]{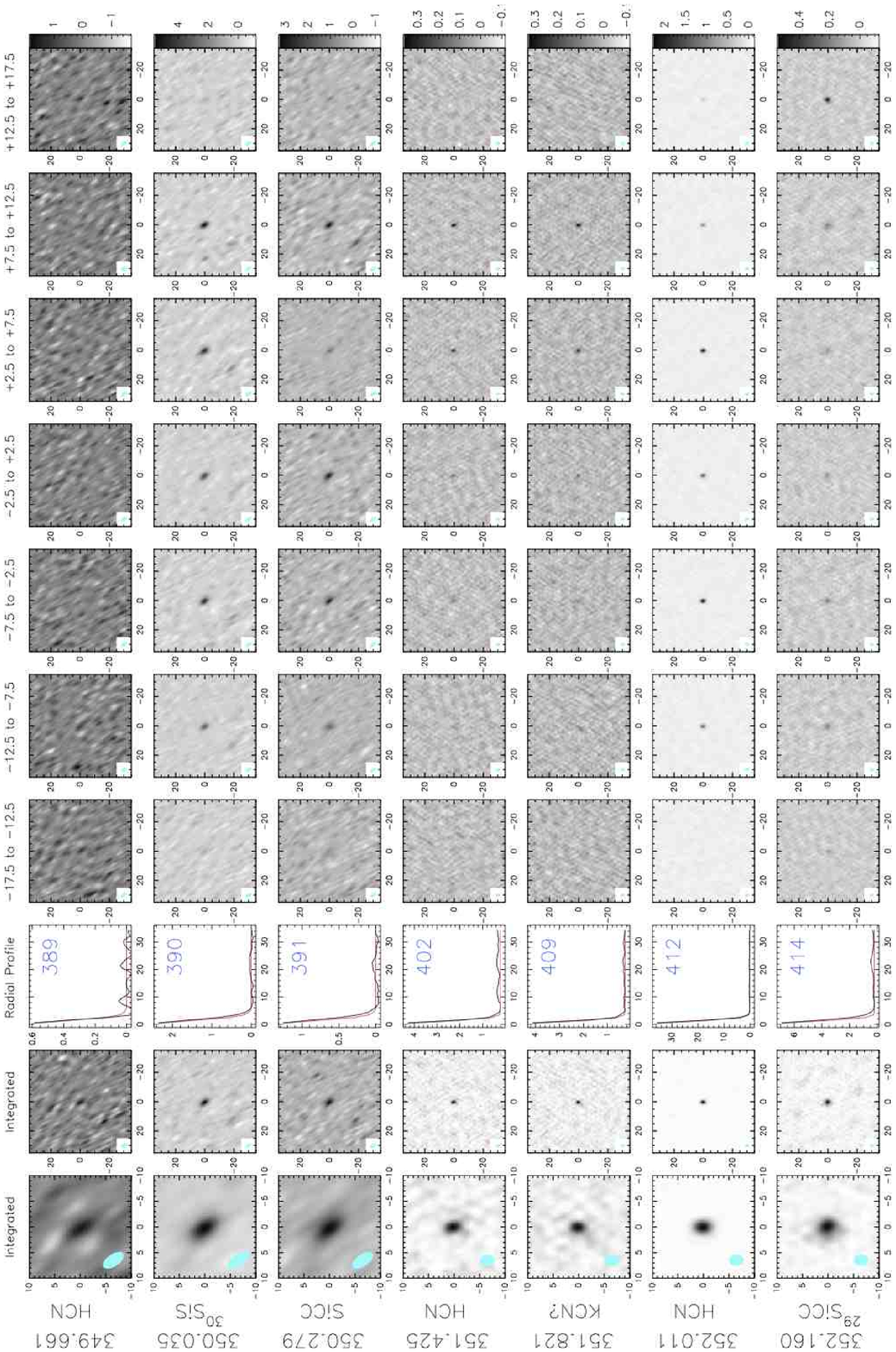}
\caption{continued.  \label{}}
\end{figure*}
\clearpage
\addtocounter{figure}{-1}

\begin{figure*}[tbH]
\centering
\includegraphics[width=6in]{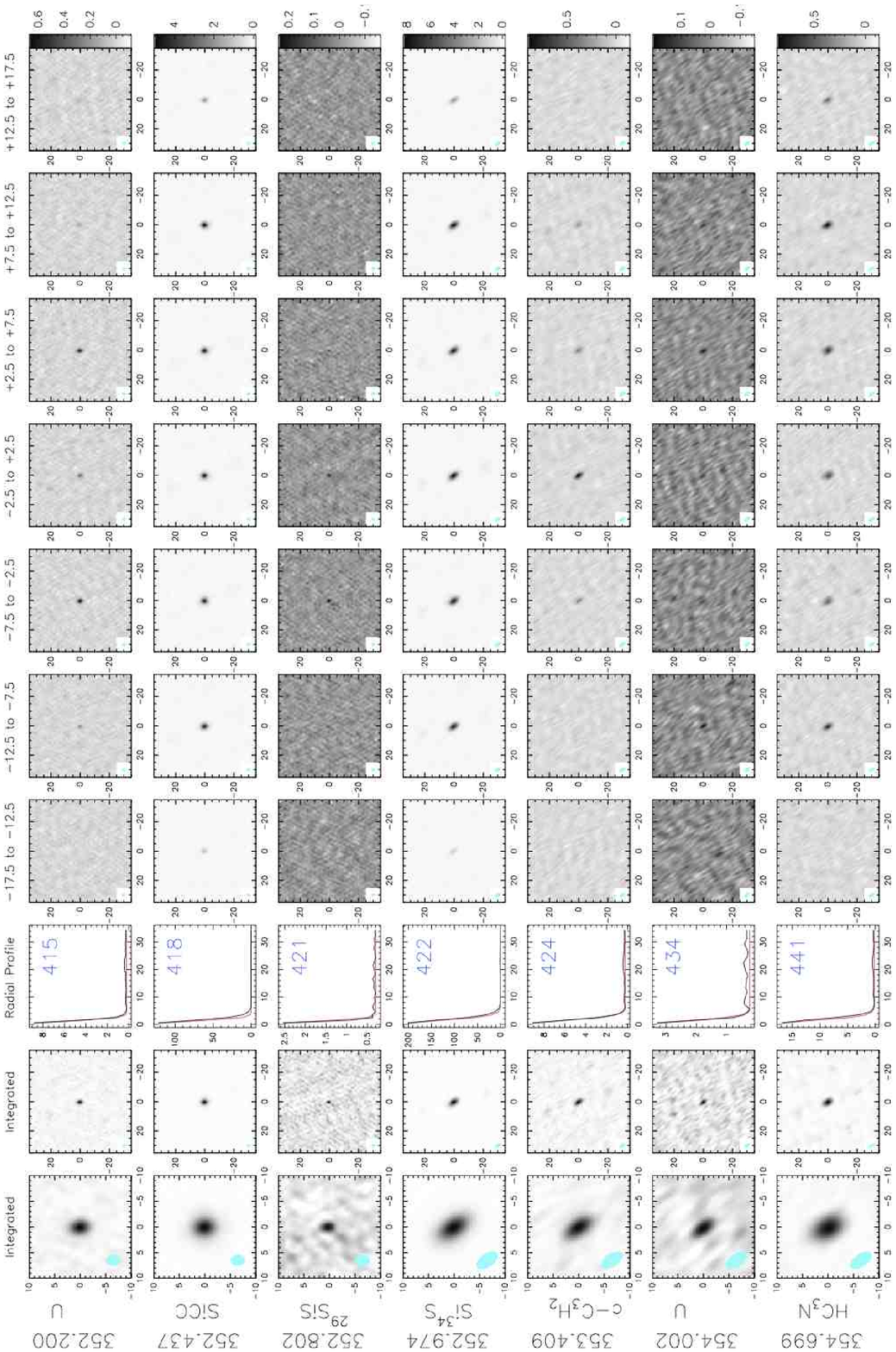}
\caption{continued.  \label{}}
\end{figure*}
\clearpage

{\bf Figure \ref{maps}:} Each row in this matrix of images shows the emission at the line with frequency (GHz) written on the left-most column. 
The first and second columns are the integrated intensity images at two different scales, to cover the inner and outer parts of the circumstellar shell. The synthesized beam is shown as light-blue filled ellipse in the lower left-hand corner of each image.  The third column shows an azimuthally averaged radial intensity profile in black. The radial profile of the beam is shown in red. Columns 4--10 are the channel maps with integrated emission over velocity ranges as indicated on top of each map.
\clearpage

\begin{figure*}[tbH]
\centering
\includegraphics[angle=-90,width=6in]{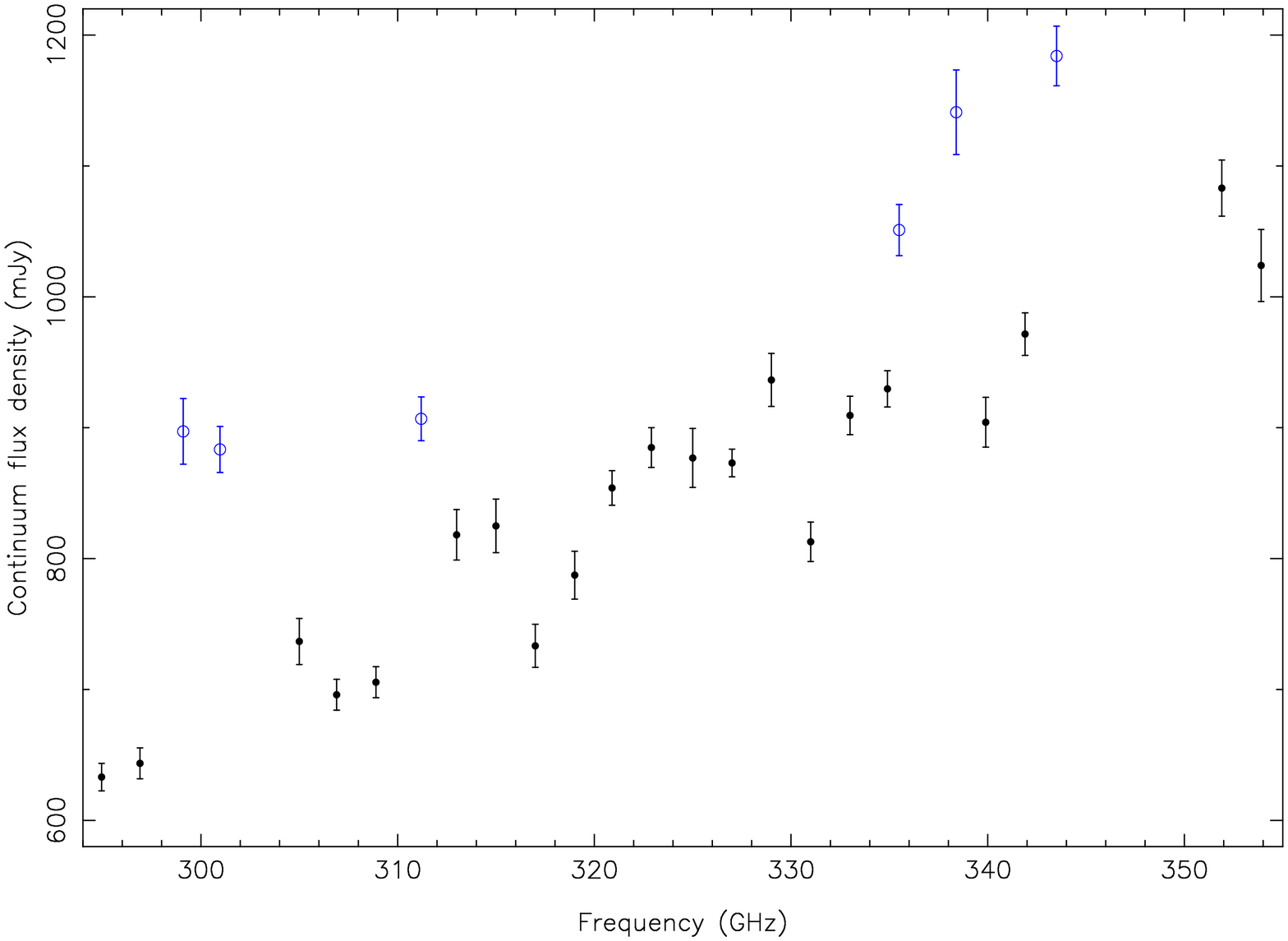}
\caption{The continuum flux density as a function of frequency is
consistent with blackbody photospheric emission (see Figure \ref{contvsfreq2}). Blue open symbols are 2007 measurements. The emission appears to
be spatially unresolved with the  3$''$ beam.  \label{contvsfreq}(See online for color)}
\end{figure*}

\clearpage

\begin{figure*}[tbH]
\centering
\includegraphics[angle=-90,width=6in]{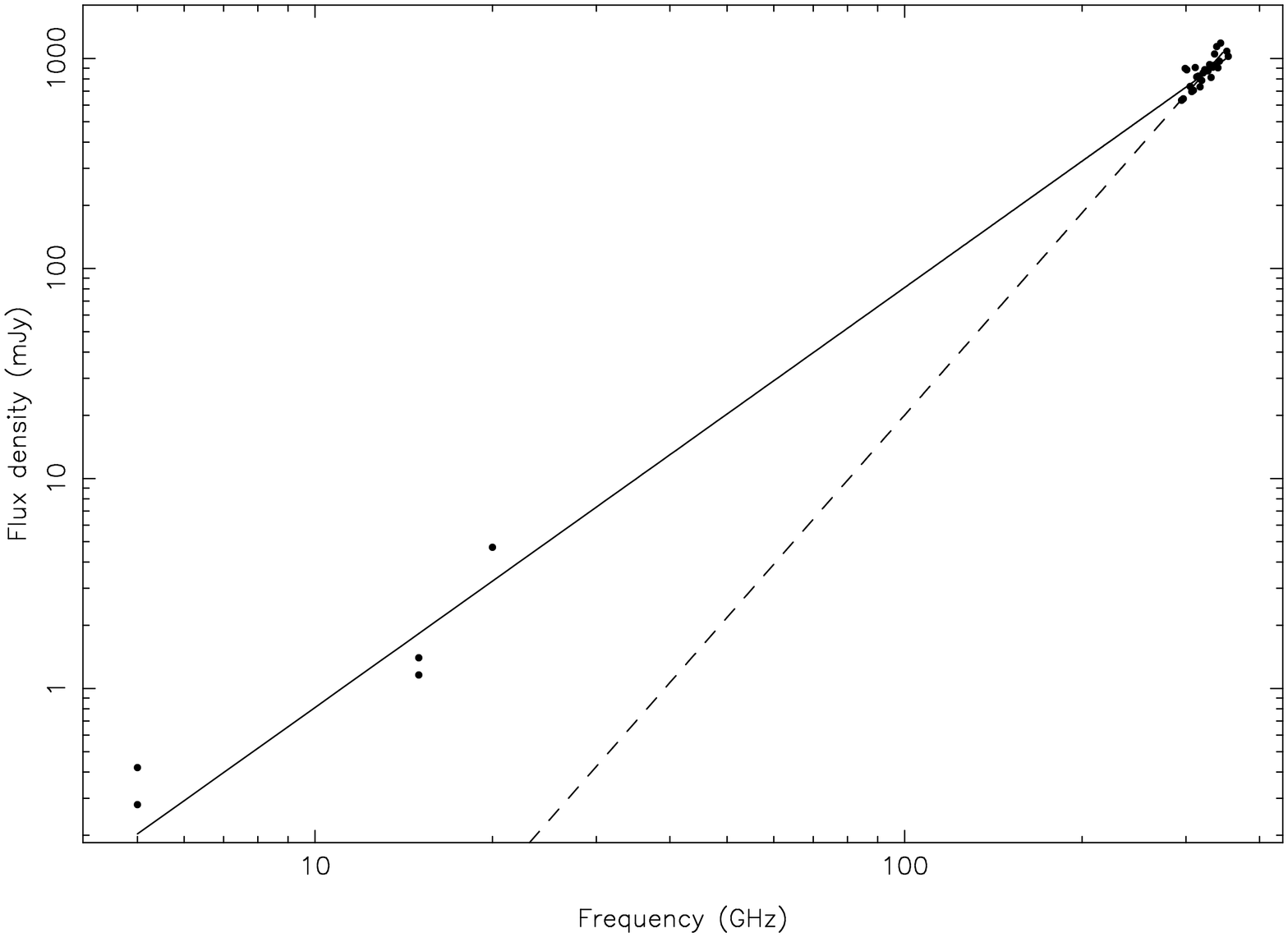}
\caption{Continuum flux density vs frequency including measurements at cm wavelengths from \cite{ReidAndMenten1997}. The solid line shows a fitted curve for $S\propto \nu^{2}$ and the dashed line for $S\propto \nu^{3.2}$. \label{contvsfreq2}}
\end{figure*}

\clearpage

\begin{figure*}[tbH]
\centering
\includegraphics[angle=-90,width=6in]{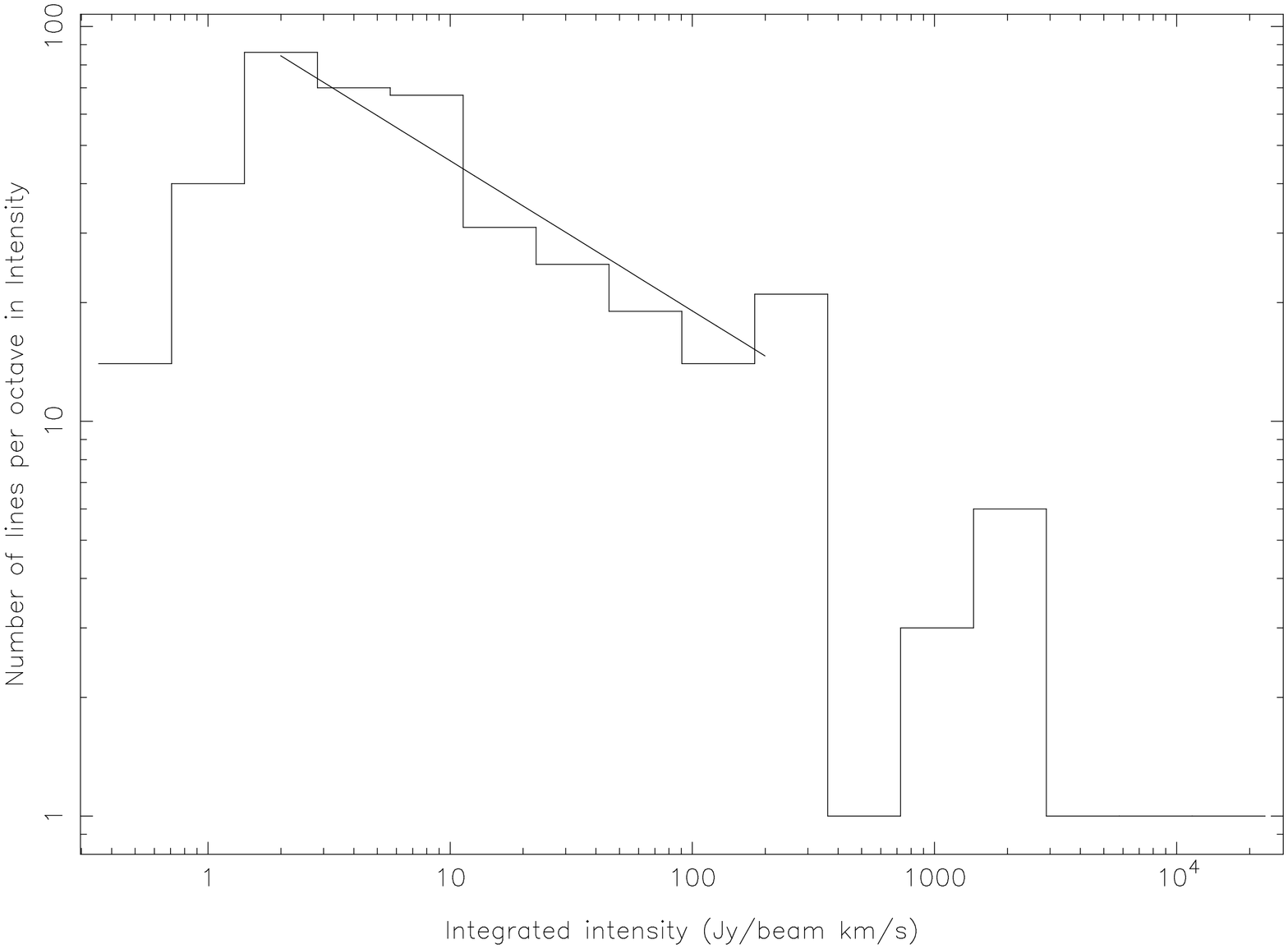}
\caption{Distribution of line intensities. The fitted line has a slope of -0.4 (excluding the strongest detected lines).  \label{weaklinesFlux}}
\end{figure*}

\clearpage

\begin{figure*}[tbH]
\centering
\includegraphics[angle=-90,width=6in]{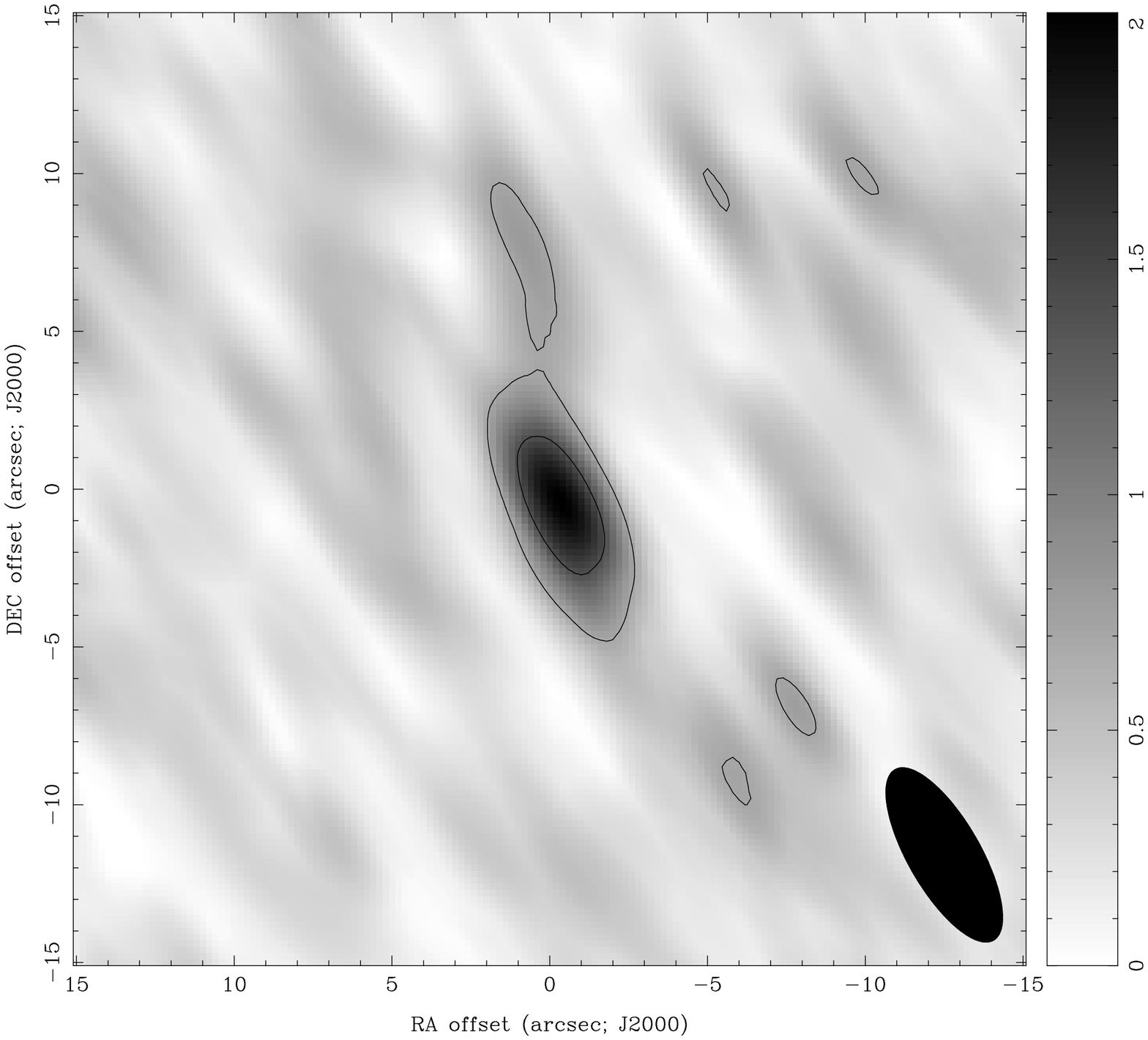}
\caption{Integrated intensity map of  $^{13}$CO $v=1$ J=3--2 emission toward IRC+10216, integrated over the velocity interval: $-$30 to $-$20 km s$^{-1}$. The synthesized beam is shown as an ellipse in the lower right corner. The starting contour level and interval is 5$\sigma$ with $\sigma=0.15$ Jy beam$^{-1}$ km s$^{-1}$. This is an example of a weak line that is not clearly seen in the full spectrum (Figure \ref{spectra}) but is clearly detected in the integrated intensity map.   \label{vibex13co3-2}}
\end{figure*}
\clearpage

\begin{figure*}[tbH]
\centering
\includegraphics[angle=-90,width=6in]{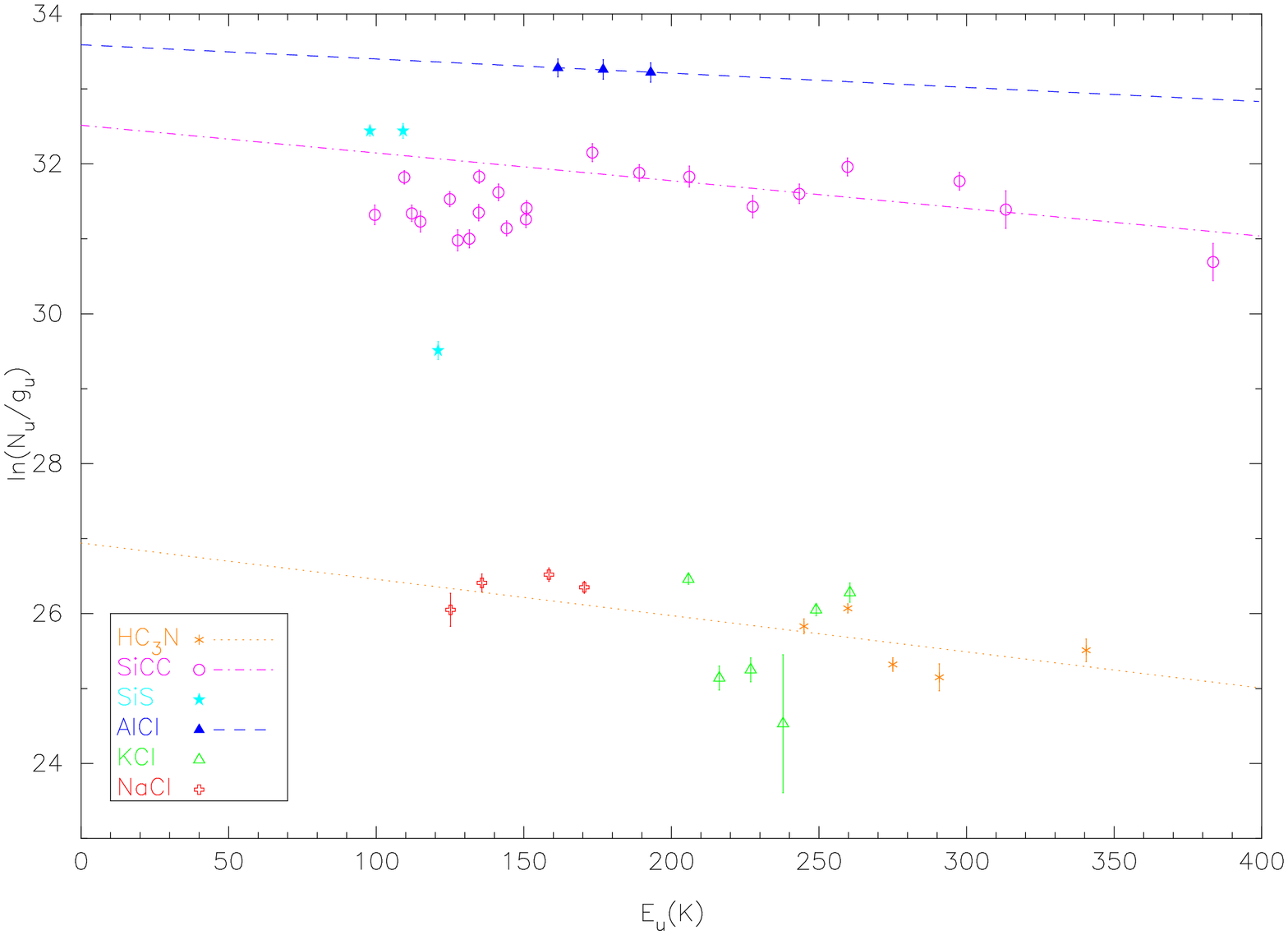}
\caption{Rotation temperature diagram. Only fits to SiCC, AlCl and HC$_{3}$N are reliable. For SiCC, lines at lower energies, the emission may be getting resolved, resulting in lower measured intensities. These points at E$_{u}<150$ K were excluded from the fit. \label{rottemp}}
\end{figure*}

\begin{figure*}[tbH]
\centering
\includegraphics[width=6in]{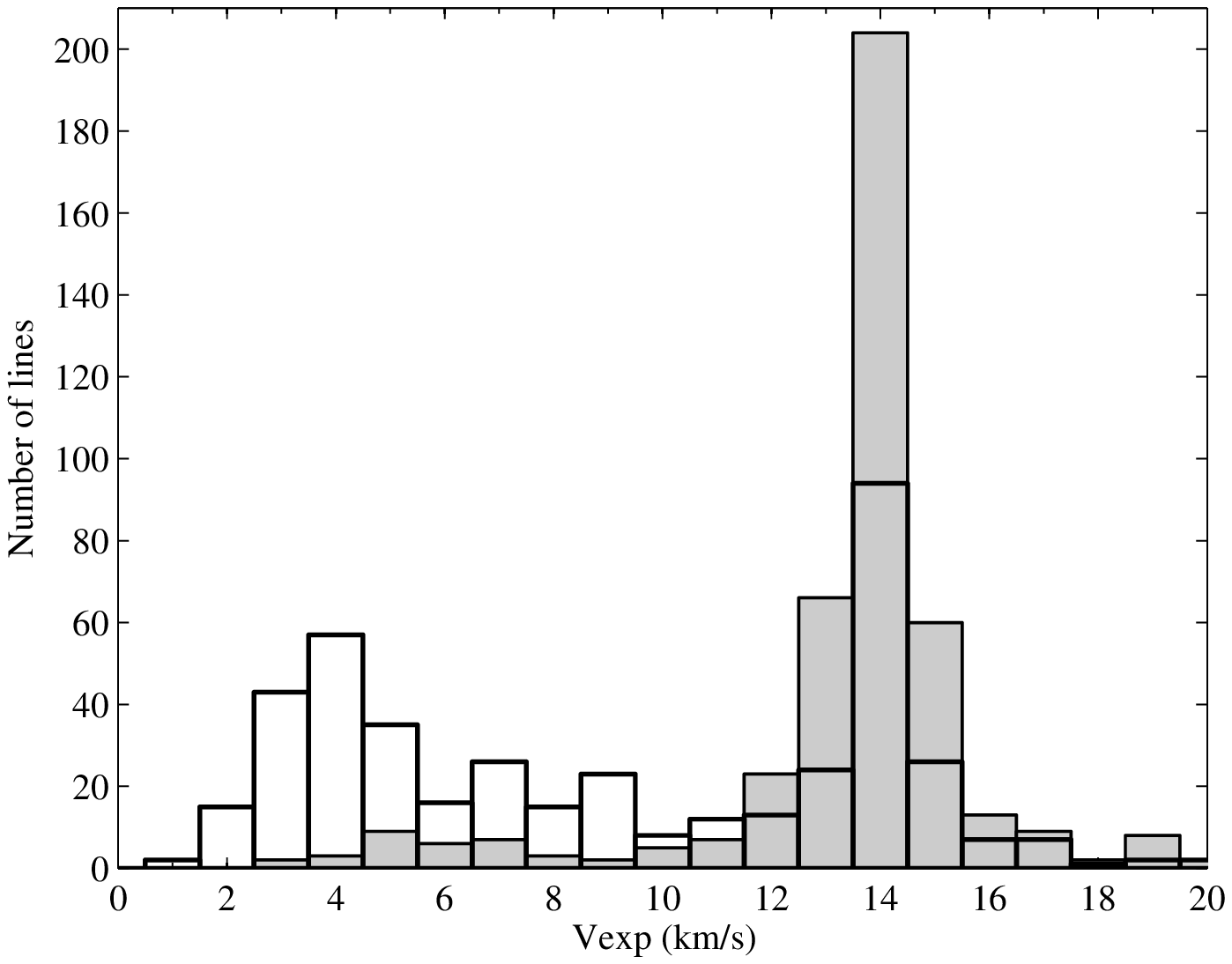}
\caption{Distribution of expansion velocities. The grey histogram shows the
data from \cite{TenenbaumEtAl2010}, representative of all previous single-dish line
surveys. The white histogram with bold outlines shows the distribution of
$V_{exp}$ from the SMA line survey. Both distributions peak at the terminal
velocity of 14 km s$^{-1}$ but a new population of narrow lines, peaking at
$\sim$ 4 km s$^{-1}$ is seen. Lines with $V_{exp}<10.0$ km s$^{-1}$ are
likely arising in the acceleration region of the inner envelope.
\label{histogramvexp}}
\end{figure*}

\begin{figure*}[tbH]
\centering
\includegraphics[width=6in]{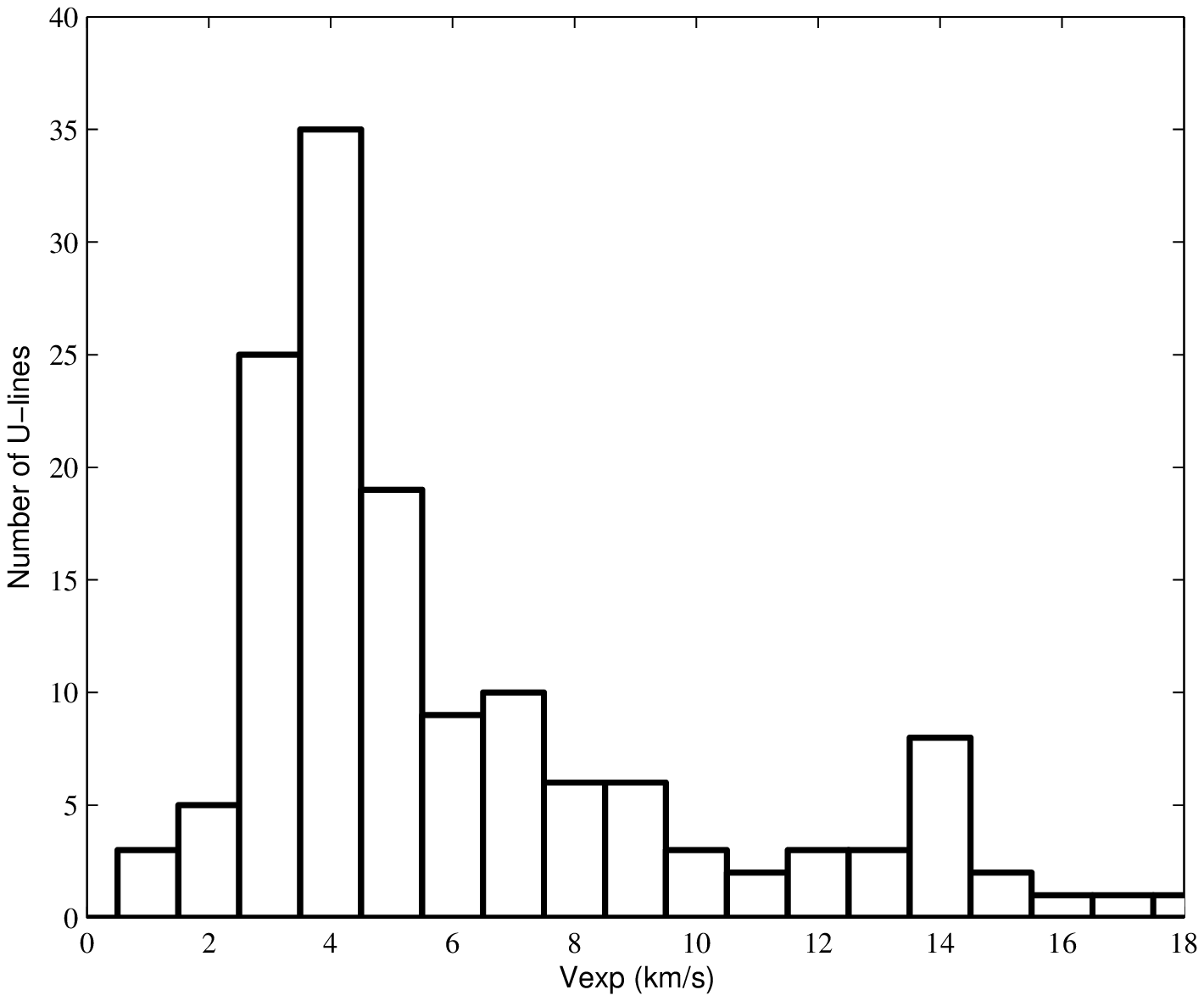}
\caption{Distribution of expansion velocities for the unassigned lines.  Most of the lines are narrow, peaking at $V_{exp}\sim4$ km s$^{-1}$. 
\label{histogramvexpUlines}}
\end{figure*}

\begin{figure*}[tbH]
\centering
\includegraphics[angle=-90,width=6in]{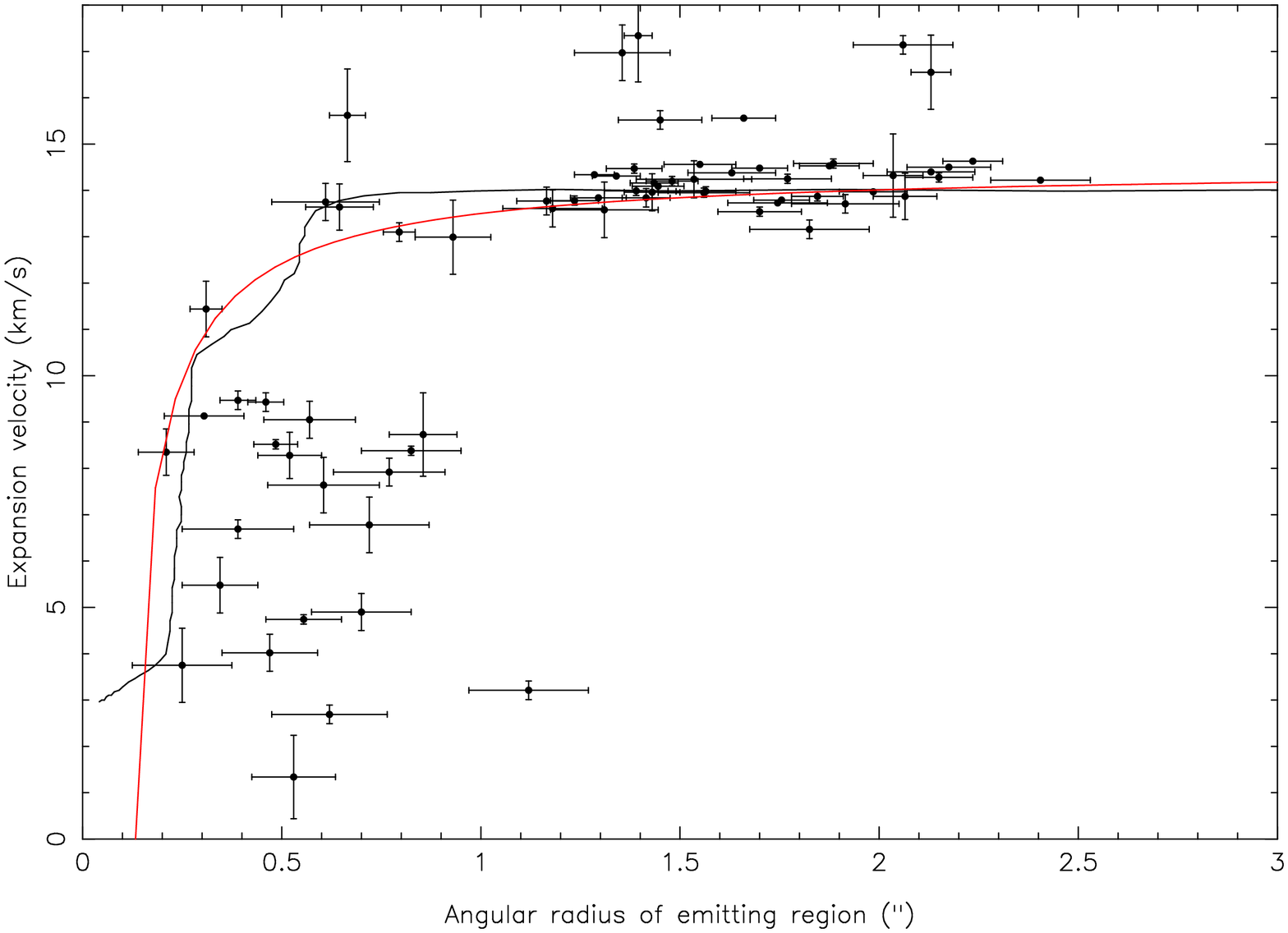}
\caption{Expansion velocity as a function of radius. The black curve is from
\cite{KeadyAndRidgway1993}. A subset of 71 out of 440 observed lines which
have reliable measurements of V$_{exp}$ and de-convolved angular sizes were
plotted. See text for details. The red curve is a plot of Equation~2. \label{vexpvsr}}
\end{figure*}

\clearpage

\end{document}